\pdfoutput=1
\documentclass{article}

\usepackage{arxiv}

\usepackage[utf8]{inputenc} 
\usepackage[T1]{fontenc}    
\usepackage{hyperref}       
\usepackage{url}            
\usepackage{booktabs}       
\usepackage{amsfonts}       
\usepackage{nicefrac}       
\usepackage{microtype}      
\usepackage{lipsum}

\usepackage{graphicx}
\usepackage{titlesec}
\usepackage{mathtools}
\usepackage{amsmath}
\usepackage[english]{babel}
\usepackage{blindtext}
\usepackage{algorithm}
\usepackage{algpseudocode}

\titleformat{\paragraph}
{\normalfont\normalsize\bfseries}{\theparagraph}{1em}{}
\titlespacing*{\paragraph}
{0pt}{3.25ex plus 1ex minus .2ex}{1.5ex plus .2ex}

\title{Control Strategies for Microgrids with Renewable Energy Generation and Battery Energy Storage Systems}

\author{
  Wenhao Zhuo\thanks{This work was in part supported by the Australian Research Council} \\
  School of Electrical Engineering and Telecommunications \\
  The University of New South Wales\\
  \texttt{w.zhuo@student.unsw.edu.au} \\
}

\begin{document}
\maketitle

\begin{abstract}
In this report, several control strategies for microgrids with renewable energy resources and battery energy storage systems are proposed. Renewable energy has the potential to reduce global carbon emissions, while higher penetration of renewable energy is hindered by the inherent variability and intermittency of renewable resources. This motivates the development of microgrids supplied by renewable energy resources. With proper control of storage units and communications with the energy market, the non-dispatchable energy can be regulated to reduce the difficulty of power scheduling in the grid operation. Battery energy storage systems (BESS) are widely used to make renewable energy more controllable and usable on demand. With suitable prediction techniques and control algorithms, BESS can store part of the generated renewable energy and supply some stored energy at different periods, so that the actual power dispatched can meet the required operating criteria. The control algorithms introduced in this report are developed to maximize the profit and minimize the energy cost in microgrids, which require predicted data on renewable power, temperature, electricity price, and load demand.
\end{abstract}


\section{Introduction} 
Due to the growing concern about sustainability and demand on energy, renewable generations are receiving more interests from governments, researchers and investors, which leads to an increase in the number of renewable power systems integrated into the current electrical grids. The~penetration of renewable energy is mostly hindered by their variability and intermittency. In order to increase the dispatchability of renewable power, the battery energy storage system (BESS) is used in many previous studies, which is essential for controlling the actual power dispatched to the local customers and the grid. Utilizing forecasting data on renewable power and power demand to arrange BESS actions over different periods, the power constraints and other operating parameters in the power grid can be satisfied. Also, in a deregulated energy market with variable electricity price, the profit of power trading with the utility grid can be maximized with appropriate charge\char`\/discharge decisions over different price intervals. Higher penetration of renewable energy can be achieved by incorporating more renewable generations into these decentralized power system, which can reduce the level of uncertainties imposed on the electrical grid, and bypass limitations due to a congested power transmission network \cite{intro1}.

Therefore, the main focus of our research are optimal energy management problems for microgrids comprising distributed renewable energy resources and battery energy storage systems. In this report, we introduce several optimal control strategies of renewable power systems to overcome two major challenges that are insufficiently considered in previous studies, namely battery lifetime degradation and computation efficiency.

\subsection{Battery lifetime degradation}

Due to the uncertainty of renewable energy, batteries used in associated optimal control problems are likely to experience more charge and discharge cycles than regular tasks, which can accelerate the degradation of batteries, leading to an increased operating cost of batteries. As a result, the operating cost of batteries will be a key factor to achieve an optimal dispatch of renewable generations. We believe that some functions reflecting the degradation of batteries due to charge/discharge actions should be considered in these problems. 

Over~the past decades, battery lifetime and degradation modeling have been investigated extensively, partly due to the increasing interest in the electric vehicle, energy arbitrage, and~renewable power applications from researchers and investors {\cite{r17}}. The~empirical models are believed to be appropriate for BESS planning and have been studied previously {\cite{r17,r18,r20}}. These models are developed based on degradation experiments, in~which batteries are cycling at different operating conditions until they reach the end of life. The~number of charge--discharge cycles that the battery can experience will be counted, and~the effects of stress factors, such as temperature, depth of discharge (DOD), average state of charge (SOC), and~operating currents on the rate of life cycle loss can be investigated.

With these models, calculating the cost of battery cycling would become identifying the number of cycles and their corresponding stress factors. For~all cycles and half-cycles (charge/discharge events), the~corresponding operating parameters such as DOD, temperature, and~current will be compared with a predefined operating condition measured from empirical experiments to determine their actual consumption on the battery life, and~the resulting cost can be identified. However, according to the authors of, in~the case of scheduling the optimal operation of BESS, the~relationship between the number of cycles and battery actions can only be expressed analytically with complicated forms and would be difficult to use with an optimization solver. In~other words, the~models used to calculate the cost of battery cycling would require some algorithms to search the operating cycles of batteries, which are analytical methods that cannot be described with mathematical equations, therefore would be difficult to include in the objective function of optimization problems. As~a result, these models are generally used for assessment rather than planning.

\subsection{Real-Time Considerations}

Since predictive estimates on a number of factors, such as renewable power, electricity price, and load demand are considered in previous studies to determine the optimal controls, the corresponding forecasting errors cannot be avoided. To improve the accuracy of results, the receding horizon approach is utilized in many previous studies {\cite{c5fengji1,r2,r3,r4,r5,r6,r8,r7}}. The basic idea is to constantly update the predicted data over a fixed future interval on every control step. The optimization will be conducted at every control step to determine an optimal control sequence over the corresponding forecasting horizon. Only the first component of the control sequence will be used as the actual input to the system in the current step, and the optimization will repeat in the following step with updated system states and predicted data. 

Nevertheless, this approach has a strict computation time requirement in real-time. The optimization algorithm should provide the solution within each control stage, while delays in communication and response time of power electronics should be considered as well. In the case of a large-scale power system, the number of components to manage, such as distributed generators, controllable loads, and storage units, can be enormous, and the evolution of the whole system will be significantly affected by the uncertainties caused by the renewable generations integrated. As a result, conventional optimization solvers such as dynamic programming, stochastic mixed integer programming are unlikely to meet the time constraint in the receding horizon control.

\subsection{Organization}

The rest of the report is structured as follows: {\bf Section 2} provides a critical review on previous studies in the field of energy management in microgrids. {\bf Section 3} presents a novel cost model of batteries that correlates the charge\char`\/discharge actions of batteries with the cost of battery life loss, which is compatible with  commercial optimization solvers and can be used in power scheduling of microgrids. {\bf Section 4} introduces an approximate dynamic programming algorithm that can be used in energy management problems in microgrids. The use of the approximate technique is to meet the time constraint in real-time for problems that are complex and stochastic. {\bf Section 5} introduces a novel control strategy of a microgrid with a distributed generator (DG), a battery energy storage system, a solar photovoltaic (PV) system and thermostatically controlled loads (TCLs). The proposed strategy is efficient and totally decentralized, where sliding mode controllers \mbox{\cite{smc1,smc2,smc3}} are utilized in the subsystems of TCLs to regulate the indoor temperature of residential buildings in the microgrid at a minimum cost. The strategy can be applied in real-time for large-scale communities. {\bf Section 6} proposes a control strategy to smooth the fluctuation of renewable energy using BESS and TCL, with the purpose of providing high-quality renewable power under low costs. {\bf Section 7} extends the energy management problem to a network of microgrids, and an approximate dynamic programming algorithm is developed to address such problem with a large number of states and stochastic variables. The last chapter concludes the report and provides directions for future studies.


\section{Literature review\label{cha:litreivew}}

Many control strategies and optimization methods, including model predictive control (MPC) \cite{r5,r8,r14,r15,finalshi1}, dynamic programming (DP) \cite{r3,r4,r7,r16}, sliding mode control (SMC) \cite{exr3,exr4}, reinforcement learning (RL) \cite{r9,r10}, particle swarm optimization (PSO) \cite{r11,exr1,exr2,exr5}, and mixed-integer linear programming (MILP) \cite{r6,r13,r15}, have been proposed for renewable power control under different conditions. The use of MPC is mainly due to forecasting errors. With real-time forecasting data within a short horizon that updated at every control step, the effect of errors can be reduced, and any mismatch in the supply and the demand can be identified promptly and solved in the following control step. In \cite{r5}, the wind power smoothing problem is formulated to optimize the maximum ramp rate and the battery state with wind power prediction. This model is further investigated in the case of frequency control due to disturbances in the supply-demand balance \cite{r8}. Authors of \cite{r14} present a MPC scheme to optimize microgrid operations while meeting changing request and operation constraints. The optimization problem is formulated using MILP that can be solved efficiently to meet the real-time operating constraints. An online power scheduling for microgrids with renewable generations, BESS, heating, and cooling units is presented in \cite{r15}. The MPC scheme is used with a feedback correlation to compensate for prediction errors. The energy management problem in microgrids can be perceived as controlling different units over multiple time periods under uncertainty, which can be considered as a Markov decision process (MDP) and solved with DP. In \cite{r3}, the energy management problem in microgrids with renewable power from six different generation sites is considered. The optimization problem is formulated based on different energy prices and solved with DP. A decentralized energy management strategy in microgrids with thermostatically controlled loads, solar power, distributed generators, and BESSs is proposed in \cite{r3}, which can determine the optimal controls for BESS and distributed generator to minimize the energy cost while maintaining the desired temperature in local buildings. An optimal dispatch strategy for grid-connected wind power plant with BESS is proposed in \cite{r7}, in which the DP algorithm used can incorporate the prediction of wind power and electricity price simultaneously to determine the optimal controls for BESS to maximize the profit. Authors of \cite{r16} develop a recursive DP algorithm to solve the optimal power flow in a microgrid considering limits on storage devices, network currents, and voltages. In the SMC scheme, a desired trajectory in the system will be defined, and the control objective is to track it. In \cite{exr3}, a decentralized SMC-based strategy to improve the performance of microgrids with renewable generations, BESS, non-linear, and unbalanced loads. The sliding surfaces used are predefined trajectories for active/reactive power to minimize fluctuations, compensate negative sequence, and harmonic currents. They further investigate the issue of stability and power-sharing in hybrid AC/DC microgrids with a similar control scheme in \cite{exr4}. The population-based algorithm, PSO, is used in some studies to optimize the power scheduling in microgrids. A day-ahead multi-objective optimization dispatch in a microgrid considering costs from batteries and carbon emissions is solved with a PSO algorithm in \cite{r11}. The algorithm is also applied in microgrids with hybrid renewable generation units to address the optimal power management problems that are complicated, which is to optimize objectives including the annual cost of the system, loss of load expected, loss of energy expected, system costs of investment, replacement, operation, and maintenance, see, e.g.(\cite{exr1,exr2,exr5}). In addition, there are some learning-based algorithms proposed for microgrids. Authors of \cite{r9} formulate the energy management problem in a microgrid as a Markov Decision Process to minimize the daily cost and address it with deep learning. A Q-learning based operation strategy to maximize the profit of a microgrid with community battery storage system is proposed in \cite{r10}, which combines both centralized and decentralized approaches to control the system units. 

As for the cost of battery degradation, battery lifetime and degradation modeling have been investigated extensively due to the increasing interest in the electric vehicle, energy arbitrage, and renewable power applications from researchers and investors {\cite{r17}}. According to {\cite{r18}}, these models can be categorized into theoretical models and empirical models. The theoretical models are constructed based on the chemical mechanism in the battery cell, such as the aging of active material, chemical decomposition, and surface film modification {\cite{r19}}. In the case of operation planning, authors of {\cite{r18}} believe that the theoretical models may not be suitable as the chemical reaction processes inside the cells can be difficult to correlate with charge and discharge actions. On the other hand, the empirical models are believed to be appropriate for BESS planning and have been studied previously \mbox{\cite{r17,r18,r20}}. These models are developed based on degradation experiments, in which batteries are cycling at different operating conditions until they reach the end of life. The number of charge-discharge cycles that the battery can experience will be counted, and the effects of stress factors, such as temperature, depth of discharge (DOD), average state of charge (SOC), and operating currents on the rate of life cycle loss can be investigated. It should be pointed out that most of the cost models of BESS in previously reviewed studies are either not considered \mbox{\cite {r3,r5,r6,r7,r8,r9,r10,r14,r15,r16}} in the optimization problem, or simple models that without consideration on the cycling model under different stress factors \mbox{\cite {r4,r11,r13}}.With these models, calculating the cost of battery cycling would become identifying the number of cycles and their corresponding stress factors. For all cycles and half-cycles(charge/discharge events), the corresponding operating parameters such as DOD, temperature, and current will be compared with a predefined operating condition measured from empirical experiments to determine their actual consumption on the battery life, and the resulting cost can be identified. However, according to the authors of {\cite{r17}}, in the case of scheduling the optimal operation of BESS, the relationship between the number of cycles and battery actions can only be expressed analytically with complicated forms and would be difficult to use with an optimization solver. 


\section{Profit Maximizing Control of Microgrids} \label{cha:chap3}

This chapter presents an optimal control strategy for grid-connected microgrids with renewable generation and battery energy storage systems (BESS), which is mostly based on our previous publication \cite{r4.1}. In order to optimize the energy cost, the proposed approach utilizes predicted data on renewable power, electricity price, and load demand within a future period, and determines the appropriate actions of BESSs to control the actual power dispatched to the utility grid. We formulate the optimization problem as a Markov decision process and solve it with a dynamic programming algorithm under the receding horizon approach. The main contribution in this chapter is a novel cost model of batteries derived from their  life cycle model, which correlates the charge/discharge actions of batteries with the cost of battery life loss. Most cost models of batteries are constructed based on identifying charge--discharge cycles of batteries on different operating conditions, and the cycle counting methods used are analytical, so cannot be expressed mathematically and used in an optimization problem. As a result, the cost model proposed is a recursive and additive function over control steps that will be compatible with dynamic programming and can be included in the objective function. 

\subsection{Introduction} \label{cha:chap3s1}

Due to the growing concern on sustainability and demand on energy, renewable generations are receiving more interests from governments, researchers and investors, which leads to an increase in the number of renewable power systems integrated into the current electrical grids. The penetration of renewable energy, however, is mostly hindered by their variability and intermittency, which motivates the development of microgrids supplied by renewable power systems. Either grid-connected or islanded, these decentralized power systems are believed to be the promising solution to achieve higher penetration of clean energy in the future \cite{r2}. Given proper control of storage units and communications with the electricity market, the non-dispatchable renewable power can be smoothed and used on demands, therefore reducing the difficulty of power scheduling in the main grid operation \cite{r3}. 

In many studies that focus on the control of renewable power systems \cite{r4,r5,r6,r7,r8,r9,r10,r11,r12,r13,r14,r15}, the battery energy storage system (BESS) is essential for controlling the actual power dispatched to the local customers and the grid. Utilizing forecasting data on renewable power and power demand to arrange BESS actions over different periods, the power constraints in the microgrid and the operating parameters in the main grid can be satisfied. Also, in a deregulated energy market with variable electricity price, the profit of power trading with the utility grid can be maximized with appropriate charge\char`\/discharge decisions over different price intervals. For instance, part of the renewable power can be used to meet the local demand, and the remaining can either be charged to the BESS or sold to the market. In addition, operators of the microgrid can purchase some energy from the main grid at a low price and stored for further use during high price intervals.

A significant challenge in these studies is the existence of forecasting errors \cite{c3exr1,c3exr2,c3exr3}. To optimize the operation of microgrids, predicted data on renewable power, load demand, and electricity price within a future period ranging from hours to days would be required, and the forecasting errors cannot be avoided. A relatively small level of errors would be acceptable, which can be addressed with some online ancillary services such as fast-response generators or operating reserves, while a large error can be detrimental to both microgrid and the main power grid. Another issue is related to battery cost\cite{c3exr4,c3exr5,c3exr6,c3exr7}. In these studies, batteries used are likely to experience more charge and discharge cycles than regular tasks, which can accelerate the degradation of batteries, leading to an increased operating cost of batteries. Therefore, the additional cost resulted from a lower than expected lifetime of batteries should not be neglected, and some functions reflecting the cost of cycling batteries due to charge/discharge actions should be included to achieve the optimal dispatch.

The cost of batteries can a key factor to achieve an optimal dispatch in microgrids with renewable generation units \cite{c3exr8,c3exr9,c3exr10,c3exr11,finalthomas1}. Over~the past decades, battery lifetime and degradation modeling have been investigated extensively, partly due to the increasing interest in the electric vehicle, energy trading, and renewable energy applications from researchers and investors {\cite{r17}}. According to {\cite{r18}}, these battery models can be categorized into theoretical models and empirical models. The~theoretical models are constructed based on the chemical mechanism in the battery cell, such as the aging of active material, chemical decomposition, and~surface film modification~ {\cite{r19}}.  Regarding the energy management problems, authors of {\cite{r18}} believe that the theoretical models may not be suitable as the chemical reaction processes inside the cells can be difficult to correlate with charge and discharge actions. On~the other hand, the~empirical models are believed to be appropriate for BESS planning and have been studied previously \cite{r17,r18,r20}. These models are developed based on degradation experiments, in~which batteries are cycling at different operating conditions until they reach the end of life. The~number of charge--discharge cycles that the battery can experience will be counted, and~the effects of stress factors, such as temperature, depth of discharge (DOD), average state of charge (SOC), and~operating currents on the rate of life cycle loss can be investigated. It should be pointed out that most of the cost models of BESS in previously reviewed studies are either not considered \cite {r3,r5,r6,r7,r8,r9,r10,r14,r15,r16} in the optimization problem, or~simple models that without consideration on the cycling model under different stress factors \cite {r4,r11,r13}.

With these models, calculating the cost of battery cycling is equivalent to determining the number of cycles and their corresponding stress factors. For~all cycles and half-cycles (charge/discharge events), the~corresponding operating parameters such as DOD, temperature, and~current will be compared with a predefined operating condition measured from empirical experiments to determine their actual consumption on the battery life, and~the resulting cost can be identified. However, according to the authors of \cite{r17}, in~the case of scheduling the optimal operation of BESS, the~relationship between the number of cycles and battery actions can only be expressed analytically with complicated forms and would be difficult to use with an optimization solver. In~other words, the~models used to calculate the cost of battery cycling would require some algorithms to search the operating cycles of batteries, which are analytical methods that cannot be described with mathematical equations, therefore would be difficult to include in the objective function of optimization problems. As~a result, these models are generally used for assessment rather than planning.

As a result, we aim to develop a control strategy of BESS in a grid-connected microgrid to optimize the cost with consideration on the cost from the battery cycling model. Based on previous studies, the optimization problem is formulated as a MDP, and a dynamic programming algorithm is utilized to optimize the overall cost in the microgrid over each planning horizon. We also employ the receding horizon approach to minimize the effect of forecasting errors and real-time mismatches; therefore only the first action solved will be used as the actual input to the system in the current stage, and the algorithm will repeat in the next step with updated system states and predicted data. 

The main contribution of our work is a novel cost model of battery based on the empirical battery life cycle models. The proposed cost model is recursive and additive over the control stages and can be included in the objective function of the optimization problem. It is also compatible with the DP algorithm that can be used to solve MDP problems, which does not require additional computations than regular DP algorithms.

\subsection{Problem statement}\label{chapter3s2}

\begin{figure}
  \centering
  \includegraphics[width=13 cm]{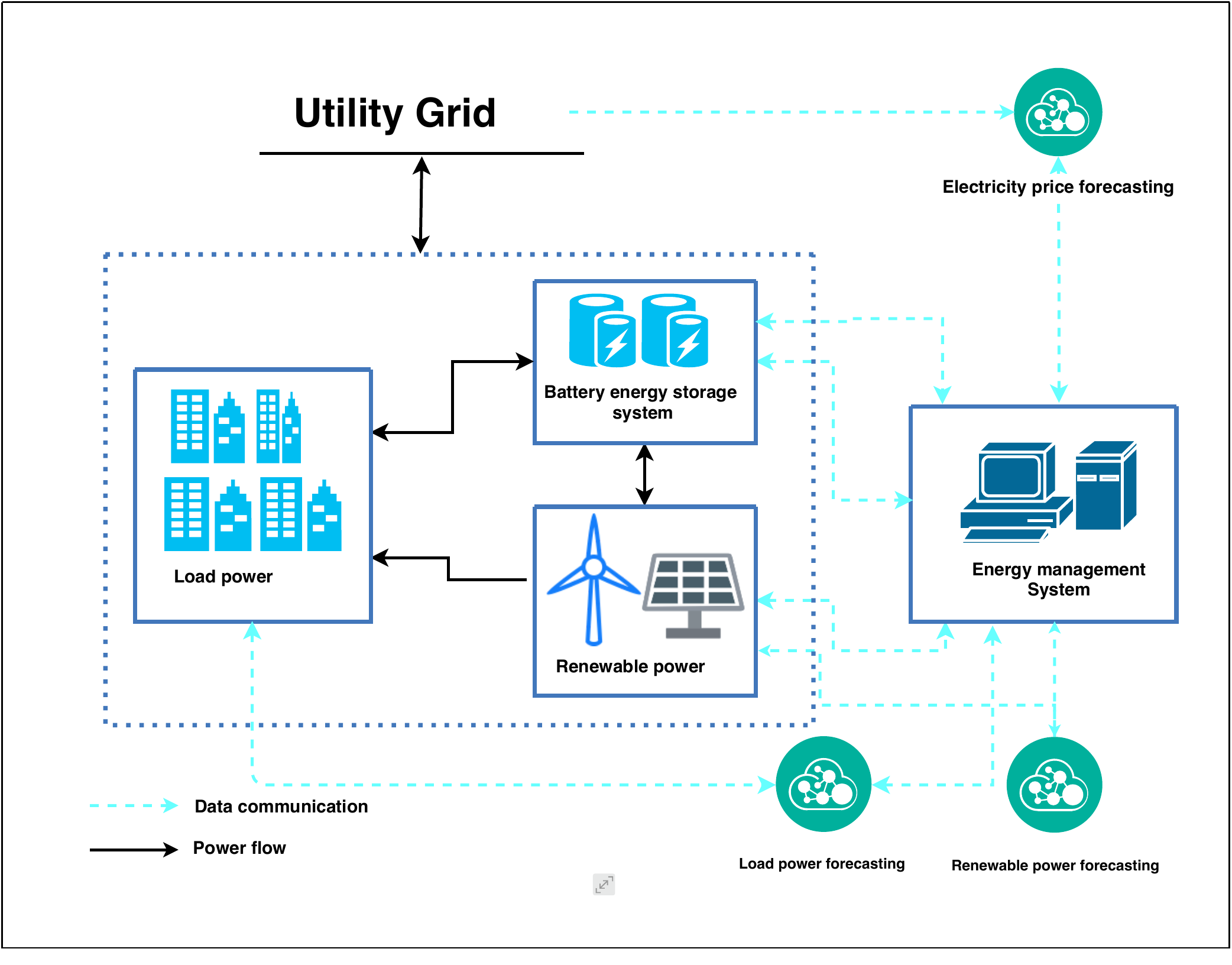}
  \caption{Configuration of the~microgrid.}
  \label{figure3.1}
\end{figure}

We consider a grid-connected microgrid consisting of a battery energy storage system (BESS) and a renewable generation unit. A~deregulated energy market environment is assumed, and~the microgrid can communicate with the utility grid to conduct energy trading by controlling the actions of the BESS. The~objective is to minimize the operating cost of BESS and maximize the profit of the microgrid. A~configuration of the microgrid can be seen in Figure~\ref{figure3.1}.

The receding horizon approach is utilized, and~predicted data on renewable power, electricity price, and~power demand over every planning horizon are updated at each control step. The~cost of power trading with the energy market, and~the cost of battery cycling over a planning horizon are used as the cost function of the problem, and~a dynamic programming algorithm is used to solve the optimization problem over each horizon. Only the first action solved in each horizon will be used as the actual input to the system, and~the algorithm will repeat in the next step with updated forecasting data and system~states. 

The energy stored in the BESS is considered as the system state, and~the charge/discharge power is the control input. We assume multiple batteries are coordinated and charged/discharged with the same amount of power, which is to avoid the long string structure that can lead to charge imbalance~\cite{r22} and increase the amount of controllable power.

\subsection{System Model and~Costs}\label{s3}
\unskip
 \subsubsection{Microgrid~Model}\label{s3.1}
 We consider our system as a discrete-time system with the sampling rate \(\delta>0\), and~most data used are piecewise constant with constant values over periods \([n\delta,(n+1)\delta)\), where $n$ is a non-negative~integer. 
 \paragraph{Predicted~Data}\label{s3.1.1}
   It is assumed that three sets of predicted estimates of renewable power, load power demand, and~electricity price can be made over every planning horizon, which is supposed to be piecewise constants. We use $P_{rn} (k)$, $P_{ld} (k)$ and~$C_{ele} (k)$ to represent the renewable power, load power demand, and~electricity price predicted over the period \([k\delta,(k+1)\delta)\) respectively. As~the model predictive control approach is utilized, these forecasting data will be updated prior to each control step, and~fast predicting technique for a short period would be required,  see, e.g.,~\cite{r24}. Also, we assume that the electricity price is identical for both selling and~buying.
 \paragraph{Battery Energy Storage~System}\label{s3.1.2}
 The dynamics of the battery~\cite{r13} used in the microgrid is described by the following equation.
\begin{equation}
  \label{eq1}  
E_B (k+1) = E_B (k)-P_B(k)\Delta \delta - d|P_B(k)\Delta \delta|,
\end{equation}
where $E_B (\cdot)$ is the energy state of the battery, $\Delta \delta$ is the factor used to convert power to energy based on the actual time of each control step, $P_B (\cdot)$ is the charge/discharge power of the BESS, $d>0$ is the charging/discharging loss factors of the BESS. Based on the model, $P_B (\cdot)>0$  indicates the discharging action and $P_B (\cdot)<0$ indicates the charging action of BESS. In~addition, we assume n batteries integrated into the microgrid, which are controlled and balanced simultaneously with the same amount of charge/discharge~power.

The following constraints limit the operation of the battery,
\begin{equation}
  \label{eq2}  
\\E_B^{min} \leq \\E_{B}(k)\leq\\E_B^{max}\ 
\end{equation}
\begin{equation}
  \label{eq3}  
\\-P_B^{min} \leq \\P_{B}(k)\leq\\P_B^{max}\ 
\end{equation}
where $E_B^{min}$ and $E_B^{max}$ are the lower and upper limits of the battery energy state, which are used to avoid overcharge and over-discharge. $P_B^{max}$ and $P_B^{min}$ are the maximum amounts of power that can be charged and discharge from the battery within a control~step.
 \paragraph{Power Balancing in the~Microgrid}\label{s3.1.3}
Let $P_G$ be the power exchanged with the energy market, and~the following equation can represent the power balancing task in the microgrid.
\begin{equation}
  \label{eq4}  
P_G (k) = P_{rn} (k)-P_{ld} (k) + n\cdot P_{B} (k).
\end{equation} 


\subsubsection{Battery Aging and Cost~Models}\label{s3.2}

\begin{figure}
  \centering
  \includegraphics[width=13 cm]{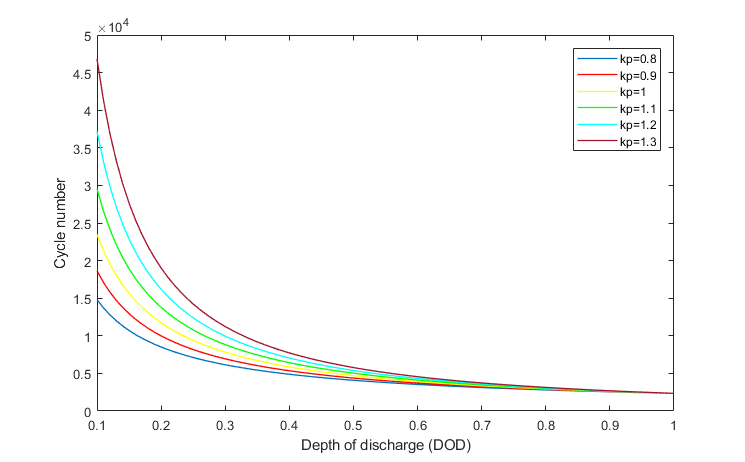}
  \caption{Maximum number of cycles the battery can experience at different~conditions.}
  \label{figure3.2}
\end{figure}

Our cost model of battery is built on the battery lifetime model used in~\cite{r17}. One significant feature of the model is to identify the ‘half-cycle’, or~the difference between two adjacent local extremes on the curve of the battery energy level. These ‘half-cycles’ contain information on the charge and discharge actions of the battery and can be used to assess its degradation. Another significant factor is the temperature, which has been studied in~\cite{r18,r20}. In~our case, we assume a constant operating temperature of the BESS controlled by specific cooling~devices.
\paragraph{Cycle Life~Model}\label{s3.2.1}
The basic cycle life model considered in~\cite{r17} is described as follows:
\begin{equation}
  \label{eq6}  
\\T_{cycle} = \dfrac{N_{d}^{fail}}{w \cdot n_{d}^{day}}\ 
\end{equation}
where $N_{d}^{fail}$ is the maximum number of charge--discharge cycles that the battery can experience at a specific DOD before its end of life, $n_{d}^{day}$ is the number of daily cycles that the battery experienced at the DOD, and~$w$ is the average number of operating days within a year. So $T_{cycle}$ denotes the estimated lifetime of the battery in years. Furthermore, $N_{d}^{fail}$ can be expressed as a function of the DOD by fitting the typical empirical data provided by the manufacturers, which is:
\begin{equation}
  \label{eq7}  
\\N_{d}^{fail} = f(d)=N_{100}^{fail}d^{-kp}\ 
\end{equation}
where $d$ is the DOD, $N_{100}^{fail}$ is the number of cycles at 100\% DOD, and~$kp$ is a constant. The~curves of cycle life versus DOD at different $kp$ values are presented in Figure~\ref{figure3.2}.

Assuming $n_d$ cycles of d DOD are experienced by the battery, its cycle life loss $Ls_{cycle}$(\%) can be described as:
\begin{equation}
  \label{eq8}  
\\Ls_{cycle} = \dfrac{n_d}{f(d)}\times100\%\ 
\end{equation}

As a result, with~the same rate of cycle life loss $Ls_{cycle}$, the~equivalent 100\%-DOD cycle number, indicated as $n_{100}^{eq}$, at~$d$ DOD with $n_d$ cycles can be derived from the following equation.
\begin{equation}
  \label{eq9}  
\\ \dfrac{n_{100}^{eq}}{N_{100}^{fail}}=Ls_{cycle}=\dfrac{n_d}{f(d)}\ 
\end{equation}

Substituting Equation~(\ref{eq7}) into (\ref{eq9}), the~equivalent 100\%-DOD cycle can be derived as:
\begin{equation}
  \label{eq10}  
\\n_{100}^{eq} = n_d\cdot d^{kp}\ .
\end{equation}

Based on the equation, the~DOD and the number of different cycles can be counted and converted to the equivalent value, and~if we assume that the battery would experience the similar pattern of operation within a certain period, the~corresponding lifetime can be estimated with Equation~(\ref{eq6}).

\paragraph{Counting Half~Cycles}\label{s3.2.2} 

  Instead of counting full cycles, authors of~\cite{r17} assume the battery completes a half-cycle between two adjacent local maximum and minimum of the energy level, which is also the switching point between a charge and a discharge action. Let $E_{max}$ be the rated energy capacity of the battery, and~$E_k$ be the energy level in the battery at the end of $k$-th half cycle, the~corresponding DOD indicated as $d_k^{half}$, can be described as:
\begin{equation}
  \label{eq11}  
d_k^{half}=|{\dfrac{E_k-E_{k-1}}{E_{max}}}|
\end{equation} 

Based on Equation~(\ref{eq11}), the~equivalent 100\%-DOD cycle number of K half cycles can be calculated~as:
\begin{equation}
  \label{eq12}
\\n_{100}^{eq} = \sum_{k=1}^{K} 0.5\cdot (d_k^{half})^{kp} \ 
\end{equation}  

\paragraph{Cycle Life Cost~Model}\label{s3.2.3} 

As the half cycle can be perceived as the individual charge/discharge action in-between local extremes, the~term $Ls_{cycle}$(\%), namely the cycle life loss percentage would be suitable to determine the battery life cost model. Considering a replacement cost $R_c$(\$) incurred at the end of the battery life, and~a cycle life loss percentage $Ls_{cycle}(w)$ within a period $w$, the~corresponding cost of battery lifetime consumption within the period will be the product of the two terms. In~our case, the~period W is the actual time of each planning horizon, and~the cycle life loss can be determined from battery actions within the horizon. For~a single half-cycle with depth $d_k^{half}$, the~cost of cycle life loss $C_{loss}$ can be derived as:
\begin{equation}
  \label{eq13}  
\\C_{loss} (d_k^{half}) = \dfrac{n_{100}^{eq}}{N_{100}^{fail}}R_c =\dfrac{0.5 \cdot (d_k^{half})^{kp}}{N_{100}^{fail}}R_c   \ .
\end{equation} 

Therefore, the~optimization problem can be formulated to minimize the cycle life cost and the energy cost to meet the power demand by making decisions on BESS actions at different electricity~prices.


\subsubsection{Redefine the Cost Model of Battery Cycle~Life}\label{s3.3} 
In this section, we propose a new model for the cost of cycle life loss that can be utilized in the dynamic programming~approach.

The dynamic programming approach is a recursive algorithm based on the Bellman principle of optimality~\cite{bk1}. It starts with the last step of the planning horizon and loops over the two adjacent states within the step. For~every former state, an~action that can optimize the cost within this control step will be determined,  and the corresponding cost will be memorized. The~algorithm will then proceed to the second last step of the planning horizon and repeat the same procedure. The~previously memorized values, which are now the cost of the latter states in the current control step, will be used to determine the optimal costs within the step. Once the algorithm reaches the first step in the planning horizon, the~optimal costs for all initial states over the horizon can be~retrieved.

One basic principle of using the approach is that the cost of the problem should be additive~\cite{bk2}. In~our case, the~cost of energy trading is related to the amount of power exchanged with the main power grid, which will be accumulated over control steps. However, the~costs from battery cycle life loss are determined by the local extreme points, which cannot be identified within every control step, as~the battery could undergo consecutive charge or discharge actions over multiple steps in practice. In~other words, the~cost of battery life loss, derivable from Equation~(\ref{eq13}), cannot be calculated from looping over adjacent states in the DP algorithm. As~a result, we propose a method to calculate the cost that can be applied in the DP~algorithm.

\begin{figure}
  \centering
  \includegraphics[width=13 cm]{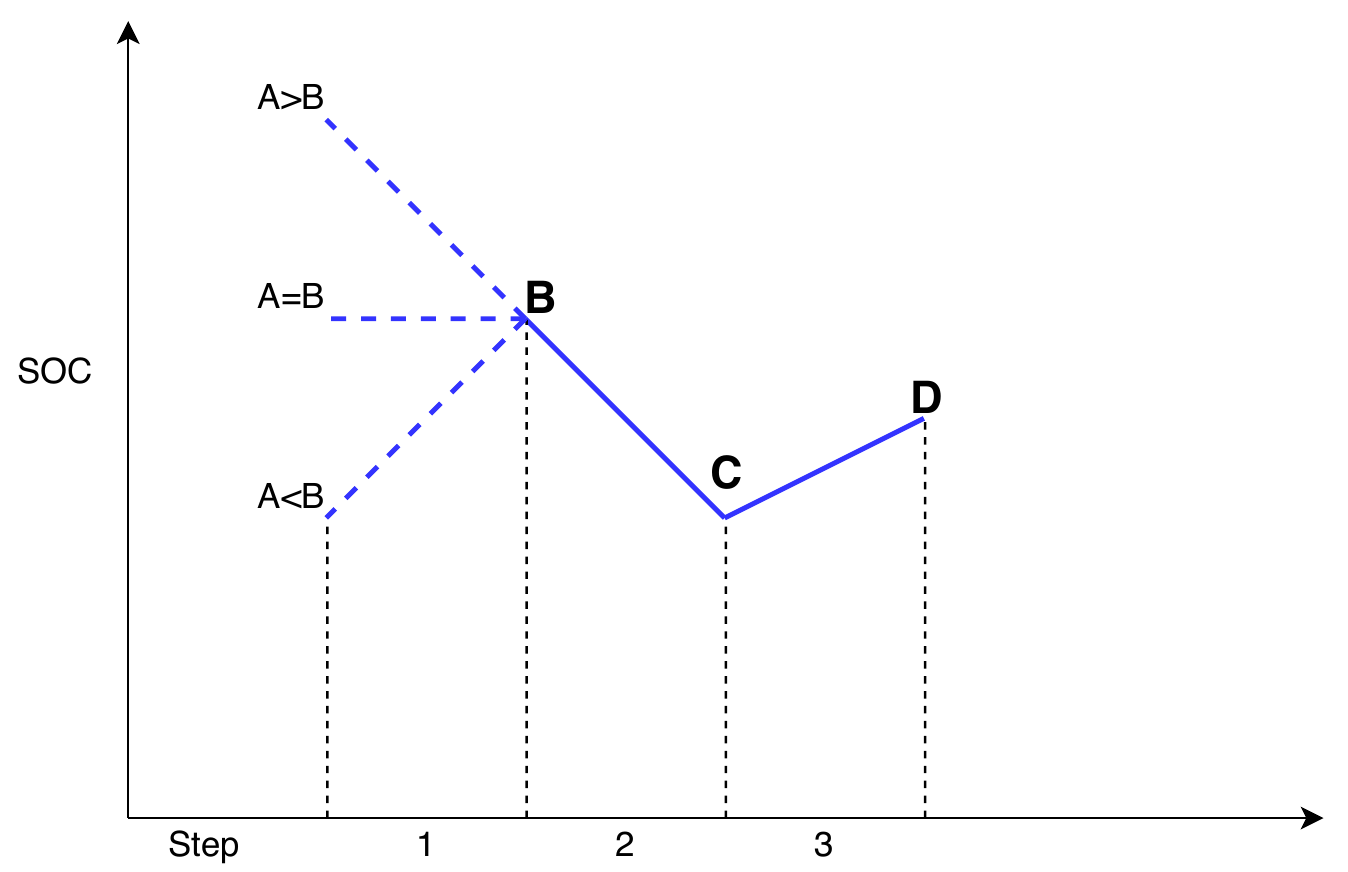}
  \caption{State of charge (SOC) profile used for~illustration.}
  \label{figure3.3}
\end{figure}

Referring to Figure~\ref{figure3.3}. Let BC and CD be the actions of BESS on step 2 and 3, and~A be the former state of SOC to be decided in step 1. Based on Equation~(\ref{eq13}), the~cost over step 2 and 3, or~the ‘cost-to-go’, will be $\dfrac {0.5R_c}{N_{100}^{fail}}( {B-C})^{kp}+\dfrac {0.5R_c}{N_{100}^{fail}}( {C-D})^{kp}$. If~$A$ is higher than $B$, the~cost over step 1, 2 and 3 will be $\dfrac {0.5R_c}{N_{100}^{fail}}( {A-C})^{kp}+\dfrac {0.5R_c}{N_{100}^{fail}}( {C-D})^{kp}$, as~$AB$ and $BC$ are two subsequent discharge actions with the local minimum $C$. On~the other hand, if~$A$ is lower than $B$, $AB$ will become a charge action, and~the cost over the three steps is $\dfrac {0.5R_c}{N_{100}^{fail}}( {A-B})^{kp}+\dfrac {0.5R_c}{N_{100}^{fail}}( {B-C})^{kp}+\dfrac {0.5R_c}{N_{100}^{fail}}( {C-D})^{kp}$. 
\smallbreak

Since only the cost within the current step will be evaluated by the dynamic programming algorithm, we propose the following method to calculate the life cycle cost: 

Considering a step k within the planning horizon of N steps, $k=0,1,\ldots ,N-1$. Based on the battery model introduced in Section \ref {s3.1}, $E_B (k)$ and $E_B (k+1)$ will be the former and latter state on step $k$, and~we define $\sigma_k$ as the ‘subsequent local extreme’ of step $k$, which is the following local extreme seen from the state $E_B (k)$, and~determined by the actions after step $k$. Then the cost of battery life cycle loss on step $k$ can be calculated as:
\begin{equation}
  \label{eq14}  
\\C_{loss} (k) = \dfrac{0.5R_c}{N_{100}^{fail}}( {\dfrac{E_B (k)-\sigma _k}{E_{max}}})^{kp}-\dfrac {0.5R_c}{N_{100}^{fail}}( {\dfrac{E_B (k+1)-\sigma _k}{E_{max}}})^{kp} \ 
\end{equation} 

  Referring to Figure~\ref{figure3.3}, it can be perceived that if $A\geq B$, $\sigma_1$ will be state $C$. Substituting the values into Equation~(\ref{eq14}), the~cost in this step will be $\dfrac {0.5R_c}{N_{100}^{fail}}({A-C})^{kp}-\dfrac {0.5R_c}{N_{100}^{fail}}({B-C})^{kp}$, and~the cost over the three steps is  $\dfrac {0.5R_c}{N_{100}^{fail}}({A-C})^{kp}-\dfrac {0.5R_c}{N_{100}^{fail}}( {B-C})^{kp}+\dfrac {0.5R_c}{N_{100}^{fail}}( {B-C})^{kp}+\dfrac {0.5R_c}{N_{100}^{fail}}( {C-D})^{kp}$, which is equal to $\dfrac {0.5R_c}{N_{100}^{fail}}(\ {A-C})^{kp}+\dfrac {0.5R_c}{N_{100}^{fail}}( {C-D})^{kp}$, the~result we discussed previously. Similarly, If $A<B$, $\sigma_1$ will be state $B$ and the cost within the step is $\dfrac {0.5R_c}{N_{100}^{fail}}({A-B})^{kp}-\dfrac {0.5R_c}{N_{100}^{fail}}({B-B})^{kp}$. The~cost over the three steps will become $\dfrac {0.5R_c}{N_{100}^{fail}}({A-B})^{kp}+\dfrac {0.5R_c}{N_{100}^{fail}}({B-C})^{kp}+\dfrac {0.5R_c}{N_{100}^{fail}}({C-D})^{kp}$,
  which coincides with the previous result.
  In the dynamic programming algorithm, an~extra array should be created to memorize the ‘subsequent local extreme’ for all states, and~a program to update this value should be included as well. We will discuss more on this in the next~section.


\subsection{Optimization~Technique}\label{s4}
\vspace{-6pt}
 
\subsubsection{Optimization~Problem}\label{s4.1} 
Based on the system defined in previous sections, the~control inputs are the BESS actions $P_B (k)$ in $n$ batteries, which determine the cost of battery life cycle cost $C_{loss}$ in Equation~(\ref{eq14}), and~the power exchanged from the microgrid with the energy market. To~state the optimization problem, we propose the following cost function based on Equations~(\ref{eq4}) and (\ref{eq14}).
\begin{equation}
  \label{eq15} 
    \sum_{k=0}^{N-1} nC_{loss}(k)-C_{ele}(k)P_G(k) 
\end{equation}

Then the optimal control problem can be stated as: given $P_{rn} (k)$, $P_{ld} (k)$, and~$ C_{ele} (k)$ for all $k$ in every planning horizon, find the control input $P_B (k)$ such that the constraints (\ref{eq2})$-$(\ref{eq6}) hold and the minimum of (\ref{eq15}) is~achieved.

To solve this problem, we introduce the Bellman function $V(k,E_B)$ as follows: For all \mbox{$k=0,1,\ldots,N-1$,} $E_B \in [E_B^{min},E_B^{max} ]$, $P_B \in [-P_B^{min},P_B^{max}]$,
\begin{align}
V(N,E_B)    := {} & 0 \quad \forall E_B\in[E_B^{min},E_B^{max}]
\end{align}
\begin{align}
V(k,E_B):= {} {V((k+1),E_B)+\min_{P_B}(nC_{loss}(k)-C_{ele}(k)P_{G}(k))}.
\end{align}

The algorithm can be solved recursively by starting from $k=N-1$ and computing $V(k,E_B)$ for all $E_B$. With~a given initial state $E_B (0)$, the~minimum of (\ref{eq15}) can be obtained when $k=0$, and~the optimal set of $P_B$ over the planning horizon of $N$ steps can be~retrieved. 


\subsubsection{Updating Local~Extremes}\label{s4.2} 

As the ‘subsequent local extreme’ $\sigma_k$, introduced in Section~\ref{s3.3}, is required to compute the cost of life cycle loss $C_{loss}$. Based on the Bellman function introduced previously, we can assign $\sigma_k$ for every $V(k,E_B )$, which can be considered as the next local extreme seen from the state $E_B (k)$. However, since $\sigma_{k-1}$ is required to calculate the values of $V(k-1,E_B)$, we will introduce a method to update $\sigma_{(k-1)}$ with $\sigma_k$ in a recursive order suitable for the dynamic programming~algorithm.

\begin{figure}
  \centering
  \includegraphics[width=12 cm]{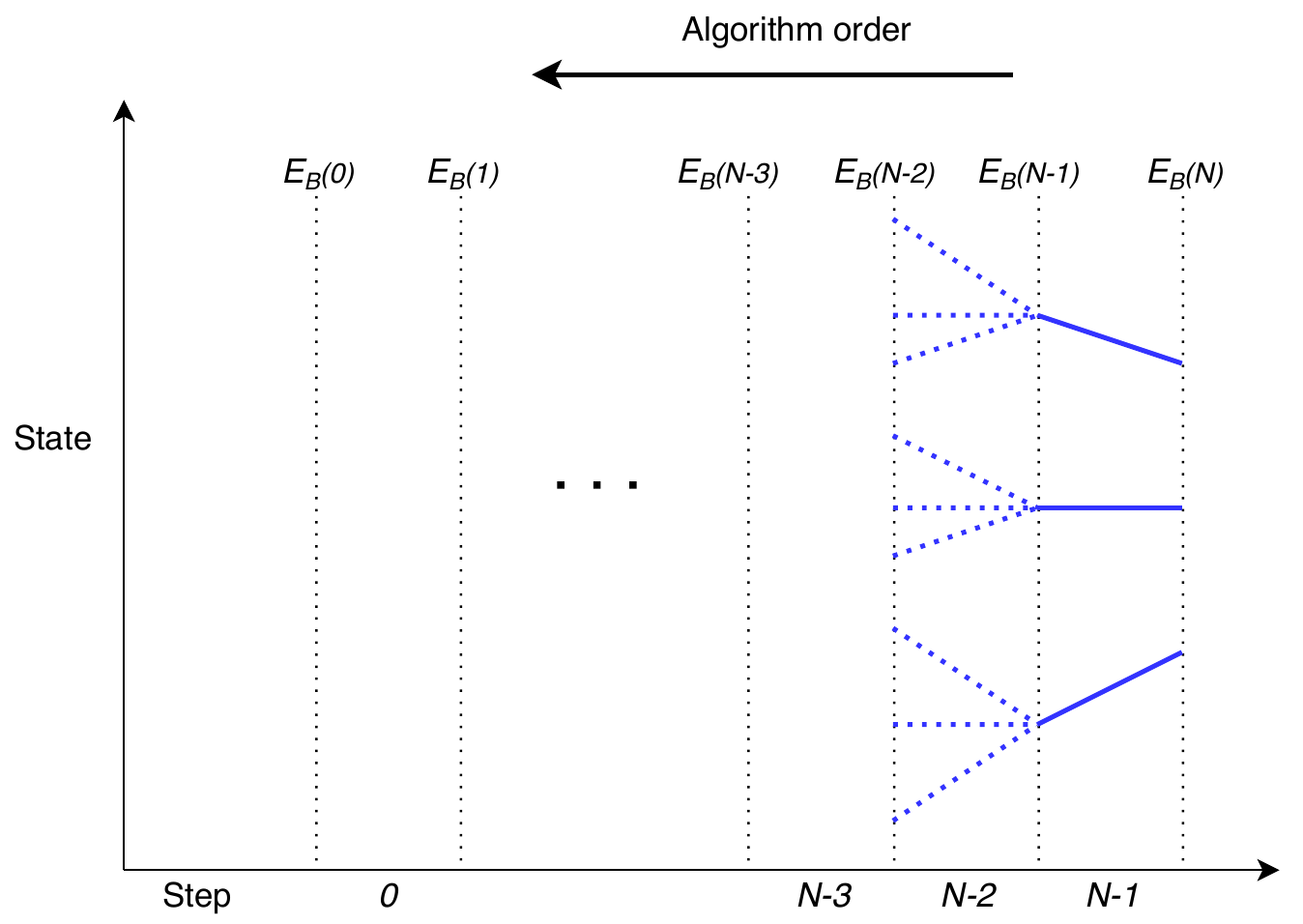}
  \caption{Dynamic programming~order.}
  \label{figure3.4}
\end{figure}

Referring to Figure~\ref{figure3.4}, as~the dynamic programming algorithm starts from the last step of the horizon, for~all $E_B (N-1)$, the~corresponding values of $\sigma_{N-1}$ will be equivalent to a state $E_B (N)$ that generates the minimal cost along the path, since $V(N,E_B )=0$ and $\sigma_N$ does not exist. Then the values of $\sigma_{N-1}$ will be memorized for the corresponding $E_B (N-1)$. In~the remaining iterations, the~values of $\sigma_k$ can be updated with the following method:

At step $k-1$, and~given $E_B (k-1)$, $E_B (k)$, and~$\sigma_k$ of $E_B (k)$ solved previously. There are two conditions. First, if~the action between $E_B (k-1)$ and $E_B (k)$ is opposite to the action from $E_B (k)$ to $\sigma_k$, the~‘subsequent local extreme’ of $V(k-1,E_B (k-1))$ will be $E_B (k)$, namely $\sigma_{k-1}=E_B (k)$. Second, if~both actions are the same type (both charging/discharging), or~the action in step $k-1$ is idling, then the ‘subsequent local extreme’ $\sigma_{k-1}$ seen from the state $E_B (k-1)$ should update to $\sigma_k$.

With this updating method, we can determine the ‘subsequent local extreme’ for every former state in every step in the dynamic programming algorithm, and~the cost of cycling can be calculated with Equation~(\ref{eq14}) to determine the minimum of (\ref{eq15}).

It can be perceived that the cost model eliminates the need to run a cycle counting algorithm along the planning horizon and decompose it into a recursive and additive function, which allows the cycle-counting-based battery cost models to be included in the objective function of the optimization~problem.  

\subsection{Simulation}\label{s5}
\unskip 
\subsubsection{Set~Up}\label{s5.1} 
The proposed strategy is tested with computer simulation. Parameters of the battery used in the simulation are summarized in Tables~\ref{t3.1} and \ref{t3.2}, which are based on the data provided in~\cite{r16,r18}. 

\begin{table}
\caption{Microgrid~parameters.}\label{t3.1}
\centering

\begin{tabular}{cccccc}
\toprule
\boldmath{$P_B^{max},P_B^{min}$}	& \boldmath{$P_G^{max},P_G^{min}$}	& \boldmath{$E_B^{min}$}& \boldmath{$E_B^{max}$}	& \boldmath{$E_{max}$} & \boldmath{$R_c$}\\
\midrule
24		& 60			& 1.25    & 11.25			& 12.5  & 2,500,000\\
MW	& MW			& MWh    & MWh 			& MWh   & \$\\
\bottomrule
\end{tabular}
\end{table}
\unskip

\begin{table}
\caption{Battery cost~parameters.}\label{t3.2}
\centering

\begin{tabular}{ccccc}
\toprule
\boldmath{$N_{100}^{fail}$}	& \boldmath{$kp$}	& \boldmath{$\Delta \delta$}& \boldmath{$d$}	& \boldmath{$n$} \\
\midrule
2347		& 1.1			& 1/12    & 0.05			& 5  \\
\bottomrule
\end{tabular}
\end{table}
\unskip

\subsubsection{Parameters and~Database}\label{s5.2} 
The battery used in~\cite{r18} can experience 3000 cycles at 80\% DOD, assuming the value of $kp$ is 1.1, the~cycles to failure at 100\% DOD is around 2347. The~cost of the battery used is 200 \$/kWh. In~addition, we choose 90\% and 10\% state-of-charge of BESS as the upper and lower bound for the BESS energy state to avoid overcharging/discharging.

It is assumed that the actual time of each control step is 5 min, as~the predictive data on renewable power, electricity price, and~load power are average values based on five-minute observations. The~planning horizon considered is two-hour, and~there will be 24 control steps on each~horizon.

We conducted a one-day simulation, and~the renewable power data is retrieved from the Woolnorth wind farm in Tasmania, Australia; the electricity price and power demand data are retrieved from the Australian Energy Market Operator. It should be noted that the load demand data is downscaled to the level of a~microgrid.

\begin{figure}
  \centering
  \includegraphics[width=13 cm]{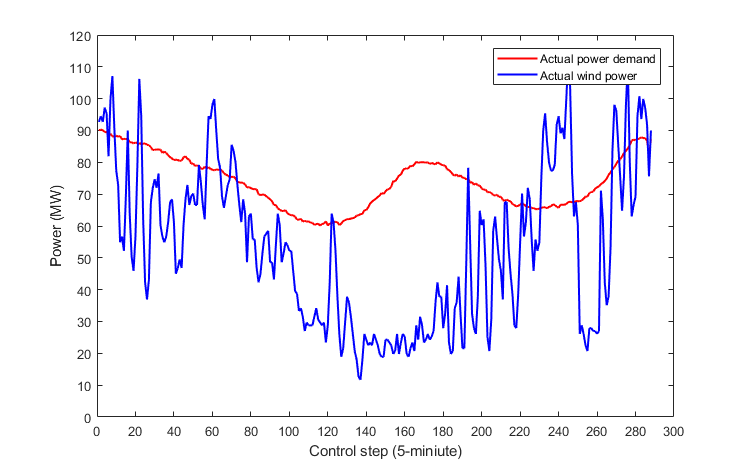}
  \caption{Actual renewable power and power demand (higher demand).}
  \label{figure3.5}
\end{figure}

\begin{figure}
  \centering
  \includegraphics[width=13 cm]{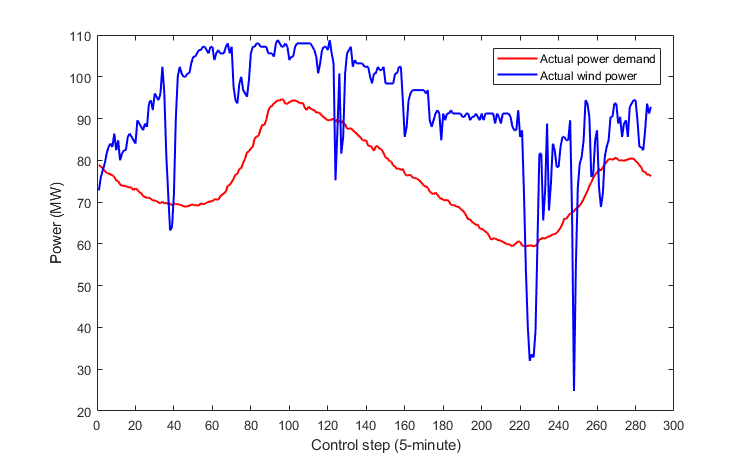}
  \caption{Actual renewable power and power demand (lower demand).}
  \label{figure3.6}
\end{figure}

We test two sets of data. The~first set has an overall higher demand than the renewable power generated, and~the total amount of power generated in the other one is higher than the demand, which can be seen in Figures~\ref{figure3.5} and \ref{figure3.6}.

\unskip

\begin{figure}
  \centering
  \includegraphics[width=13 cm]{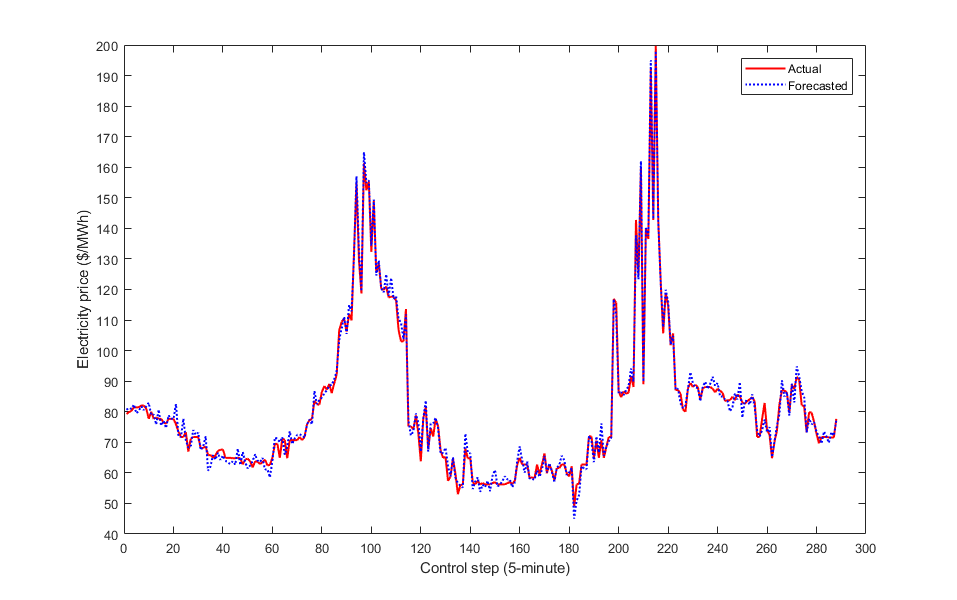}
  \caption{Actual and predicted electricity~price.}
  \label{figure3.7}
\end{figure}

As the data used are actual observations, we include some errors in the electricity price, renewable power, and~load power demand based on the short-term forecasting technique presented in~\cite{r24} that produces a normalized mean absolute error that ranges between 5\% and 14\%. As~a result, we include the normally distributed random errors with zero mean and the standard deviation equal to 5\% of the mean value of the actual data. The~actual electricity price and the one with errors are illustrated in Figure~\ref{figure3.7}.

Due to the errors in renewable power and load power, the~actual power dispatched to the grid will deviate from the results solved by the algorithm, whereas the cost of battery life cycle loss will not be affected by these errors as the state transition in the battery is deterministic. In~our simulation, the~data sets with errors are used in the algorithm, and~the results are assessed with the actual data~sets.


\subsubsection{Simulation~Results}\label{s5.3} 

\begin{figure}
  \centering
  \includegraphics[width=13 cm]{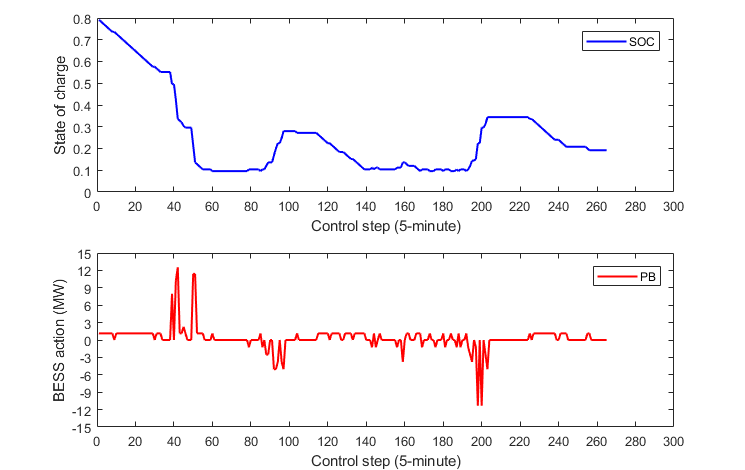}
  \caption{Battery energy storage system (BESS) actions and SOC (higher demand).}
  \label{figure3.8}
\end{figure}

\begin{figure}
  \centering
  \includegraphics[width=13 cm]{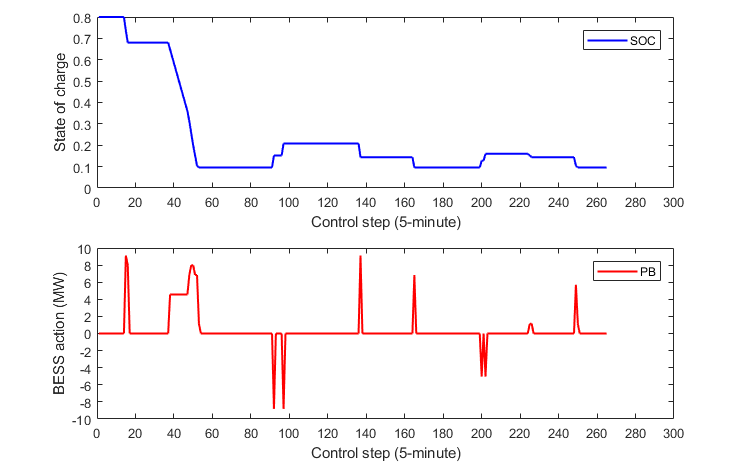}
  \caption{BESS actions and SOC (lower demand).}
  \label{figure3.9}
\end{figure}

With an initial state of 80\% SOC (10 MWh), and~the electricity data in Figure \ref{figure3.7}, the~control inputs of the BESS and the change in the battery SOC solved under the two data sets can be seen in Figures~\ref{figure3.8} and \ref{figure3.9}.

\unskip

The number of ‘half-cycles’ identified from the first data set was 21, with~a long half cycle from the initial state to the lower limit and several small cycles near the lower limit, whereas the number of ‘half-cycles’ identified from the second data set was five. Most of the charging actions, namely the negative values in BESS actions in both data sets can be observed around step 90 to 100, and~step 200 to 210, which are periods that have relatively higher values of electricity prices than others. As~the state of charge before these two periods are close to the lower limit, those spikes in the electricity prices would lead to some charging actions solved by the algorithm to achieve higher~profit. 

The costs calculated based on the simulation results are summarized in Table~\ref{t3.3}.

\begin{table}
\caption{Costs solved with both data~sets.}\label{t3.3}
\centering

\begin{tabular}{ccccc}
\toprule
\textbf{}	& \textbf{BESS Cost}	& \textbf{Energy Trading
Cost
}& \textbf{Overall
Cost}	& \textbf{Cost without BESS} \\
\midrule
Higher demand		& 3849.07			& 410,851.24    & 414,700.31			& 433,108.81 \\
Lower demand	& 2548.24			& $-$394,509.25    & $-$391,961.01 			& $-$369,880.45   \\
\bottomrule
\end{tabular}
\end{table}

The BESS cost is calculated using Equation~(\ref{eq13}), and~the energy trading cost is determined with $\sum(-C_{ele}(k)P_G(k))$ over the simulation period with forecasting errors. Also, we include the results solved with the case of balancing the power differences by exchanging power with the utility grid only, which is the cost without BESS in the table, solved with $\sum(-C_{ele}(k)(P_{rn}(k)-P_{ld}(k)))$.

As the objective function in our problem is to minimize the cost function (\ref{eq15}), a~lower cost indicates a better result. In~the case of higher demand, the~overall cost 414,700.31 is the cost spent on buying energy from the energy market, and~a reduction of 18,408.5, which is around 4.5\% of the daily cost, can be achieved. In~the lower demand condition, the~negative sign in the cost indicates the profit gained from selling the excessive power, and~an increase of 22,080.56 can be observed, which is around 5.6\% of the daily~profit.


\subsubsection{Discussions}\label{s5.4} 

Compared with some population-based heuristics optimizers, such as differential evolution (DE) and PSO \cite{exr1,exr2,exr3,exr4,exr5}, DP could be inefficient due to the well-known ‘curse of dimensionality’. Nevertheless, in the case of optimal control considering battery cost from cycling, these population-based optimizers can be costly for a large size of population, as a cycle counting algorithm will be invoked each time the entities/particles calculate the value of cost function. Also, these algorithms would require some adjustments in their parameters and initial positions to obtain the best results, which could be undesirable for real-time applications. In addition, the major drawback of these population-based algorithms is the lack of a solid mathematical foundation to assess the result \cite{dis1}. In comparison, the DP algorithm can guarantee the global optimum in the planning horizon, which could be more reliable and consistent in real-time operating. Also, our proposed cost model of battery eliminates the need to execute the cycle counting algorithm at the cost of extra storage space, which reduces the computation complexity and can be used as the benchmark to assess those population-based algorithms.

The proposed approach is a high-level control scheme, and the basic idea is to determine the optimal actions of BESS to maximize the profit in the microgrid. Although it is beyond the scope of this paper to investigate the low-level control, we would like to emphasize the importance of power electronics in the distributed generators (DGs)/batteries structure. In the microgrid environment, a large amount of DG units (e.g., solar panels, microturbines), storage units, and non-linear loads will be integrated, and a network of power inverters connected in parallel will be necessary in order to obtain good power sharing \cite{dis2} and stabilize system frequency \cite{dis3}. In our case, the distributed BESS is used, and multiple batteries connected in parallel should be coordinated and synchronized, which raises concerns on power sharing and frequency. We propose the high-level power scheduling control to optimize the cost, and a low-level control on the power inverters would be necessary, such as the hierarchical droop control for parallel-connected inverters introduced in \cite{dis4,dis5,dis6,dis7}.

\subsection{Summary}

A control strategy to maximizing the profit in a microgrid with a renewable power system and a battery energy storage system is introduced in this chapter. The~predicted data on renewable power, load power and electricity price are used to determine the suitable control inputs of the system, and~we used the receding horizon approach to alleviate the effects of forecasting errors by updating the prediction constantly. We also considered the cost of battery cycling and developed a recursive cost model. A~dynamic programming algorithm is used to solve the optimization problem over each planning horizon, and~the cost model is compatible with the algorithm. We tested our algorithm with actual data, and~simulation results have shown significant improvements in different~conditions.


\section{Approximate Dynamic Programming for Control of Microgrids \label{cha:ADPwind}}

This chapter introduces an approximate dynamic programming algorithm that can be used for optimal control problems in microgrids, some of the models and ideas introduced are based on our previous publication \cite{zhuo1,zhuo2}. The objective is to maximize the income and minimize the battery lifetime consumption, and the proposed algorithm can solve the optimal charge and discharge actions of the BESS within a future period. The rolling horizon approach is utilized, and the algorithm works with both short-term and long-term predicted data that are updated in every control step. The use of an approximate technique is to avoid the long computation time resulted from a large-scale BESS, and to meet time constraints in the rolling horizon approach. The algorithm is tested with practical wind power and electricity price data.

\subsection{Introduction} \label{c4s1}

Motivated by the growing demand on electricity and concerns on sustainability, grid-connected renewable power systems are receiving more interests from governments, investors and researchers, which increases the number of wind turbines and solar panels integrated into the electrical grid. This tendency in the energy market has provided opportunities for the development of microgrids \cite{c4exr1,c4exr2,c4exr3}, namely decentralized power systems with distributed renewable generation plants, energy storage units and localized loads that can be either grid-connected or islanded \cite{c4exr4}. A grid-connected microgrid is independent of the faults in the main grid \cite{c4exr5}, and reliability of the overall power system can be improved. In addition, it is believed that microgrids with clean energy systems will be a promising solution to achieve higher penetration of renewable energy in the future \cite{c4exr6}. Since connecting the variable and intermittent renewable power directly into the main grid will pose significant challenges in the grid operation, proper control in small or medium-scale microgrids will be necessary to smooth out these non-dispatchable power and utilized for local demands\cite{c4exr7,c4exr8}.

In previous studies \cite{r3,r5,r6,r7,r8,c4khalid1,c4khalid2,c4arash}, battery energy storage systems (BESS) are used in microgrids with renewable power systems to improve the dispatchability of wind power. With the help of different control algorithms, BESS can store part of the generated wind energy and supply some stored energy at different periods, such that the actual power dispatched can meet specific operation criteria and reduce the amount of energy curtailed. Among these studies, the profit potential of using BESS in wind farms are receiving attention from researchers. In countries with deregulated energy markets, qualified private participates with generation units can sell electricity from their generators and receive payments from energy market operators. As the electricity price varies based on factors such as available generators online and power demands, proper control of BESS at different price periods can produce higher profit for individual participants.

Nevertheless, some issues are insufficiently considered in these studies, one of which is the battery lifetime\cite{c4exr9,c4exr10}. As stated in multiple papers, the costs from BESS are relatively high for renewable energy system applications currently. To increase the profit, batteries will experience more charging and discharging cycles than regular tasks, and their aging process can be accelerated due to stress factors such as variable state of charge level, charge/discharge currents. Therefore, the additional cost from a lower expected life expectancy of batteries should not be neglected, and it should be included in the control algorithm to avoid the unprofitable charge and discharge decisions. 

The other issue is related to the prediction error\cite{c4exr11,c4exr12}. In some studies, the solution to reduce the effects of errors in wind power forecasting is the rolling horizon technique. The algorithm will use data with a relatively short forecasting horizon and solve a set of optimal control inputs over the horizon. Only the first component will be used as the actual input to the system in the current step, and the algorithm will repeat in the next step with updated system states and predicted data. However, as the algorithm must provide the solution before the current step expires, optimization technique such as dynamic programming is unlikely to meet the time constraints in a large-scale wind farm and BESS.

In this chapter, a wind power dispatch strategy based on an approximate dynamic programming algorithm is proposed. The use of an approximate technique is to reduce the computation time and meet the requirement for online scheduling tasks. The rolling horizon approach is utilized for long-term simulation, and the algorithm will compute a sequence of charge and discharge actions for the BESS that can maximize the income within a control horizon. In addition, a battery lifetime estimation function \cite{c4blife} is included in the algorithm. It can estimate the battery life expectancy based on discharge actions within a control horizon, which is used to compute the operational cost from BESS.


\subsection{Problem Statement} \label{c4s2}

We consider a wind farm integrated with a large-scale BESS that can charge and discharge part of the generated wind energy to control the actual power dispatched to the grid. The objective is to maximize the profit gained by energy trading based on wind power and electricity price forecasting within a further period. 

A discrete-time control system model is considered in the problem. Prior to each discrete-time interval, the predicted data on wind power and electricity price will be received by the algorithm. In this study, the discrete-time interval is 5 minutes with a six-step forecasting horizon of 30 minutes. As the Australian National Electricity Market adopts a half-hour trading period with six 5-min dispatch intervals, and we use the 5 minutes average data for wind power forecasting. In addition, a long period of predicted data on electricity price in the future will be received by the algorithm at each control step, which is used to estimate the value of remaining energy stored in the battery at the end of each control horizon.

The objective function of the algorithm to be maximized is the difference between the profit gained by selling electricity and the operational cost from the BESS estimated within a prediction horizon. We use the battery lifetime estimation function introduced in \cite{c4blife} to calculate the operational cost, in which the energy discharged through the battery are counted and compared to a predefined amount of energy throughput that the battery can experience before its end of life. For one discharge action, its actual value used in the function will be adjusted by comparing its condition to the predefined one, which is used to reflect the influence of different stress factors on the battery lifetime consumption \cite{c4exr13}. Furthermore, we included an estimated value of the remaining energy stored in the battery at the end of each control horizon. This value is computed with the average electricity price within a further period, and we assume the remaining energy will be discharged throughout this period.


\subsection{Proposed Methodology} \label{c4s3}

\subsubsection{Battery Lifetime Estimation Function} \label{c4s3.1}

The battery lifetime estimation function proposed can estimate the battery lifetime resulted from a series of discharge events. Cases of its applications in microgrids and electric vehicles can be viewed in \cite{c4conf1,c4conf2}. Assuming $n$ discharge events with corresponding values of discharge current, depth of discharge and discharge time, the function is:

\begin{equation}
  \label{c4eq1}  
\\L_{time} = \dfrac{L_RD_RC_R}{\sum_{i=1}^{n} d_{eff}}=\dfrac{L_RD_RC_R}{\sum_{i=1}^{n} \dfrac{L_R}{L_A} \dfrac{C_R}{C_A}d_{actual}}T\ 
\end{equation}

and the parameters are summarized below:

\begin{itemize}
  \item $L_R$: Cycle life of the battery at rated depth of discharge $D_R$.
  \item $L_A$: Cycle life of the battery at the depth of discharge of one discharge action.
  \item $C_R$: Ampere hour capacity of the battery at rated discharge current.
  \item $C_A$: Ampere hour capacity of the battery at the discharge current of one discharge action.
  \item $D_R$: Rated depth of discharge at which rated cycle life was determined.  
  \item $d_{eff}$: Effective ampere hour discharge as adjusted for DOD and rates of discharge.
  \item $d_{actual}$: Ampere hour energy discharged from some discharge actions.  
  \item $T$: Operation period of some of discharge actions.
  
\end{itemize}

The numerator term is the overall amount of Ah throughput that can be discharged by the battery until its end of life at rated conditions. The denominator term is the sum of energy throughput of some discharge events within the period $T$, which has been adjusted to an effective value with respect to rated conditions using the formula $\dfrac{d_{eff}}{d_{actual}} =\dfrac{L_R}{L_A} \dfrac{C_R}{C_A}$. 

For one discharge decision, $L_A$ and $C_A$ can be calculated with two graphs that are generally available in the battery data sheet, namely the cycles to failure versus depth of discharge graph, and the ampere-hour capacity at different discharge currents chart. As a result, depth of discharge and discharge current can be used to determine $L_A$ and $C_A$ with suitable curve fitting functions. The following two figures are the corresponding data of the battery used in the simulation.

 \begin{figure}
  \centering
  \includegraphics[width=13 cm]{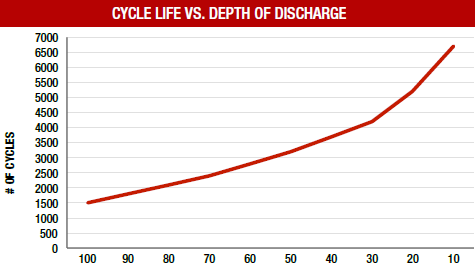}
  \caption{Cycle life versus depth of discharge curve of the battery.}
  \label{c4figure1}
\end{figure}

 \begin{figure}
  \centering
  \includegraphics[width=13 cm,height=8 cm]{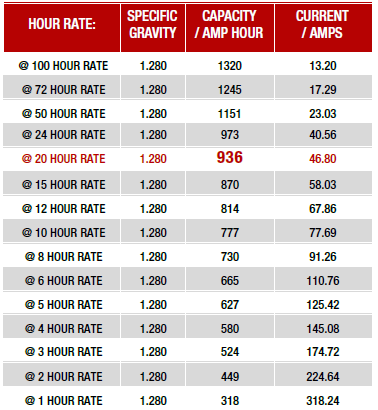}
  \caption{Discharge current on Ampere-hour Capacity of the battery.}
  \label{c4figure2}
\end{figure}

Based on equation \ref{c4eq1}, discharge events with deeper depth of discharges than the rated one will increase the effective Ampere-hour throughput $d_{actual}$, leading to a reduction in the estimated battery lifetime $L_{time}$. Same applies to the discharge current.

In our case, discharge events within each control horizon are used in the function to calculate an estimated battery lifetime, and the operational cost can be determined by dividing the price of battery cells to this battery lifetime and multiplying the duration of control horizon. Let $P_B$ be the price of the battery, this operational cost $C_{opr}$ is: 

\begin{equation}
  \label{c4eq2}  
\\C_{Opr} = \dfrac{P_B}{L_{time}}T=\dfrac{P_B\sum_{i=1}^{n}\dfrac{L_R}{L_A}\dfrac{C_R}{C_A}d_{actual}}{L_RD_RCR}\
\end{equation}


\subsubsection{Control System Model} \label{c4s3.2}

Let $p(k)$ be the wind energy generated from $k$ to$k+1$, $g(k)$ be the energy dispatched to the grid within $k$ to $k+1$, and let $m(k)$ denote the trading price of electricity between $k$ to $k+1$. Assuming the available amount of energy stored in the battery at time $k$ is the control system state $x(k)$, and energy conversions are lossless. The following discrete-time dynamic system model will be satisfied:

\begin{equation}
  \label{c4eq3}  
x(k+1) = x(k)+p(k)-g(k),
\end{equation}

Define the difference between wind energy and energy to the grid as the control input $u(k)$ for the system, the system model will be: 

\begin{equation}
  \label{c4eq4}  
x(k+1) = x(k)+u(k),
\end{equation}

So $u(k)>0$ indicates a charge action, and $u(k)<0$ indicates a discharge action. Based on this system model, the operational cost  $C_{opr}$ can be redefined as follows: 

Within a control horizon of $N$ steps starting from $0$ to $N-1$, the set of controls at each step is $[u(0),u(1),\cdots,u(N-1)]$, and $C_{opr}(k)$ is the estimated operational cost from control $u(k)$ and $k\in{0,1,\cdots,N-1},$ then

\begin{align}
C_{opr}(k)= {} & 0 \quad \forall u(k)\geq 0 \label{c4eq5}
\end{align}
\begin{align}
C_{opr}(k)= {} & \dfrac{P_B\sum_{i=1}^{n}\dfrac{L_R}{L_A}\dfrac{C_R}{C_A}d_{actual}(k)}{L_RD_RCR}\ \quad \forall u(k)<0  \label{c4eq6}
\end{align}

Since the estimation function deals with discharge events only, charge actions with currents lower than the value recommended by manufacturers will not incur this operational cost. Furthermore, values of $L_A (k)$, $C_A (k)$ and $d_{actual} (k)$ can be calculated with corresponding discharge current, depth of discharge and discharge time at time $k$.

Therefore, the objective function of the optimization problem will be:

\begin{equation}
  \label{c4eq7}  
\max_{u(k)}\quad \sum_{k=0}^{N-1} m(k)g(k)-C_{opr}(k)+V_{rm} 
\end{equation}

With the following constraints:

\begin{align}
g(k)\geq 0 \label{c4eq8}
\end{align}
\begin{align}
LB \leq x(k) \leq UB \label{c4eq9}
\end{align}
\begin{align}
-r_d \leq x(k+1)-x(k) \leq r_c \label{c4eq10}
\end{align}

The first constraint ensures the system will not charge from the grid. In the second constraint, $LB$ and $UB$ are the lower and upper limits of the battery energy state, which are used to avoid over and deep discharge. $r_c$ and $r_d$ are the maximum amounts of energy that can be charged and discharge from the BESS within a control step.

In the objective function \ref{c4eq7}, $V_{rm}$ is the estimated value of the remaining energy stored in the BESS at the end of control horizon, which is:

\begin{equation}
  \label{c4eq11}  
V_{rm}=m_{rm}(x(N-1)+u(N-1)-LB)-C_{rm}
\end{equation}

$m_{rm}$ is the average electricity price in the next 20 hours, and $C_{rm}$ is the operational cost to discharge the remaining energy to the lower limit of the battery state LB at the 20-hour discharge rate. As the ampere-hour capacity of the battery used in the simulation is rated at 20-hour, we choose a 20-hour forecasting horizon, and a relatively long prediction horizon can provide a better estimate of what the electricity price would be in the future.


\subsubsection{Approximate Dynamic Programming} \label{c4s3.3}

The approximate dynamic programming algorithm used in our problem is introduced in this section. Define $J_k (x_k)$ as the maximum cost starting from time $k$ onwards to the last step of the control horizon with the state $x(k)$ at $k$, the Bellman equations of the problem will be:

\begin{align}
J_N(X_N)= {} & m_{rm}(x(N-1)+u(N-1)-LB)-C_{rm}  \label{c4eq12}
\end{align}
\begin{align}
J_N(X_k)= {} & \max_{u(k)} (m(k)(p(k)-u(k))-C_{opr}(k))+J_{k+1}(x_{k+1}) \quad k=0,1,\cdots,N-1.    \label{c4eq13}
\end{align}

Unlike some dynamic programming algorithms that set $J_N (x_N)$ to zero, we use it to represent the estimated value of remaining energy stored at the end of control horizon. The approximate dynamic programming algorithm we used is based on the algorithm stated in \cite{c4powel1}, which is illustrated as follows:

\begin{itemize}
  \item[] \textbf{Step $0$}. Initialization:
  \begin{itemize}
  \item[] Step $0a$. Initialize the algorithm to compute $\bar{J}_k^0 (x_k)$ for all states $x_k$.
  \item[] Step $0b$. Choose an initial state $x_0^1$.
  \item[] Step $0c$. Set $n=1$.
  \end{itemize}
  \item[] \textbf{Step $1$}. Choose the predicted sets $p_k^n$ and $m_k^n$.
  \item[] \textbf{Step $2$}. For $k=0,1,2,\cdots,N-1$ do:
  \begin{itemize}
  \item[] Step $2a$. Solve $\hat{v}_k^n=\max_{u(k)} (m_k^n(P_k^n-u_k)-C_{opr}(k)+\bar{J}_{k+1}(x(k+1)))$, and let $u_k^n$ be the value of $u_k$ that solves the maximization problem.
  \item[] Step $2b$. Update $\hat{J}_k^{n-1}(x_k)$ using:
  \begin{itemize}
  \item[] $\hat{J}_k^n(x_k)=(1-a_{n-1})\hat{J}_k^{n-1}(x_k^n)+a_{n-1}\hat{v}_k^n, \quad \forall x_k=x_k^n$
  \item[] $\hat{J}_k^n(x_k)=\hat{J}_k^{n-1}(x_k), \quad$ otherwise. 
   \end{itemize}
   \item[] Step $2b$. Compute $x_{k+1}^n=x_k^n+u_k^n$.
  \end{itemize}
  \item[] \textbf{Step $3$}. Let $n=n+1$. If $n<maxInteration$, go to \textbf{step 1}.
\end{itemize}

$\hat{J}_k^n(x_k)$ represents an estimated value of $J_k(x_k)$ at iteration $n$. By replacing the true value  $J_k(x_k)$ with this value, it is unnecessary to loop over all possible states recursively and the computation time is significantly reduced. $p_k^n$ and $m_k^n$ are predicted samples of wind power and electricity price. $a_(n-1)$ is the step size, or the learning rate schedules used to smooth between old and new estimates, and we use the Harmonic step size sequence in our simulation.

There are two main challenges in this algorithm. The first one is Step $0a$, in which estimated values of $J_k (x_k)$, are required before the algorithm is executed. The challenge is finding approximations that are good enough to achieve the optimization purpose. The second challenge is Step $2a$. For systems with large control inputs or high dimensional ones, solving this maximization problem will be compute-intensive.

We use the following propositions to overcome the challenges described in the last section. 

\textbf{Proposition 1}: Assume $u^{\star}(k)$ is one optimal solution within a control horizon with electricity price $m(k)$ and state $x(k)$. For $u^{\star}(k)<0$, namely the optimal decision at $k$ is to discharge the BESS with energy equivalent to $-u^{\star}(k)$, then $-u^{\star}(k)m(k)-C_{opr} (k)>0$.

As the term $-u^{\star}(k)m(k)-C_{opr} (k)>0$ indicates the extra income gained by discharging the battery, it must be positive to achieve the optimal solution. In the case of charging, $u(k)$ will be positive and the operational cost $C_{opr}(k)$ will not incur. 

\textbf{Proposition 2}: If there exists $u_b (k)\leq 0$ such that $-u_b(k)m(k)-C_{opr} (k)>0$ is maximized, the optimal solution $u^{\star}(k)\geq u_b (k)$.

In this proposition, we assume an action $u_b (k)$ can maximize the extra income, and we believe that the optimal decision is to discharge an amount of energy that is equal or smaller than this value. To prove this, one can assume a decision $u(k)<u_b (k)$, or discharging more than $-u_b (k)$. As discharging more will lead to a state $x(k+1)$ with a deeper depth of discharge, the operational cost from discharging in $k+1$ will be higher than we start discharging with the state achieved by $u_b (k)$. And since $-u_b(k)m(k)-C_{opr}(k)$ is maximized, $u_b (k)$ will always be a better decision than $u(k)$.

On the other hand, there will be cases that an action $u(k)>u_b (k)$, can be a better decision than $u_b (k)$. This happens when electricity prices after k are higher, and discharging more during those price periods with lower operational costs may be more profitable. Therefore, we might choose to discharge less than $-u_b (k)$ so that we can start discharging with a higher depth of discharge during those high price periods. 

This proposition can be used to determine $\hat{J}_k^0 (x_k)$ by choosing $u_b (k),u_b (k+1),\cdots, u_b (N-1)$ as solutions from control stage $k$ to the last stage $N-1$. In this way, it is unnecessary to compute all $\hat{J}_k^0 (x_k)$) before we execute the main algorithm; we only compute $\hat{J}_k^0 (x_k)$) when we visit some state $x_k$ in Step $2a$. If this value has been solved before or updated in Step $2b$, we just use that value stored in the memory. Furthermore, $u_b (k)$ can be used to limit the action space to solve Step $2a$ in the algorithm, so that we can avoid visiting states that are not optimal.


\subsection{Simulation} \label{c4s4}
\subsubsection{Curve Fitting Functions} \label{c4s4.1}

As stated in Section \ref{c4s3.1}, two functions are required to approximate the two curves in Figure \ref{c4figure1} and \ref{c4figure2}, and the Matlab Curve Fitting Toolbox is used. For the life cycle versus depth of discharge curve (Figure \ref{c4figure1}), the polynomial fitting is adopted, and for the current on Ampere-hour capacity curve (Figure \ref{c4figure2}), we use the exponential fitting. Results are shown below:

 \begin{figure}
  \centering
  \includegraphics[width=13 cm]{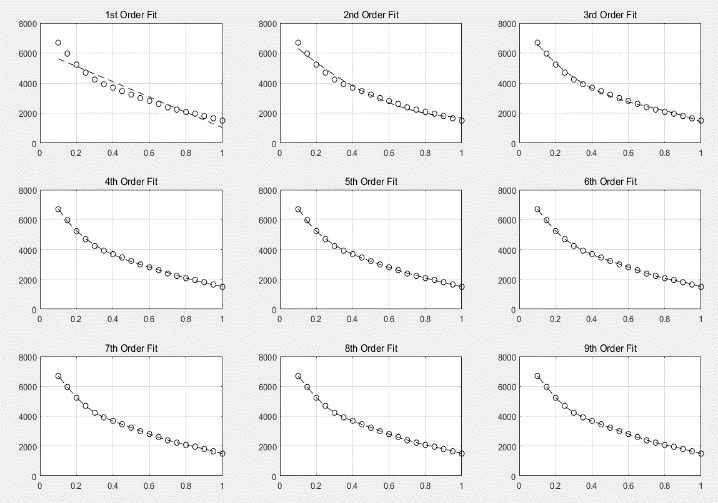}
  \caption{First to ninth order polynomial fitting.}
  \label{c4figure3}
\end{figure}

 \begin{figure}
  \centering
  \includegraphics[width=13 cm,height=8 cm]{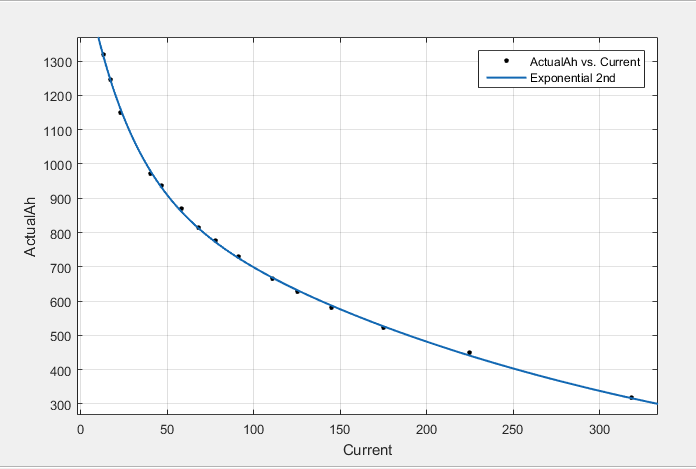}
  \caption{Second order exponential fitting.}
  \label{c4figure4}
\end{figure}

Base on the results, the $4^{th}$ order polynomial fitting function is used in the algorithm, and the $2^{nd}$  order exponential function is used to fit the current on Ampere-hour capacity curve. The two functions are:

\begin{equation}
  \label{c4eq14}  
L_A=17612(D_A)^4-48325(D_A)^3+49771(D_A)^2-26417D_A+8898 
\end{equation}

\begin{equation}
  \label{c4eq15}  
C_A=638.5e^(-0.03876I_A)+975.9e^(-0.003531I_A)
\end{equation}

where $D_A$ and $I_A$ are the depth of discharge and the discharge current of certain discharge action. 


\subsubsection{Battery Specifications} \label{c4s4.2}

The battery data used is summarized in the following table. We consider 10 batteries connected in series to construct a $320volt$, $936Ah$ battery, which is $299.52kWh$. The maximum discharge current is limited to $1C$, which is $318.24A$, and the maximum charge current is $20\%$ of the rated Ampere-hour capacity ($936Ah$) of the battery.

\begin{table}
\caption{BESS and system parameters}\label{c4t1}
\centering

\begin{tabular}{ccccc}
\hline
$L_R$	& $D_R$	& $C_R(Ah)$ & $LB(Wh)$ & UB(Wh) \\
\hline
1500		& 1			& 936    & 89856			& 269568 \\
\hline
$r_d(Wh)$	& $r_c(Wh)$			& $V_b(volt)$    & $\delta h$(hour) 			& $PB$(\$)   \\
\hline
8486	& 4992	& 320    &  1/12			& 44928   \\
\hline
\end{tabular}
\end{table}


\subsubsection{Simulation and Discussion} \label{c4s4.3}

\begin{figure}
  \centering
  \includegraphics[width=13 cm,height=8 cm]{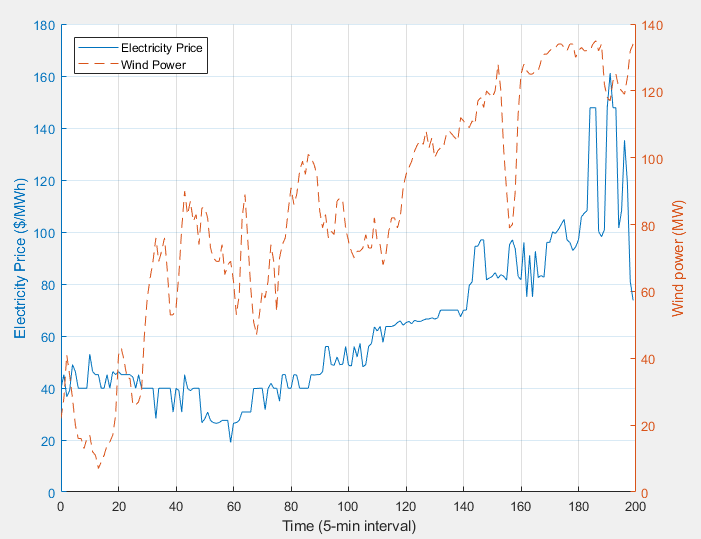}
  \caption{Wind power and electricity price.}
  \label{c4figure5}
\end{figure}

The wind power and electricity price data are shown in Figure\ref{c4figure5}. As indicated in the \cite{c4windprediction}, the short-term wind power forecasting errors are normally distributed, and the proposed prediction technique can produce results with normalized mean absolute error ranges between 5\% and 14\%. Therefore, we refer to the study to include some errors in our simulation. 

\begin{figure}
  \centering
  \includegraphics[width=13 cm,height=8 cm]{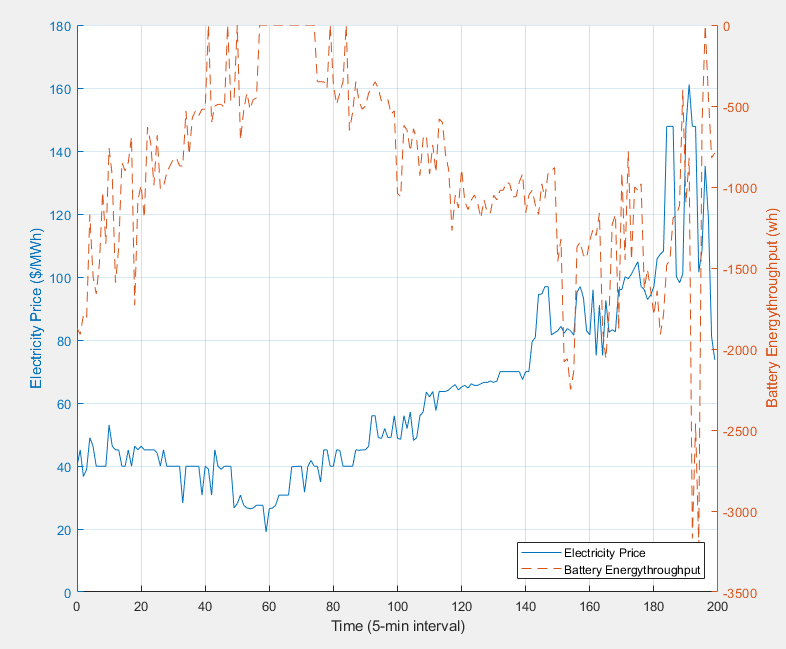}
  \caption{Battery actions and electricity price.}
  \label{c4figure6}
\end{figure}

The control inputs solved from our algorithm are shown in Figure \ref{c4figure6} with a comparison to the electricity price. It can be observed that some zero actions are obtained between control stages 40 to 80, where electricity prices are lower than 40\$ $ MWh$. This indicates that any discharge decisions taken on those periods would be unprofitable.

\begin{figure}
  \centering
  \includegraphics[width=13 cm,height=8 cm]{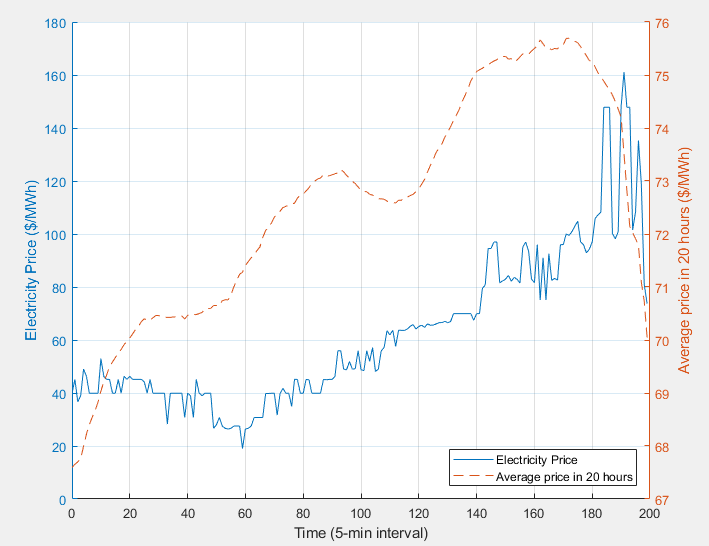}
  \caption{Average electricity price in 20hours and current electricity price.}
  \label{c4figure7}
\end{figure}

The average electricity price in the next 20 hours is shown in Figure \ref{c4figure7} with the current electricity price, and its effect on the battery actions can be perceived. For instance, a decrease in the average electricity price can be perceived after stage 190, and with an increase in the current price after stage 190, it will generate a large discharge action during that stage.

\begin{figure}
  \centering
  \includegraphics[width=13 cm,height=8 cm]{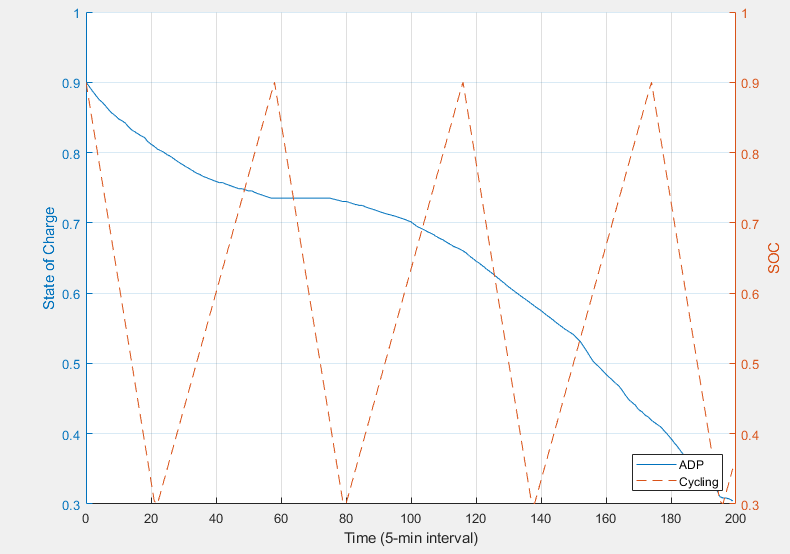}
  \caption{Change in SOC under two strategies.}
  \label{c4figure8}
\end{figure}

We further compare our algorithm with the strategy of cycling the battery at maximum charge and discharge rate between the upper and lower limits. Variations in the state of charge under two methodologies are presented in Figure \ref{c4figure8}.

It can be perceived that in the cycling strategy, the battery experienced more than three full charge and discharge cycles within the simulation duration, while it executed only one full discharge action based on the proposed ADP algorithm. 

The additional income gained from discharging the BESS over the simulation period is $12.8623$ in our strategy, whereas this value is $-12.9312$ in the cycling strategy. The negative value solved in the cycling strategy is due to some charging actions made at high electricity price periods. As our battery rating $(299.52kWh)$ is relatively small compared to the level of wind power (40 to 140$MW$), the additional income is insignificant. In practice, one can utilize multiple batteries of controlled by independent power convention systems construct a large-scale BESS to increase the overall income.


\subsection{Summary} \label{c4s5}

An economic dispatch strategy for wind power in deregulated energy markets is introduced in this chapter. An approximate dynamic programming algorithm is used to determine the control actions for the battery energy storage system at different price intervals. It can maximize the income and minimize the consumption of battery lifetime based on predicted data of wind power and electricity price in both short and long period. The rolling horizon approach is used to maintain a low level of forecasting error, and we use the approximate dynamic programming to avoid the curse of dimensionality in large-scale BESS systems. The algorithm has been proven to be beneficial with practical wind and electricity data.
\section{Decentralized Optimal Control of Microgrids\label{cha:chap5}}

Constructing microgrids with renewable energy systems could be one feasible solution to increase the penetration of renewable energy. With proper control of the battery energy storage system (BESS) and thermostatically controlled loads (TCLs) in such microgrids, the~variable and intermittent energy can be smoothed and utilized without the interference of the main power grid. In this chapter, a decentralized control strategy for a microgrid consisting of a distributed generator (DG), a battery energy storage system, a solar photovoltaic (PV) system and thermostatically controlled loads is proposed. The~control objective is to maintain the desired temperature in local buildings at a minimum cost. Decentralized control algorithm involving variable structure controller and dynamic programming is used to determine suitable control inputs of the distributed generator and the battery energy storage system. The~model predictive control approach is utilized for long-term operation with predicted data on solar power and outdoor temperature updated at each control step. This chapter is mostly based on our previous work \cite{r4}.


\subsection{Introduction}

The inherent variability and intermittency of renewable power could be the major issue to hinder its penetration. One feasible solution is to construct more microgrids integrated with battery energy storage systems (BESSs) and renewable energy systems (RESs), which allows the renewable energy to be smoothed and used locally without interfering with the utility power grid \cite{r3}. With the help of predicted renewable energy, electricity price, and load demand, proper control of BESS at different time intervals could minimize the overall energy cost. The~paper \cite{r3} proposed an energy trading strategy for a microgrid. They considered different electricity prices from six RESs at different dispatching intervals to determine the actions of a BESS to meet the power demand at a minimum cost. An energy management strategy for a microgrid community is introduced in \cite{r13}, which consists of multiple microgrids integrated and regulated with a community energy management system. The~authors utilize day-ahead forecasting data to plan the desired controls for all units in those microgrids, including output power of distributed generators, charging/discharging power of BESSs and power exchanged between them. An energy scheduling model of microgrids with renewable generations and BESSs is proposed in {\cite{c5exr1}}, in which the storage system is considered as the spinning reserve to smooth the power fluctuations due to the intermittent renewable generators, and the authors utilized chance-constrained programming to solve the stochastic optimization problem efficiently. A bi-level energy scheduling strategy for microgrids with battery swapping stations of electric vehicles is introduced in {\cite{c5exr2}}, in which the net costs of the microgrid is minimized in the upper-level, and the profits of the battery swapping stations are maximized based on real-time pricing.

Thermostatically controlled loads (TCLs), such as air conditioners, heaters, and ventilation systems, are some of the most widely used electrical appliances \cite{c5kemeng1}. In microgrids with intermittent energy, they could provide the required ancillary services for renewable generation balancing, which could be more flexible than conventional generators that are often constrained by their capabilities~\cite{c5r2}. Furthermore, in energy markets with demand response programs, due to the inherent flexibility of TCLs, they are well-suited for consumers to adjust energy usage in response to electricity prices or incentive payments \cite{c5r3}.

In this chapter, we focus on the control of TCLs in the microgrid environment. The~primary objective is to maintain the desired temperature at residential buildings in a cost-effective manner. These TCLs are powered by a distributed generator (DG) and a solar power system, and a BESS is used to regulate the actual power dispatched to them. We consider an islanded microgrid environment, and there will be no power exchanged from the microgrid to the unity grid. As a result, over periods without solar energy, the~TCL units will be supplied only by the generator and the BESS, so there will be a non-zero minimum output requirement for the generator to avoid start-up or shut-down cost. In addition, as~the residential TCLs are generally constructed in a large quantity with relatively small capacity in practice, it can be complicated and inefficient for the microgrid operator to control each TCL directly. Therefore, the~aggregation model of distributed TCLs is used\cite{c5r6,c5fengji1,c5r8}.

Authors of \cite{c5r6} propose a two-stage optimization model for a group of TCLs to smooth the power fluctuation resulting from PV systems in distribution networks. In the first stage, the~regulation capability of TCL groups is determined from the aggregation of user inputs including the change in the temperature set points and maximum allowed temperature, which is then used to calculate the required regulation power to smooth the net exchange power fluctuation at a minimum cost. In the second stage, the~regulation task is decomposed to each TCL group. A short-term operation model of the microgrid with high penetration of solar power and TCLs is introduced in \cite{c5fengji1}, which schedules the distributed generator, BESS and TCLs in the microgrid based on solar power forecasting. The~receding horizon optimization technique is utilized to alleviate the effects of forecasting errors. The~use of aggregated TCLs in the wholesale electricity market is discussed in \cite{c5r8}, and an aggregator determines the volume of electricity exchanged with the market in the next day based on renewable generation and load forecast. The~objective is to maintain the suitable water temperature for prosumers at a minimum energy cost, and a model predicted control optimization is applied intraday to minimize the imbalance between the forecast data and the actual measurements. 

Most of these studies can be perceived as the multi-objective control problem, with temperature control from TCLs being one primary target. Others can be the overall optimization in different environments such as microgrid and deregulated energy market. Since predicted data on renewable power, electricity price and load demand are used in their algorithms, one of the main challenges in these studies can be the errors in forecasting. To address this, the~receding horizon approach is utilized in most studies \cite{r4.1,c5exr3,c5exr4,c5exr5,c5exr6,c5exr7,c5exr8,c5exr9,c5exr10,zhuo1,zhuo2,zhuo3,zhuo4,zhuo5,zhuo6}, as uncertainties can be reduced to a relatively low level with constantly updated forecasting data. Nevertheless, this approach has a strict requirement for the computation time as the set of results over a forecasting horizon should be solved before the first control step over the horizon expires. In practice, delays due to forecasting techniques, data communications, and~response time in power electronics can further degrade the performance of the algorithm.

As a result, a novel decentralized algorithm for optimal control of microgrids with high solar power penetrations and TCLs is developed. We aim to solve our problem in one stage optimization. With the aggregation of TCLs, only the outdoor temperature is required to calculate the power demand, which reduces the computation time spent on the TCL model. As other units in the microgrid, such as renewable energy system, BESS, and distributed load, have cost models related to their control signals and states, the~objective function of the optimization problem can be an additive cost over the forecasting horizon. Therefore, we formulate the problem in a cost-to-go equation, or the dynamic programming type equation to solve the sets of controls for both BESS and DG based on forecasting data over each horizon. As the model predictive control approach is applied, only the first component of both sets will be used as the actual inputs to the system in the current stage, and the algorithm will repeat in the next step with updated system states and predicted data. A critical feature of our approach is that TLCs are controlled in a totally decentralized fashion
and only the indoor temperature in the corresponding area is needed to control each TCL. Moreover, control of the distributed generation unit and charge/discharge of the BESS does not require any information on the current indoor temperatures.


\subsection{Problem Statement}\label{c5s2}

\begin{figure}
  \centering
  \includegraphics[width=13 cm,trim= 0 20 0 40,clip]{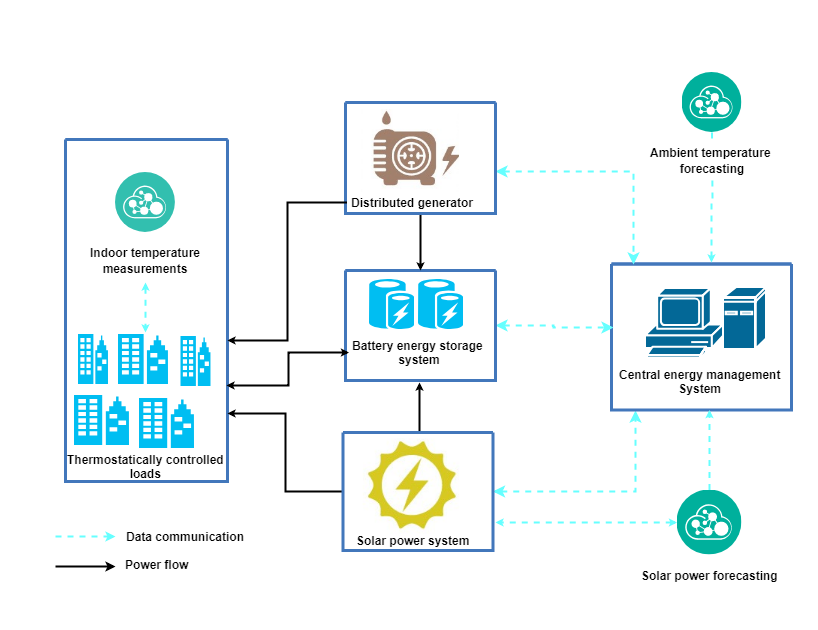}
  \caption{Configuration
  of the microgrid.}
  \label{chap5figure1}
\end{figure}

We consider a microgrid consists of a distributed generator (DG), a battery energy storage system (BESS), a solar power generation unit, and a group of thermostatically controlled loads (TCLs). An~islanded microgrid environment is considered, and the TCLs are supplied by the solar power system, DG and BESS without power exchanged from the microgrid to the unity grid. The~main objective is to minimize the overall cost required to keep the desired building temperature. The microgrid under consideration can also be viewed as an example of networked control systems, see, e.g.,\mbox{\cite{network1,network2,network3,network4,network5,network6,network7,network8,network9,network10}}. The~configuration of the microgrid is presented in Figure \ref{chap5figure1}.

A decentralized control scheme is used, in which the central energy management system would receive predicted data on solar power and outdoor temperature over a forecasting horizon, and the control inputs for the DG, BESS, and TCLs over the period are then solved based on the predicted data. The~DG and the BESS have cost models related to their output power and charging/discharging power, which are controlled following the predicted solar power to meet the power demand from TCLs. In~the corresponding areas of TCLs, the~indoor temperature is controlled to slide along the setpoint. In addition, the~receding horizon approach is utilized for long-term operation, with predicted data updated at each control step.

Since the objective is to minimize the overall energy cost, the~additive costs from both DG and BESS over each forecasting horizon is used as the objective function.  Based on predicted solar power and outdoor temperature, a dynamic programming (DP) algorithm is used to solve two sets of optimal controls for both DG and BESS, which are the output power of DG and the charging/discharging power of BESS respectively. As the receding horizon approach is used, only the first component of both sets will be used as actual inputs to the system at the current stage, and the algorithm will repeat in the next control step with updated system states and predicted data.


\subsection{System Model}\label{c5s3}

The control system model of the microgrid is introduced in this section.

\subsubsection{Distributed Generator}\label{c5s3.1}

The cost of the generator used in the microgrid is described by the following function:
\begin{equation}
  \label{c5eq1}  
C_{G}(P_{G}(t)) = aP_{G}(t)^2 + bP_{G}(t) +c
\end{equation}
where \(P_{G}(t)\) is the power output of DG at time $t; a, b,$ and $c$ are generation cost coefficients of DG. We consider our system as a discrete-time system with the sampling rate \(\delta>0\), So the function \(P_{G}(t)\) is supposed to be piecewise constant with constant values over periods \([k\delta,(k+1)\delta)\). Hence, the~cost of generation over time interval \([k\delta,(k+1)\delta)\) is 

\begin{equation}
  \label{c5eq2}  
\sum_{k=k_0}^{N-1}  aP_{G}(k\delta)^2 + bP_{G}(k\delta) +c
\end{equation} 

In addition, the~power output of DG should always satisfy the constraints: 

\begin{equation}
  \label{c5eq3}  
\\P_G^{min} \leq \\P_{G}(t)\leq\\P_G^{max}\
\end{equation} 
with some given constants \(0< P_G^{min}\leq\ P_G^{max}\).

\subsubsection{Battery Energy Storage System}

We refer to the battery model introduced in \cite{c5fengji1}, and the dynamics of BESS used in this study is described by the following equation:

\begin{equation}
  \label{c5eq5}  
x((k+1)\delta) = x(k\delta)-P_B(k\delta)\Delta t + d|P_B(k\delta)\Delta t|,
\end{equation}
where \(x(\cdot)\) represents the available energy stored in BESS, $\Delta t$ is the factor used to convert power to energy based on the actual time of each control step, \(P_{B}(\cdot)\)  is the charging/discharging power of the BESS, \(d > 0\) is the charging/discharging loss factors of the BESS. Based on the model, \(P_{B}(\cdot) > 0\) indicates the discharging action and \(P_{B}(\cdot) < 0\) indicates the charging action of BESS. 

Based on the cost model introduced in \cite{c5kemeng2}, the~operational cost of BESS over the time interval \([k\delta,(k+1)\delta)\) is modelled as:
\begin{equation}
  \label{c5eq6}  
\sum_{k=k_0}^{N-1}  \gamma_{1}|P_B(k\delta)\Delta t|+\gamma_{2}x(k\delta), 
\end{equation}
where \(\gamma_{1}>0\) and \(\gamma_{2}>0\) are some given constants calculated based on the battery life degradation. In addition, the~following constraints should be satisfied: 
\begin{equation}
  \label{c5eq7}  
\\P_B^{min} \leq \\P_{B}(t)\leq\\P_B^{max}\ 
\end{equation}
\begin{equation}
  \label{c5eq8}  
\\x^{min} \leq \\x(t)\leq\\x^{max},\ 
\end{equation}
where $0<x^{min}<x^{max}$ are the upper and lower limits of the BESS energy state.

\subsubsection{Solar Power Generation Unit}

As the receding horizon approach is applied, it is assumed that prior to every control horizon, a predictive estimate $P_s(t)$ of the power output from the PV system can be made over periods \([k\delta,(k~+~1)\delta)\), which is supposed to be piecewise constant with constant values. {In addition, the~maximum power that can be supplied from the solar power system is $P_s^{max}$.} Since the receding horizon approach is used, some fast forecasting methods would be required to produce accurate results over a relatively short period, such as the techniques introduced in \cite {r24,c4windprediction}.

\subsubsection{Thermostatically Controlled Loads}

The microgrid also contains $n$ thermostatically controlled loads, labelled by $i=1,2,\ldots,n$. Based on previous research, we use the following continuous time equation to describe their~dynamics:
\begin{equation}
  \label{c5eq9} 
 \dot{T_i^{in}}(t)=\alpha_i(T^{out}(t)-T_i^{in}(t)-\beta_is_i(t)P_i)
\end{equation}
where $T^{out}(t))$ is the outside temperature at time $t$, $T_i^{in}(t))$ is the indoor temperature of the TCL $i$ at time $t$. $P_i$ is the rated power of the TCL $i$, $s_i(t)$ is the ON/OFF state of the TCL $i$ at time $t$ (0-OFF, 1-ON),  $\alpha_i>0$, $\beta_i>0$ are given constants. One of the control goals is to keep the indoor temperature in a comfortable range: 
\begin{equation}
  \label{c5eq10} 
  \\D_1 \leq \\T_i^{in}(t)\leq\\D_2,          \quad\forall t,  i,\ 
\end{equation}
where $D_1<D_2$ are given constants. Moreover, we assume that
\begin{equation}
  \label{c5eq11} 
  \\T^{out}(t)\geq\\D_2,          \quad\forall t,\ 
\end{equation}

Furthermore, all initial indoor temperatures satisfy

\begin{equation}
  \label{c5eq12} 
  \\T_i^{in}(0)\geq\\D_2,          \quad\forall t,\ 
\end{equation}

It is also assumed that we have a predictive estimate $\hat{T}^{out}(t))$ of the outside temperature $T^{out}(t))$, the~function $\hat{T}^{out}(t))$ is piecewise constant with constant values over periods \([k\delta,(k+1)\delta)\). Therefore, the~total TCL unit power consumption over the time interval \([k_0\delta,(N-1)\delta)\) is defined as:

\begin{equation}
  \label{c5eq13} 
  \sum_{i=1}^{n}(P_i\int_{k_0\delta}^{(N-1)\delta} s_i(t) dt)
\end{equation}


\subsection{Optimization Problem}\label{c5s4}

{The system control inputs include the DG power output $P_G(k\delta)$, the~charging/discharging power of the BESS $P_B(k\delta)$, and the switching functions $s_i(t)$ of the TCLs. In addition, the additive costs from both DG and BESS can be considered as the output of the system.}

Among these inputs, the~switching function $s_i(t)$ varies based on the measurement of the indoor temperature $T_i^{in}(t)$ only. The~control variable $P_G(k\delta)$ and $P_B(k\delta)$ are selected based on measurements of $x(k\delta)$, predictive estimates $\hat{T}^{out}(t))$  and $P_s (k\delta)$ of the outside temperature and the power output from the PV system respectively. Therefore, the~following constraints between generated and consumed power should hold:
\begin{equation}
  \label{c5eq14} 
  \sum_{i=1}^{n}P_is_i(t)\leq\\P_G(t)+P_s(t)+P_B(t)\leq\epsilon\sum_{i=1}^{n}P_is_i(t),
\end{equation}
where 
$\epsilon>1$ is a constant describing the maximum tolerance level of the overall power on the microgrid. As stated in previous sections, there is a non-zero minimum output requirement for the DG, so there will be cases where part of the generated solar power is curtailed to satisfy this constraint. In practice, this~excessive power can be balanced with other storage units or local loads in the microgrid, whereas we consider it as a part of the cost function to penalize unprofitable charging/discharging decisions in the optimization problem.

To state the problem, we combine the cost functions \ref{c5eq2} and \ref{c5eq5} into one cost function:

\begin{equation}
\begin{split}
  \label{c5eq15} 
    \sum_{k=k_0}^{N-1} aP_{G}(k\delta)^2+bP_{G}(k\delta)+c+\gamma_{1}|P_B(k\delta)\Delta t|+ \gamma_{2}x(k\delta)+C_{cur}(k\delta)P_{cur}(k\delta)  
\end{split}
\end{equation}  

The term $C_{cur}(k\delta)P_{cur}(k\delta)$ represents the cost of the excessive power curtailed to meet the constraint \ref{c5eq13}, where $P_{cur}$ is the amount of power curtailed, and $C_{cur}$ is the corresponding electricity price. Based on the system model defined, there are two scenarios in which this cost will be incurred, mainly due to the storage space and the maximum charging speed, which are: 

\begin{itemize}
\item The sum of solar power and the minimum power output of DG are higher than the power demand from TCLs, while the energy stored in the BESS is close to the upper limit, which cannot accommodate the excessive energy.
\item The sum of solar power and the minimum power output of DG are much higher than the power demand from TCLs, and the actual power on the microgrid is limited by the maximum charging speed of BESS, which cannot meet the constraint \ref{c5eq13}.
\end{itemize}

Let $d_{max}^c$ denote the maximum power that can be charged to the BESS within a control step, $P_{G}^{min}$ denote the minimum output of DG, and $P_{tcl} (K)$ denote the power demand from TCLs at control step $K$. For the above two conditions occurred at the step $K$, the~amount of power curtailed will be:

\begin{equation}
\label{c5eq16}
\begin{gathered}
 P_{cur}(K)=P_{G}^{min}+P_{s}(K)-\epsilon P_{tcl}(K) \\
 P_{cur}(K)=P_{G}^{min}+P_{s}(K)-d_{max}^c
\end{gathered}
\end{equation} 

Furthermore, we assume that the maximum output from DG $P_{G}^{max}$ will always be higher than the power demand from TCLs. 

{\bf Problem Statement:} Then, the~constrained optimal control problem is stated as follows: find control inputs $s_i(t)$, $P_G(k\delta)$ and $P_B(k\delta)$ such that the constraints (\ref{c5eq3}), (\ref{c5eq6}), (\ref{c5eq7}), (\ref{c5eq9}), (\ref{c5eq10}) and (\ref{c5eq11}) hold and the minimum of (\ref{c5eq12}) is achieved. Moreover, over all such control inputs, we find the control inputs such that the minimum of the cost (\ref{c5eq14}) is achieved. This problem can be perceived as the double optimization control problem, which can be solved as follows: 

Introducing the following switching rule for $s_i(t)$:

\begin{equation}
\label{c5eq17}
\begin{gathered}
 s_i(t)=0 \quad \textnormal{if} \quad T_i^{in}(t)<D_2   \\
 s_i(t)=1 \quad \textnormal{if} \quad T_i^{in}(t)\geq\ D_2
\end{gathered}
\end{equation} 

It is clear that for some large enough $k_0$, the~controller
(\ref{c5eq17}) will keep the variables $T_i^{in}$
at the level $D_2$ which is the maximum comfortable temperature. Hence, this controller will deliver the minimum
of the cost (\ref{c5eq12}) for large enough $k_0$. Now
to find the control input that delivers the minimum of the cost~(\ref{c5eq14}), for all $k=k_0,k_0+1,\cdots,N$, $x\in[P_B^{min},P_B^{max}]$, introduce the Lyapunov-Bellman equation $V(k\delta,x)$ as follows

 \begin{align}
 \label{c5eq18}
V(N\delta,x)    := {} & 0 \quad \forall x\in[P_B^{min},P_B^{max}]
\end{align}
\begin{align}
\label{c5eq19}
V(k\delta,x)    := {} & V((k+1)\delta,x)+\min_{P_G, P_B\in\Omega_k}(aP_{G}(k\delta)^2+bP_{G}(k\delta)+c+ 
                \gamma_{1}
	                |P_B(k\delta)|\\
                {} & +\gamma_{2}x(k\delta)+C_{cur}(k\delta)P_{cur}(k\delta))\nonumber  
\end{align}
where $\Omega_k$ is the set of $(P_G, P_B)$ such that $P_G\in[P_G^{min},P_G^{max}], P_B\in[P_G^{min},P_G^{max}]$ and 
 \vspace{6pt}
\begin{align}\label{c5eq20}
P_G+P_B\geq {}-\hat{P_s}(k\delta)+\sum_{i=1}^{n} (\dfrac{\hat{T}^{out}(k\delta)-D_2}{\beta_i}) \\[2pt]
\label{c5eq21}
P_G+P_B\leq {}-\hat{P_s}(k\delta)+\epsilon \sum_{i=1}^{n}(\dfrac{\hat{T}^{out}(k\delta)-D_2}{\beta_i}) 
\end{align}

Moreover, the~optimal value $P_G (k\delta)$ and $P_B (k\delta)$ are the values $(P_G,P_B)$ for which the minimum in~(\ref{c5eq18}) is achieved.

It follows from the Bellman optimality principle (dynamic programming principle), see, e.g., \cite{bk1}, that the controller defined by (\ref{c5eq18}) and (\ref{c5eq19}) delivers the minimum of the cost (\ref{c5eq15}). Hence, we have derived the following statement.

\textbf{Proposition 1}:  There exists an integer $K$ such that for all $K<k_0<N$, the~control inputs defined by (\ref{c5eq17})--(\ref{c5eq19}) is the solution to the double optimization control problem over the time~interval~$[k_0,N\delta]$.

\textbf{Remark}: The proposed system with the controller (\mbox{\ref{c5eq17}}) can be considered as a hybrid dynamical system or switched controlled system \mbox{\cite{hybrid1,hybrid2,hybrid3,hybrid4}}. The controller (\mbox{\ref{c5eq17}}) is a sliding mode controller. The~optimal control input corresponding to the sliding mode requires infinitely fast switching. In practice, ON-OFF switching in TCLs cannot be too fast, which results in some difference between the optimal and actual values of the control variables. Equations (\ref{c5eq18}) and (\ref{c5eq19}) are standard dynamic programming type equations. Since $x,P_G,P_B$ are scalar variables, these equations are computationally easy.

{\bf Model Predictive Control (MPC):} We apply the model predictive control approach as follows. Let an integer $N_0 > 0$  be our MPC horizon. For any $k_0 \geq 0$, $N:=k_0+N_0$ and any $x_0 \in  [x^{min},x^{max}]$, we find the optimal control inputs  
 which deliver the minimum in the double optimization 
 control problem and use the standard MPC scheme.

It should be pointed out that the proposed controller (\ref{c5eq17})--(\ref{c5eq19}) is decentralized. In (\ref{c5eq17}), the~controller for each TCL requires only information on the indoor temperature in the corresponding area (room, building). The~controller
(\ref{c5eq18}) and (\ref{c5eq19}) does not require any information on indoor temperatures and TCLs' controllers. Moreover, the~proposed control algorithm is easily scalable. The~dynamic programming part algorithm stays the same as the number of TCL units increases. The~variable structure control part of the algorithm is fully decentralized so as the number of TCL units increases, the~time required by the algorithm stays the same. 


\subsection{Simulation}\label{c5s5}
\vspace{-8pt}
\subsubsection{Set Up}\label{c5s5.1}

\begin{figure}
\centering 
\includegraphics{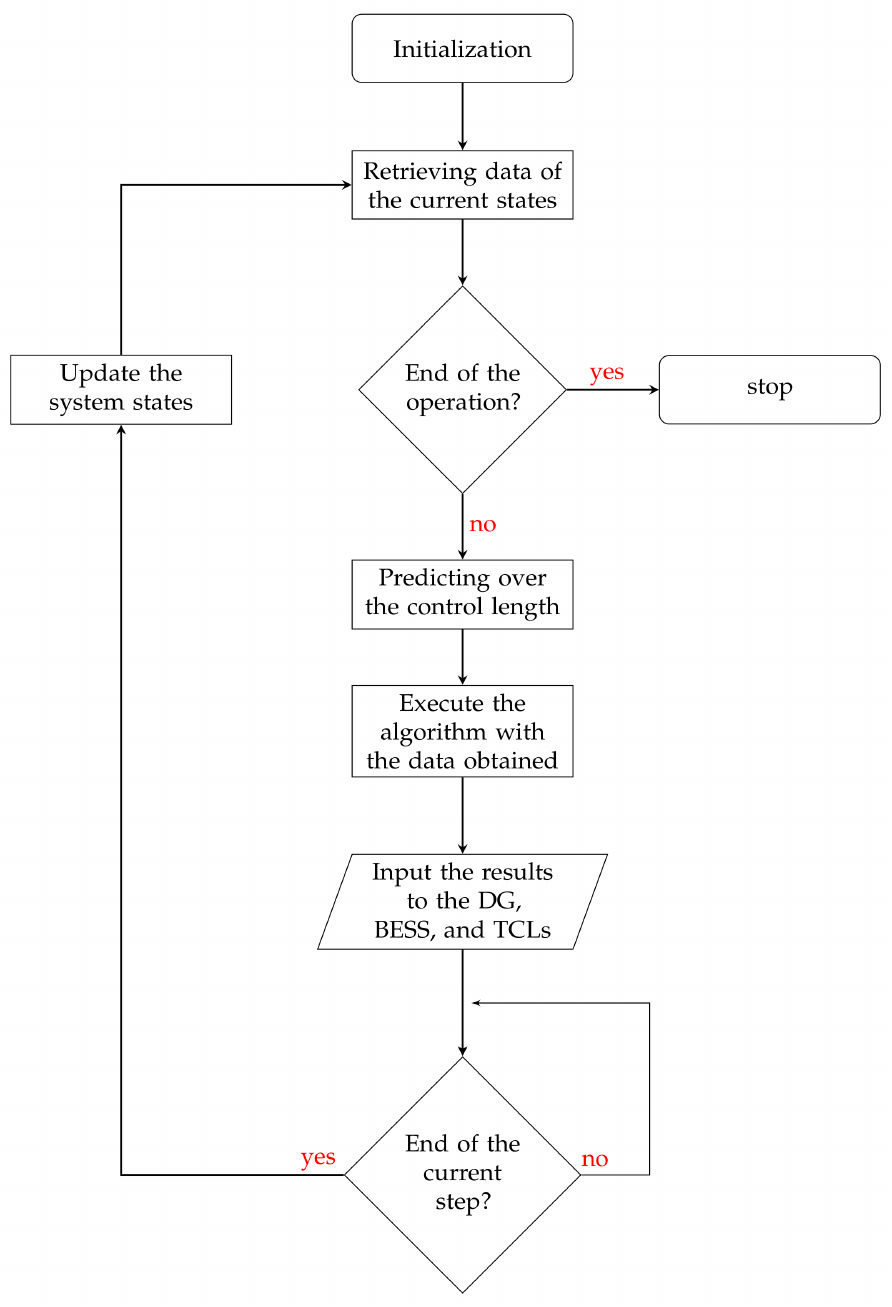}
\caption{Flow 
chart of the proposed algorithm.}
\label{c5f2}
\end{figure}

The proposed strategy is tested with computer simulation, and the process of algorithm is described with the flow chart on Figure \ref{c5f2} with the parameters of BESS, DG, and TCLs used in the simulation summarized in Table \ref{c5t1}. To verify its effectiveness, a comparison study is conducted with the control strategy introduced in \cite{c5fengji1}.
 
\begin{table}
  \centering
  \caption{System
  parameters.}
  \label{c5t1}
\begin{tabular}{cccccc}
\toprule
DG   & $P_G^{min}$   & $P_G^{min}$   & $a$     & $b$     & $c$  \\
     & 50 kW        & 500 kW       & 0.01  & 0.1   & 0  \\ \midrule
BESS & $P_B^{min}$   & $P_B^{max}$   & $\gamma_1$    & $\gamma_2$    & $C_{cur}$ \\
     & 120 kw & 0     & 0.008 & 0.008 & 60 /kWh \\   \midrule
TCL  & $n$     & $D_2$    & $\beta_i$    & $\epsilon$     &    \\
     & 3200  & 23    & 300   & 1.05  & \\ 
      \bottomrule
  \end{tabular}
\end{table}

\subsubsection{System Parameters and Database}\label{c5s5.2}

The rated capacity of BESS used in the simulation is 240 kWh. We~choose 0.9 and 0.1 state-of-charge of BESS as the upper and lower bound for the BESS energy state to avoid overcharging/deep discharging. In addition, it is assumed that the actual time of each control step is 10 min, so the predictive data on PV power and outdoor temperature are average values based on 10-min observations, and the factor $\Delta t$ to convert power to energy is $1/6$. Furthermore, we consider a control horizon of one hour and there will be six control steps in each horizon. 

\begin{figure}
  \centering
  \includegraphics[width=13 cm,trim=0 0 0 2,clip]{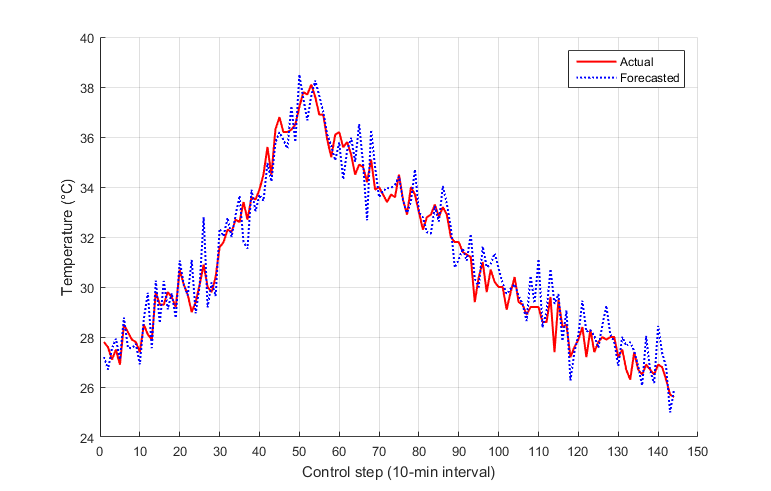}
  \caption{Actual and forecasted outdoor temperature.}
  \label{chap5fig3}
\end{figure}

\begin{figure}
  \centering
  \includegraphics[width=13 cm,trim=0 2 0 2,clip]{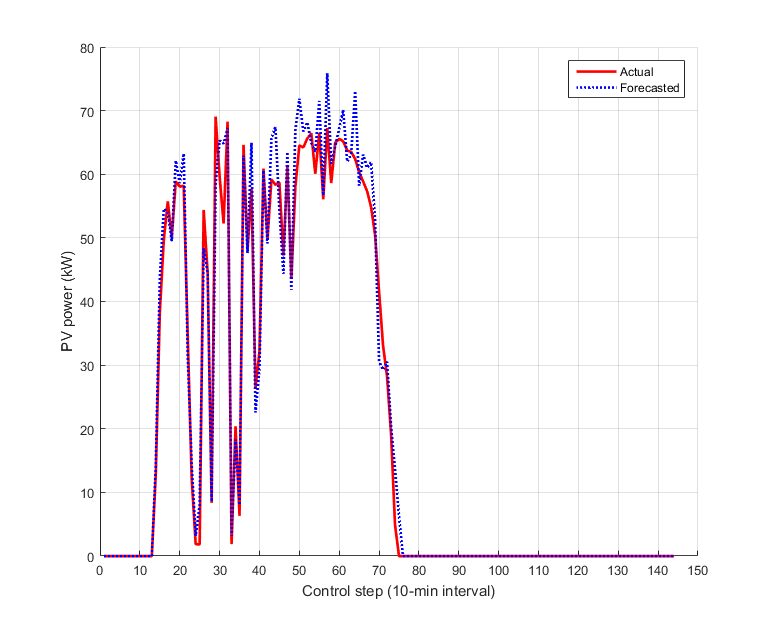}
  \caption{Actual and predicted solar power.}
  \label{chap5fig4}
\end{figure}

The solar power data is retrieved from the St. Lucia Concentrating Array located in UQ Photovoltaic Sites, and the outdoor temperature data used is based on temperature observations of Sydney. As these are actual data, we include some forecasting errors, which are normalized mean absolute errors that ranges between 5\% and 14\%. Furthermore, the~24-h solar power and outdoor temperature data are recorded from 5 a.m. on the current day to 4.50 a.m. the next day. As the sunset time is around 5 p.m. in the simulation data, there is no solar power generated afterward. PV and temperature data used in our simulation are summarized in Figures \ref{chap5fig3} and \ref{chap5fig4}.

\unskip

\subsubsection{Simulation Results}
\begin{figure}
  \centering
  \includegraphics[width=13 cm,height=8 cm]{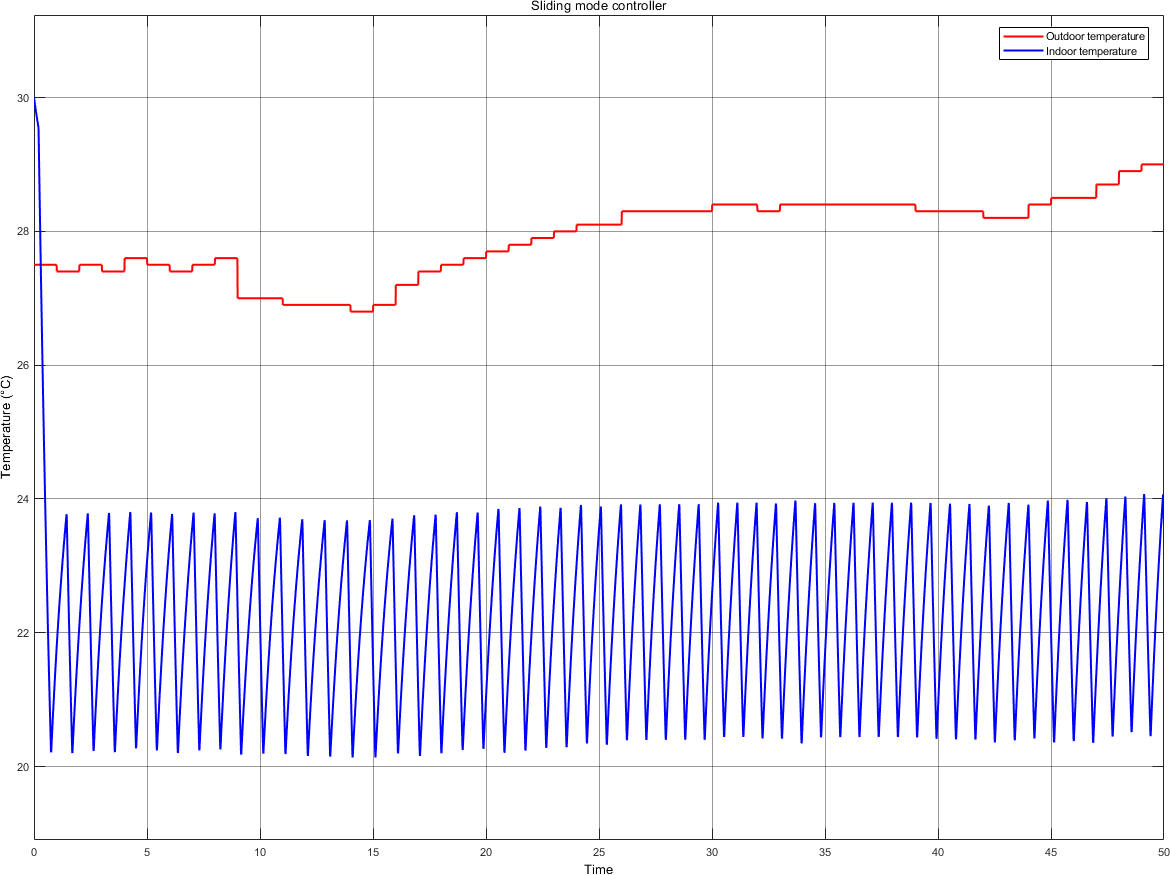}
  \caption{Ambient and controlled indoor temperature.}
  \label{chapfig5}
\end{figure}

\begin{figure}
  \centering
  \includegraphics[width=13 cm]{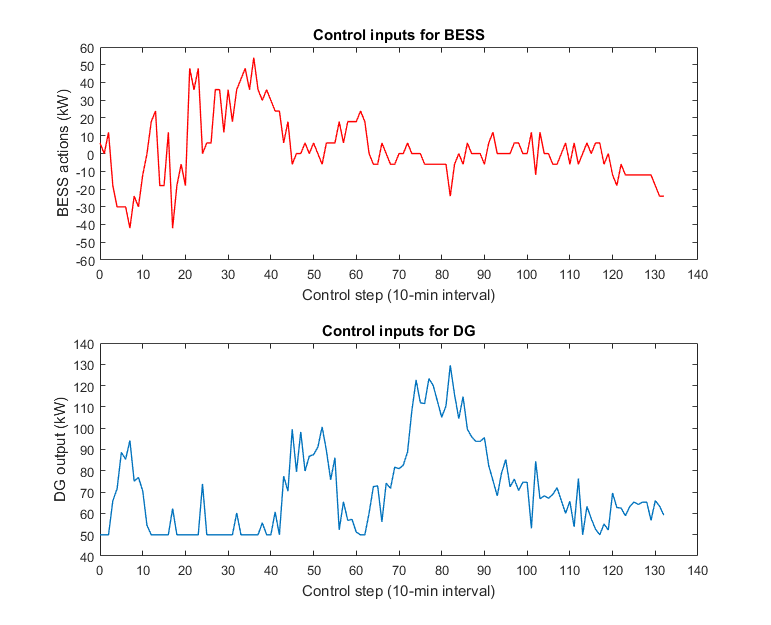}
  \caption{Simulation results of battery energy storage system (BESS) action and distributed generator (DG) output.}
  \label{chapfig6}
\end{figure}

The indoor temperature controlled with the variable structure controller is illustrated in Figure~\ref{chapfig5}. We include a 0.2 per step time delay in the controller, which is 2 minutes in practice. With a set point temperature of 23 Celsius, the~outdoor temperature and the corresponding indoor temperature over the first 50 control steps are shown in Figure \ref{chapfig6}. In addition, with an initial battery energy state of 120 kWh, the~control inputs for the BESS and the DG solved with the proposed strategy are summarized in Figure \ref{chapfig6}. Based on the obtained results, one can observe that most of the charging actions are solved over the initial steps of the simulation where the temperature is relatively lower than other periods. As a result, the~excess solar power would be charged into the BESS. In addition, the~outdoor temperature gradually increases and reaches the peak at step 50, and the majority of the discharging actions of the BESS can be observed over this period. The~corresponding operational cost calculated with the simulation results is around 8606.98. 

\unskip

Since the rolling horizon approach is utilized, we record only the first components in the sets of BESS actions and the corresponding output power of DG. It should be pointed out that the forecasted PV and temperature data are used to compute these actions; while the system states are updated with the actual PV and temperature data. Furthermore, as the initial state is on 50\% of the battery state-of-charge, there is no power curtailment during the one-day simulation.


\subsubsection{Comparison Study}\label{c5s5.4}

To evaluate the effectiveness of our algorithm, we compare it with the control scheme introduced in \cite{c5fengji1}, in which a differential evolution (DE) algorithm is used. The~DE algorithm is a population-based heuristics optimizer, and the choice of parameters in the algorithm, such as the population size, crossover rate, and mutation factor, could affect the optimization performance. Therefore, this could increase the difficulty of implementing it in practice with the rolling horizon approach, as the microgrid operators are required to adjust those parameters for different sets of data. The~DE algorithm can be used to solve the minimization problem $min_{x_1,\dots x_n}f(x_1,x_2,\cdots,x_n)$, where $x_i$ are the variables.

\begin{figure}
  \centering
  \includegraphics[width=13 cm,height=10.5cm]{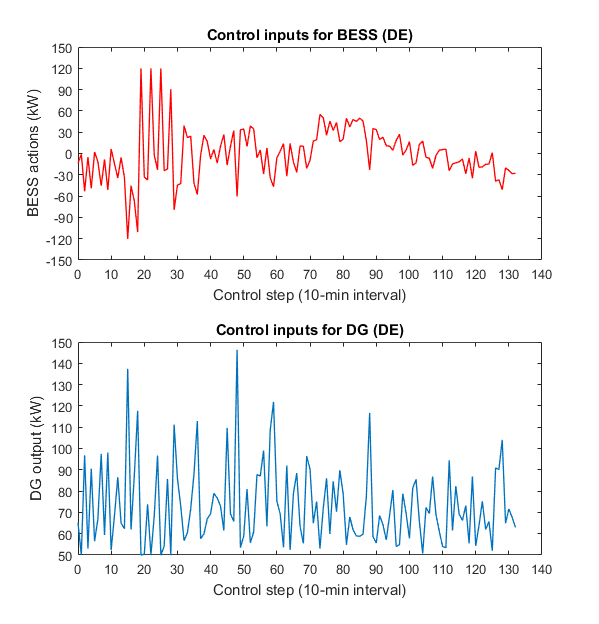}
  \caption{Control inputs for BESS and DG solved with DE algorithm.}
  \label{chap5fig7}
\end{figure}

In our case, the~variables are control inputs for BESS. The~mutation factor and crossover rate used in our simulation are both 0.9, and the maximum number of iteration is 50. Furthermore, we test the DE algorithm with the same simulation scenario considered in Section \ref{c5s5.2}, and the results obtained with the population size of 60 is presented in Figure \ref{chap5fig7}. It should be pointed out that the results obtained with the DE algorithm are inconsistent, and we need to repeat the algorithm multiple times to achieve the best result. This could be undesirable for real-time scheduling. In our case, the~corresponding operational cost over the one-day simulation is around 9527.65, which is much higher than the result solved with the proposed algorithm in this paper. Furthermore, we test the DE algorithm under different population sizes, and results are summarized~in~Table~\ref{c5t2}.

\unskip
\begin{table}
  \centering
  \caption{{Cost} of system units under different conditions.}
  \label{c5t2}
\begin{tabular}{cccc}
\toprule
                                                    & \textbf{BESS Operation Cost \boldmath{(\$)}} & \textbf{DG Operation Cost \boldmath{(\$)}} & \textbf{Total }  \\ \midrule
DE  NP:60 & 8541.54                                                              & 986.11                                                             & 9527.65 \\ 
DE NP:120 & 8389.07                                                              & 1003.39                                                            & 9392.46 \\ 
DE NP:180 & 8209.49                                                              & 814.89                                                             & 9024.39 \\ 
Proposed  scheme   & 8181.21                                                              & 425.76                                                             & 8606.98 \\ \bottomrule
\end{tabular}
\end{table}

It can be perceived that the results can be improved at higher population sizes. By further increasing the population size, (e.g., up to 100 times of the variables) we could obtain results that are close to the one solved by the proposed scheme, but it would require multiple runs of the algorithm with adjustments on the parameters. In addition, it should be pointed out that the conditions to curtail the excessive energy are not considered in \cite{c5fengji1}. As a result, during our simulation, we found that with some sets of simulation data, such as a relatively low initial BESS state, the~DE algorithm can generate results containing errors because the constraints set in the algorithm code cannot be satisfied. Since the model predictive control approach is used in our scheme, we believe that the dynamic programming algorithm will be a more appropriate choice as it does not require any adjustments in the code during the real-time operation. In addition, the~actual level of reduction could be more significant if the unused energy stored in the BESS at the end of the process is considered.


\subsubsection{Discussion}

The proposed method can achieve better results than the one used for comparison. In~consideration of the practical application, the~proposed approach could be more reliable since the adjustment of parameters in the algorithm is unnecessary. Furthermore, the~decentralized control scheme is critical to the real-time energy scheduling in the microgrid. Based on the previous sections, it~can be perceived that the control of TCLs is only based on the indoor temperature in the corresponding area, and the operation of the generator and batteries are independent of the current measurements of indoor temperature. In order words, direct controls and communications between the central energy management system and the distributed TCLs are not required, which significantly improve the efficiency of the system, and the cost of constructing the whole system can be reduced.

\subsection{Summary}
A microgrid power scheduling strategy is introduced in this chapter. The~objective was to minimize the cost required to keep the room temperature under the desired level. A decentralized control scheme was proposed. The~controller for each TCL needs to measure only indoor temperature in the corresponding area. The~control input for the BESS charge/discharge does not need to have any measurements from TCL units, so the proposed
system does not require any significant data communication subsystem. The~conducted computer simulations showed
that the proposed control scheme significantly outperforms other control algorithms. The~proposed method is fully decentralized, computationally efficient and easily scalable.


\section{Renewable Energy Fluctuation Smoothing\label{cha6}}

 Due to inherent variability and intermittency of renewable energy, high penetration of renewable energy in the current electricity market could be difficult to achieve. Use of energy storage units and controllable loads are some of the feasible solutions to regulate renewable energy. The basic idea is to store/supply/consume part of the generated renewable energy, so that the actual power dispatched to the utility grid can meet the ramp rate requirements. In this chapter, we propose a wind power smoothing strategy with the coordination of the battery energy storage system (BESS) of thermostatically controlled loads (TCLs). We considered a group of cooling TCLs, and the unsmoothed wind power will be regulated by them to minimize its fluctuation first, followed using a dynamic programming-based algorithm to determine a set of BESS actions that adjust the actual power delivered to the grid and minimize the operational cost of BESS. The rolling horizon approach is utilized with predicted wind power and ambient temperature data updated at each control step. This chapter is partly based on my previous publication \cite{zhuo4}.

\subsection{Introduction} \label{c6s1}
\unskip
In recent years, an increasing number of wind power systems have been integrated into the existing power grid as a measure to reduce emissions of greenhouse gases. Except for the capital costs, there are basically no generation costs and carbon emissions during the process of wind power production. As a result, it becomes one of the fastest growing renewable energy and attracts attentions for both researchers and investors \cite{c6r1}. Nevertheless, this growing use of renewable energy is adding significant amounts of uncertainty to the grid management and generation scheduling \cite{c6r2}. Due to the inherent variability and intermittency of wind energy, grid operators must change their current approaches to unit commitment and economic dispatch, and assign appropriate online generators in response to what produced by wind turbines to maintain network stability and meet reliability parameters \cite{c6r3}.

One feasible solution to smooth wind power fluctuation is utilizing energy storage units to control the actual power dispatched to the grid, which include battery energy storage systems (BESSs), flywheel energy storage system, and energy capacitor systems \cite{c6exr1,c6exr2,c6exr3}. In other studies, the use of thermostatically controllable loads (TCLs) can balance the variability of renewable energy as well \cite{c5r2,c6exr4}. Residential appliances such as heaters, ventilation systems, and air-conditioning systems are some of the widely used TCLs, which are flexible loads suitable to regulate the output power of renewable energy systems \cite{c6exr5,c6exr6}. Also, these TCLs account for a large proportion of overall electricity consumption, which have the potential to provide ancillary services such as load following and frequency regulation \cite{c6exr7,c6exr8}. 

In most of these studies, the control schemes are based on predicted data of renewable power. In \cite{r5} and \cite{c6exr9}, the authors propose a model predictive control (MPC)-based wind power smoothing strategy with battery energy storage system (BESS). The predicted wind power data over a future horizon are used to solve a set of BESS actions that optimizes the ramp rate of the wind power dispatched. Since the MPC is implemented, only the first component in the solution set will be used as the actual input to the system at the current stage, and the controller will repeat in the next stage with updated forecasting data and system states. A state-of-charge-based adaptive power control strategy for smoothing power fluctuations of hybrid wind solar system is introduced in \cite{c6exr10}, in which predicted data of wind and solar power are used to determine a target power of BESS that keeps the battery state-of-charge within a certain range to prevent a forced shutdown of BESS. Regarding the use of TCLs, authors in \cite{c5r6} propose a real-time two-stage optimization model to smooth the power fluctuations with TCL groups. The regulation capacity of TCLs is examined and the optimal fluctuation ranges are determined based on forecasting data in the first stage; and the coordination among different TCL groups is solved in real-time during the second stage. 

It is stated that without proper considerations in the BESS capacity and operation strategy, the cost of BESS can be expensive for renewable energy applications \cite{c4khalid1}. In the case of renewable power smoothing, batteries are expected to experience more charging and discharging cycles than some regular tasks, and their aging process can be accelerated \cite{c6exr11}, which could incur additional costs from the reduced life expectancy of batteries. On the other hand, TCLs lack the ability to store and supply power for a long period of time, so they can have a poor performance in smoothing the wind power on extreme ramping events, such as a sudden decrease in the wind power generation. 

As a result, in this chapter, we propose a wind power smoothing strategy with the coordination of BESS and TCLs. The main reason for using both components is to reduce the operational cost of BESS. TCLs can be used to consume part of the generated wind power, and actions of BESS required to smooth power fluctuations can be reduced, which is likely to increase the life expectancy of batteries. In addition, compared with utilizing BESS and TCLs individually, this structure can be less vulnerable to ramping events. In the case of a sudden increase or decrease of wind generation, the BESS could be limited by its allowable state-of-charge and rate of charge/discharge that would require power curtailment and the use of other online ancillary services. Hence, TCLs can serve as good compensation to the BESS in those extreme cases. 


\subsection{Problem Statement} \label{c6s2}

We consider a wind farm integrated with a battery energy storage system (BESS) and a group of thermostatically controlled loads (TCLS).  BESS and TCLs are used to adjust the actual power dispatched to the utility grid, so that the ramp rate requirements can be satisfied. In addition, the operational cost of BESS should be minimized, and the TCL units are required to keep the desired room temperature.

A discrete-time system model is considered in our work, and most data used in the algorithm are piecewise functions with constant values over each control step. Since forecasting data are included in the algorithm, based on those previous studies, we utilize the rolling horizon approach to constantly update the prediction on wind power and ambient temperature at each control step, which is to reduce the uncertainties due to errors in forecasting. 

The TCL units we considered are air-conditioners, which are responsible for maintaining a desired indoor temperature based on the predicted outdoor temperature. They formulated as an aggregation model rather than directly control individual unit, as the practical TCL groups are usually constructed in large quantity with small capacity. In this paper, we propose an aggregated TCL model, in which only the outdoor temperature is required to compute the power consumption, which reduces the computation complexity of the TCL model, and will be compatible with the rolling horizon approach that has a strict requirement for the computation time.   

The cost function of the optimization problem is the additive operational cost of BESS over each control horizon, and we use a dynamic programming algorithm to solve the minimization problem. As the rolling horizon approach is used, a set of optimal actions of BESS will be solved at each control step, and only the first element will be used as the actual input to the system and the algorithm will repeat in the next control stage with updated system states and predicted data.


\subsection{System Model} \label{c6s3}

Our system is considered as a discrete-time system with the sampling rate $ \delta>0$, so most data used in the algorithm are piecewise constants with constant values over periods $[k\delta,(k+1)\delta)$, which include the predicted wind power, ambient temperature, and charging/discharging power of BESS.

\subsubsection{Thermostatically Controlled Loads } \label{c6s3.1}

Suppose there are $n$ TCLs integrated with the wind power system, based on the TCL model introduced in \cite{c5fengji1}, we use the following continuous equation to describe their dynamics:

\begin{equation}
  \label{c6eq1} 
 \dot{T_i^{in}}(t)=\alpha_i(T^{out}(t)-T_i^{in}(t)-\beta_is_i(t)P_i)
\end{equation}

where $T^{out} (t)$ is the outdoor temperature at time $t$, and $T_i^{in} (t)$ is the indoor temperature of the TCL $i$ at time $t$. $P_i$ is the rated power of the TCL $i$, $s_i (t)$ is the ON$/$OFF state of the TCL $i$ at time $t$ (0$-$OFF, 1$-$ON),$\beta_i>0$ is a given constant. One of the control objectives is to keep the indoor temperature in a comfortable range: 

\begin{equation}
  \label{c6eq2} 
  \\D_1 \leq \\T_i^{in}(t)\leq\\D_2,          \quad\forall t,  i,\ 
\end{equation}
where $D_1<D_2$ are given constants. Moreover, we assume that
\begin{equation}
  \label{c6eq3} 
  \\T^{out}(t)\geq\\D_2,          \quad\forall t,\ 
\end{equation}

And all initial indoor temperatures satisfy

\begin{equation}
  \label{c6eq4} 
  \\T_i^{in}(0)\geq\\D_2,          \quad\forall t,\ 
\end{equation}

Suppose that we have a predictive estimate $\hat{T}^{out}(t)$ of the ambient temperature $T^out (t)$, and the function $T^out (t)$ is piecewise constant with constant values over periods $[k\delta,(k+1)\delta)$. Therefore, the total TCL unit power consumption over the time interval $[k_0 \delta,(N-1)\delta)$ is defined as 

\begin{equation}
  \label{c6eq5} 
  \sum_{i=1}^{n}(P_i\int_{k_0\delta}^{(N-1)\delta} s_i(t) dt)
\end{equation}

\subsubsection{Battery Energy Storage System} \label{c6s3.2}

We refer to the BESS cost model introduced in \cite{c5kemeng2}, and the operational cost of BESS is described by the following equation: 

\begin{equation}
  \label{c6eq6}  
 C_B(t)=\gamma_{B}|P_B(k\delta)\Delta t|+\gamma_{B}x(k\delta), 
\end{equation}
where $P_B$ is the BESS discharging/charging power, $E_B (t)$ is the energy stored in BESS at $t$, $\Delta\delta$ is the factor based on the actual time of each control step to convert power to energy, and $\gamma_{B}$ is calculated as follows:

\begin{equation}
  \label{c6eq7}  
 \gamma_{B}=\dfrac{IC_B}{E_{R,B}LCN}    
\end{equation}
where $IC_B$ is the investment cost of BESS, $E_{R,B}$ is the rated capacity of BESS, and $LCN$ is the number of battery life cycles before its end of life.

The dynamics of the BESS is described with the following equation:

\begin{equation}
  \label{c6eq8}  
E_B(t+1) = E_B(t)-P_B(t)\Delta t + d|P_B(t)\Delta t|,
\end{equation}

Based on the equation, it can be perceived that $P_B (t)>0$ indicates the discharging action and $P_B (t)<0$ indicates the charging action of BESS. Furthermore, $d>0$ is the charging/discharging loss factor of the BESS. 

The BESS is subjected to the following constraints:

\begin{equation}
  \label{c6eq9}  
\\P_B^{min} \leq \\P_{B}(t)\leq\\P_B^{max}\ 
\end{equation}
\begin{equation}
  \label{c6eq10}  
\\E_B^{min} \leq \\E_B(t)\leq\\E_B^{max},\ 
\end{equation}

where $P_B^{min}<P_B^{max}$ are the maximum discharging/charging rates of BESS, and $0\leq E_B^{min}\leq E_B^{max}$ are the lower and upper limits for the state-of-charge of BESS.


\subsubsection{Wind Power System} \label{c6s3.3}

It is assumed that a predictive estimate $P_w (t)$ of the power output from the wind power system can be made over periods $[k\delta,(k+1)\delta)$, which is supposed to be piecewise constant with constant values. Based on the models defined in previous sections, the actual power dispatched to the grid $P_G (t)$ can be described as:

\begin{equation}
  \label{c6eq11}  
P_G(t+1)=P_w(t)+P_B(t)-\sum_{i=1}^{n} P_is_i(t)
\end{equation}

Suppose that the BESS will not charging from the grid, then $P_G (t)\geq 0,\forall t$. Also, we use $\Delta P_G$ to indicate the rate of change of the dispatch power, which will be:

\begin{equation}
  \label{c6eq12}  
\Delta P_G(t) =P_G(t+1)-P_G(t)
\end{equation}

And the constraint used to smooth the wind power is presented as follows:

\begin{equation}
  \label{c6eq13}  
RR_{min} \leq \Delta P_G(t)\leq RR_{max}
\end{equation}

where $RR_{max}$ and $RR_{min}$ are the ramp rate limits on the utility grid. 


\subsection{Optimization Problem} \label{c6s4}

Based on system models proposed in the previous section, the control inputs include the charging/discharging power of the BESS $P_B (t)$, and the switching functions of the TCLs. The switching function $s_i (t)$ changes based on the measurements of the indoor temperature $T_i^in (t)$, and the corresponding TCL power can be formulated as a function of the ambient temperature and a set point temperature. The calculation of $P_B (t)$ is based on the predictive output of the wind power system, and states of the BESS.

A two-stage control scheme is developed. In the first stage, part of the generated wind power will be consumed by the TCLs to minimize the fluctuation and the indoor temperature should be controlled within the desired range. In the second stage, BESS actions are determined to regulate the actual power dispatched to the electric grid. 
\subsubsection{Stage I} \label{c6s4.1}

Suppose there are $N$ steps in the predictive horizon, and prior to each control horizon, two sets of predictive data, namely the wind power and ambient temperature, are provided. Also, initial values of energy stored in BESS $E_B (0)$, and dispatched wind power $P_G (0)$ in the previous control interval are known.

First, we introduce the following switching rule for $s_i (t)$:

\begin{equation}
\label{c6eq14}
\begin{gathered}
 s_i(t)=0 \quad \textnormal{if} \quad T_i^{in}(t)<D_S(t)   \\
 s_i(t)=1 \quad \textnormal{if} \quad T_i^{in}(t)\geq\ D_S(t)
\end{gathered}
\end{equation} 

where $D_2\leq D_s \leq D_1$. Also, we define a new term $Pw_{tcl}(t)$  as the wind power regulated by the TCL power, which is:

\begin{equation}
  \label{c6eq15}  
Pw_{tcl}(t)=P_w(t)-\sum_{i=1}^{n} (\dfrac{\hat{T}^{out}(t)-D_s(t)}{\beta_i})
\end{equation}

Given predicted $\hat{T}^{out}(t)$ and $\hat{T}^{out}(t)$ over the control horizon of $N$ stages, a set of optimal $D_s^{x} (t)$ should be solved for the following minimization problem:

\begin{equation}
  \label{c6eq16}  
\min \sum_{t=0}^{N-2} (Pw_{tcl}(t+1)-Pw_{tcl}(t))^2
\end{equation}

This problem can be solved with some simple nonlinear programming solver such as $fmincon$ in MATLAB. And the corresponding set of $Pw_{tcl}^{\star}(t)$, determined by the set of $D_s^{\star}(t)$, is then used in the second step.


\subsubsection{Stage II} \label{c6s4.2}

The main objective of our problem is to minimize the additive cost of BESS over each predictive horizon, so the optimization problem can be formulated as:

\begin{equation}
  \label{c6eq17}  
\min \sum_{t=0}^{N-1} \gamma_{B}|P_B(k\delta)\Delta t|+\gamma_{B}x(k\delta)
\end{equation}

subjected to constraints (\ref{c6eq2}), (\ref{c6eq3}), (\ref{c6eq4}), (\ref{c6eq9}), (\ref{c6eq10}), (\ref{c6eq11}), (\ref{c6eq12}) and (\ref{c6eq13}). To solve this problem, we introduce the Lyapunov-Bellman function $V(k\delta,x)$ for all $k=k_0,k_0+1,\cdots,N, x\in[P_B^{min},P_B^{max}]$, which will be:

\begin{equation}
\label{c6eq18}
\begin{gathered}
V(N\delta,x):=0    \\
V(k\delta,x):=V((k+1)\delta,x)+\min_{P_B} (\gamma_{B}|P_B(k\delta)\Delta t|+\gamma_{B}x(k\delta))
\end{gathered}
\end{equation} 

where $P_B$ is subjected to the grid power constraint

\begin{equation}
  \label{c6eq19}  
P_G(t+1)=P_B(t)+Pw_{tcl}^{\star}(t),
\end{equation}

and ramp rate constraints (\ref{c6eq12}) and (\ref{c6eq13}).

It can be perceived that the controller (\ref{c6eq14}) is a sliding mode controller. The optimal control input corresponds to a sliding mode which requires infinitely fast switching. In practice, ON-OFF switching in TCLs cannot be too fast, which results in some difference between the optimal and actual values of the control variables. The equation (\ref{c6eq18}) is dynamic programming type equations. Since  $P_B$ is a scalar variable, these equations are computationally easy.


\subsection{Simulation} \label{c6s5}

\subsubsection{Database} \label{c6s5.1}

Our strategy is examined with the simulation software MATLAB. Parameters of BESS and TCLs used in the simulation are summarized in the following Table.

\begin{table}
\caption{System parameters.}\label{c6t1}
\centering

\begin{tabular}{ccccc}
\hline
\textbf{BESS} & $P_B^{max}$	& $P_B^{min}$ & $\gamma_B$	& $\Delta\delta$ \\
\hline
{}		      & $30MW$		&  $0$    & $0.048$	& $1/6$\\
\hline
\textbf{TCL}  & $n$			& $\beta_i$    & $D_1$ 			& $D_1$   \\
\hline
{}		& $320$			& $300$    & $25^\circ C$	& $20^\circ C$\\

\hline
\end{tabular}
\end{table}

The rated capacity of BESS used in the simulation is $240MWh$. To avoid over charging and/or discharging, we  choose 70\% and 30\% state-of-charge as the upper and lower bound for the BESS. Also, the actual time of each control step is set to 10 minutes with a predictive horizon of one hour, which means there are 6 control steps in each horizon. We also consider a maximum ramp rate of $2MW$ per minute, so the ramp rate limits $RR_{max}$ and $RR_{min}$ are $20MW$ and $-20MW$ respectively.

The wind power data is retrieved from the $140MW$ Woolnorth wind farm in Tasmania Australia, and the ambient temperature data used is based on 10-minute temperature observations of Darwin.We include some forecasting errors in the simulation, which are normalized mean absolute errors that ranges between 5\% and 14\%. The wind power and temperature data used in our simulation are presented in the following two figures.

\begin{figure}
  \centering
  \includegraphics[width=13 cm,trim=0 0 0 2,clip]{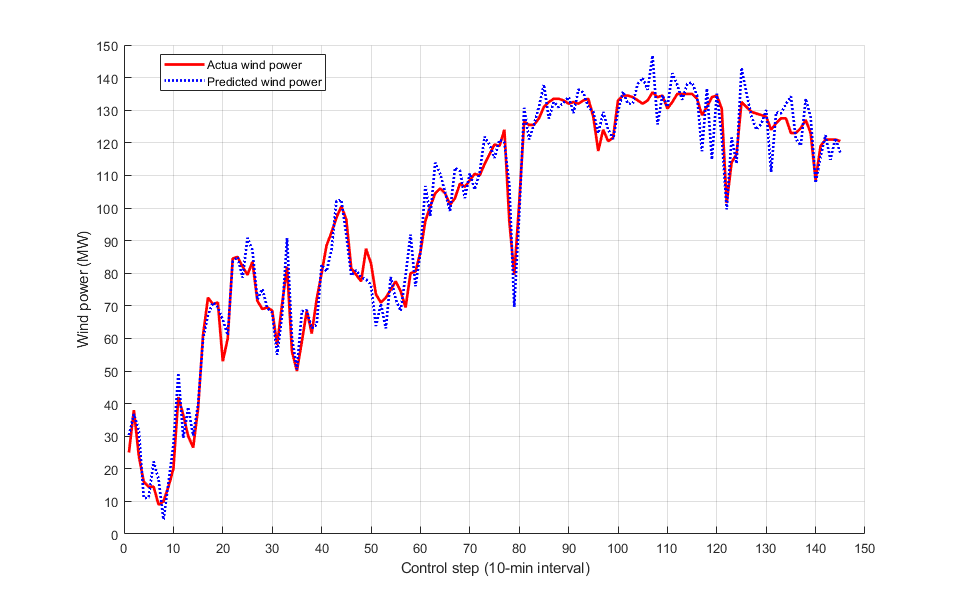}
  \caption{Actual and predicted wind power.}
  \label{chap6fig1}
\end{figure}
\unskip
\begin{figure}
  \centering
  \includegraphics[width=13 cm,trim=0 2 0 2,clip]{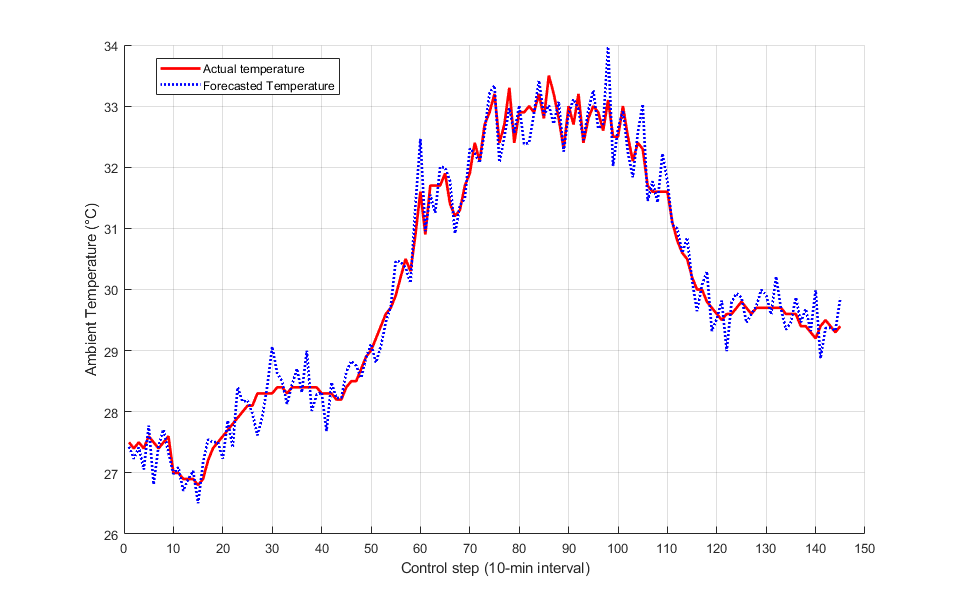}
  \caption{Actual and predicted temperature.}
  \label{chap6fig2}
\end{figure}

\subsubsection{Simulation Results} \label{c6s5.2}

The following figure illustrates how the indoor temperature varies with the sliding mode controller. We include a 0.2 per step time delay in the sliding mode controller, which is 2 minutes in practice. The set point is $25^\circ C $, and the corresponding results over the first 50 control steps is shown below 

\begin{figure}
  \centering
  \includegraphics[width=13 cm,trim=0 0 0 2,clip]{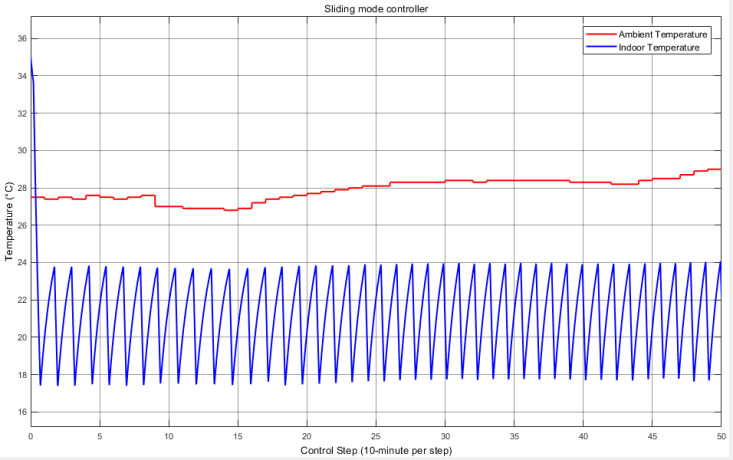}
  \caption{Outdoor and indoor temperature over the first 50 control steps.}
  \label{chap6fig3}
\end{figure}

\begin{figure}
  \centering
  \includegraphics[width=13 cm,trim=0 0 0 2,clip]{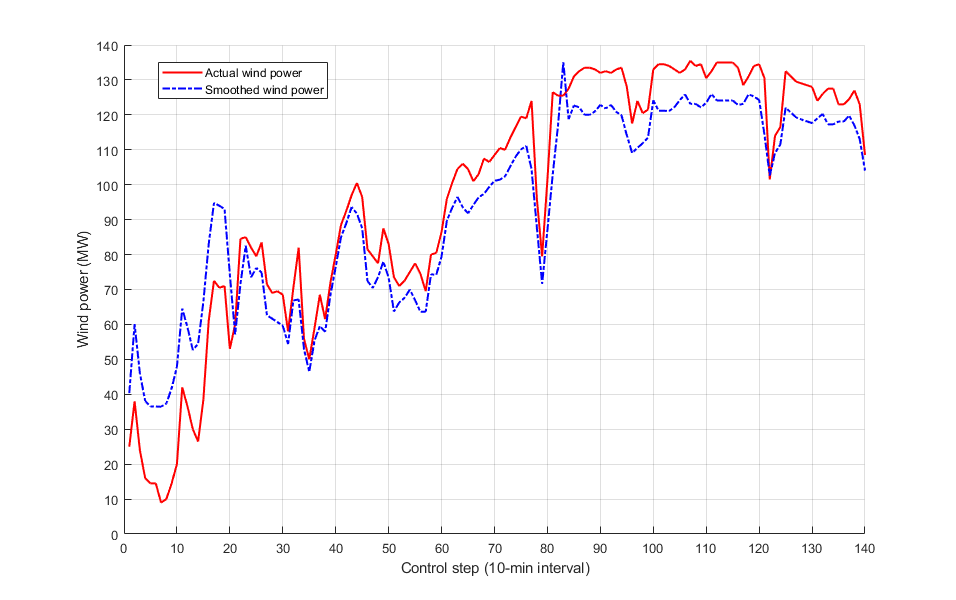}
  \caption{Actual wind power and Smoothed one.}
  \label{chap6fig4}
\end{figure}

\begin{figure}
  \centering
  \includegraphics[width=13 cm,trim=0 2 0 2,clip]{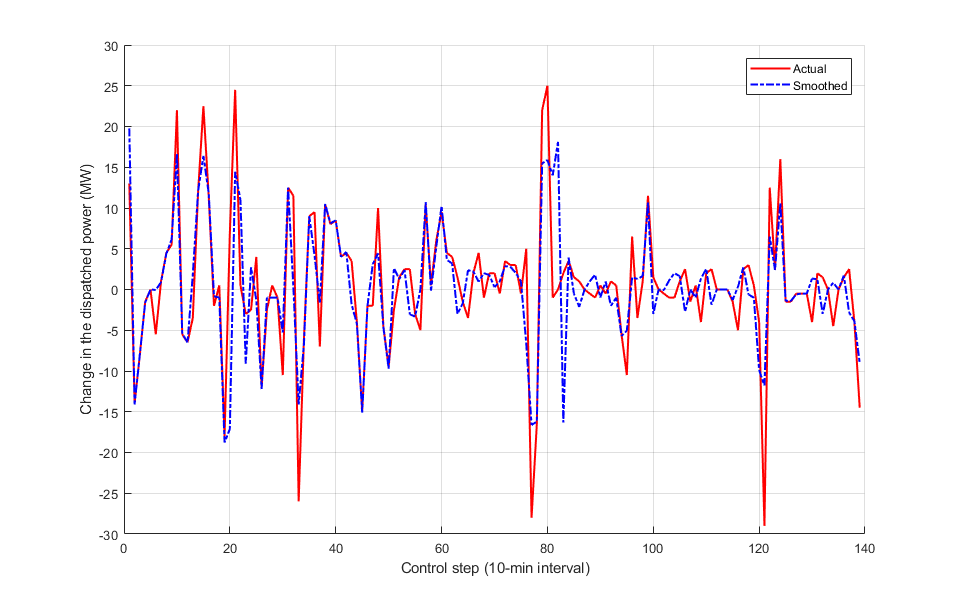}
  \caption{Changes in the dispatched power.}
  \label{chap6fig5}
\end{figure}

With an initial battery energy state of $120MWh$, the smoothed wind power is shown in Figure \ref{chap6fig4} with a comparison to the actual wind power, and the corresponding rate of change in the dispatched power is presented in Figure \ref{chap6fig5}. It should be noted that the forecasted wind and temperature data are used compute the actions for BESS and TCLs, and the system states are updated with the actual data sets.

\unskip

It can be perceived that the ramp rate is kept within the 20MW per step limit in our proposed strategy, and the corresponding operational cost of BESS is 211.056.

The aggregation model of TCL groups can be perceived as a fast-response controllable load, which can be implemented with a large quantity of air-conditioners. As only the predicted ambient outdoor temperature and wind power are required to determine the actions of an aggregated model, additional sensors or microcontrollers are not required for the individual TCL, which can save the costs for components and maintenance. This model could have a good performance in the presence of forecasting error, as the deviation in the controlled temperature due to errors are generally unnoticeable.

The optimal BESS capacity can be difficult to determine, as in practice, factors such as capital and maintenance costs, BESS model, variations in the power output during different seasons can have a huge impact on the overall operational cost. Therefore, further research would be required in this field.
\subsection {Summary}

A wind power smoothing control strategy is introduced in this chapter. The objective is to minimize the operational cost of the storage unit while meeting the grid ramp rates required to keep the room temperature under a desired level. A TCL group is used first to minimize the rate of change in the generated wind power and actual power dispatched to the grid will be adjusted by a BESS. We assess the algorithm with actual wind power and temperature data. Results show a significant improvement in the ramp rate.
\section{Microgrid Network Control\label{cha7}}

Constructing microgrids with renewable energy systems and storage units can be a feasible solution to smooth the fluctuation of renewable power and increase its penetration. In recent years, research on microgrids has proven that the overall electricity and operational costs in microgrids can be further reduced with a networked microgrid system. One major issue in controlling microgrid with renewable power system is the uncertainties due to errors in renewable power and load forecasting, and the receding horizon approach is one of the most used methods in many studies to improve the accuracy of predicted data. However, optimization problems in a microgrid network can be considered as controlling a high number of components, such as distributed generators, battery energy storage systems and controllable loads, over multiple time periods under uncertainties, which could be challenging to implement the receding horizon approach. In this paper, we propose an approximate dynamic programming-based power management strategy in a microgrid network system, and the objective is to minimize the overall cost over a control horizon with predicted renewable power, electricity price and load demand. The network consists of several microgrids and a centralized energy management system. Battery energy storage systems are used to control the exchanged power of microgrids with the network, and the centralized energy management system will communicate with the utility grid to balance the power on the network. The use of the approximate dynamic programming algorithm is to address the ‘curse of dimensionality’ due to high dimensional state and control spaces in the network so that the rolling horizon approach can be used in practice. Some of the algorithms and models used in this chapter are based on my previous publication \cite{zhuo5}.

\subsection{Introduction} \label{c7s1}

The growing concern on climate change, energy reserve and energy price has led to an increased interest in the construction of renewable energy systems (RESs) worldwide \cite{c7r1}. The penetration of renewable power, however, is mainly hindered by the inherent variability and intermittency of renewable sources. A feasible solution is to construct more microgrids (MGs) integrated with battery energy storage systems (BESSs), which allows the renewable energy to be smoothed and used regionally without interference the main power grid \cite{c7r2}. In countries with deregulated energy markets, grid-connected microgrids with BESSs can participate in energy trading. With appropriate control algorithms and forecasting techniques, the overall electricity cost in the microgrid can be significantly reduced \cite{c7r3,c7r4,c7r5,c7r6}.

In recent years, the microgrid network has received an increasing attention from some researchers \cite{r13,c7r7,c7r8,c7r9,c7r10,c7r11,c7r12,c7r13,c7r14,c7r15,c7r16,c7r17}. The basic idea is to connect multiple microgrids to form a networked system and managed by an agent, or a centralized energy management system (CEMS) that is responsible for the energy trading with the utility grid. In comparison with the standalone microgrid, the grid-connected microgrid network can decrease the overall operational and energy costs. Authors in \cite{r13} introduce a two-stage energy management strategy for a microgrid network, in which day-ahead forecasting data are utilized to optimize the operation of all components in the network. The mixed integer linear programming is used, and the optimal controls of generators, batteries and controllable loads are solved first, which are then used to help the centralized energy management system to plan the energy transaction with the utility grid. A mathematical formulation for a centralized energy management system in the microgrid network is proposed in \cite{c7r7}, a model predictive control algorithm is used to solve unit commitment and optimal power flow problems on the network. A hierarchical power scheduling approach to manage power trading, storage and distribution in a smart power grid with cooperative microgrids is introduced in \cite{c7r8}, in which the problem is formulated as a two-tier convex optimization problem to maximize user utility and minimize the power transmission cost from the centralized energy management system.

Those problems can be considered as the economic dispatch problems with renewable power, where errors in the forecasting are unavoidable. In many papers \cite{c7exr1,c7exr2,c7exr3,c7exr4,c7exr5,finalkhalid1}, the measure to alleviate the effects due to forecasting errors is utilizing the rolling horizon approach. In other words, predicted data on renewable power and load demand within a relatively short period are used to optimize the dispatch over the period. Only actions solved in the first step will be used as the actual inputs to the current stage, and the optimization algorithm will repeat in the next control stage with updated forecasting data and system states. Nevertheless, this approach has a strict computation time requirement in practice. The optimization algorithm should provide the solution within each control stage, while delays in communication and response time of power electronics should be considered as well \cite{c7exr6,c7exr7,c7exr8}. In the case of a networked microgrid system, the number of components to control can be enormous, and the evolution of the whole system will be significantly affected by the uncertainties caused by a large number of renewable power systems integrated \cite{c7exr9,c7exr10,finalshi2}. As a result, optimization algorithms such as dynamic programming, stochastic mixed integer programming are unlikely to meet the time constraint in the rolling horizon approach.

In this chapter, we propose an approximate dynamic programming (ADP) algorithm-based energy management strategy for a microgrid network. The use of the approximate technique is to address the energy management problem that can be large, complex and stochastic in a networked microgrid system \cite{c7powelbook1}. It is also used to reduce the computation time and meet the requirement for online scheduling with rolling horizon approach. The algorithm will utilize forecasting data on electricity price, renewable power and load power demand to compute a sequence of actions for distributed generators (DGs), battery energy storage systems (BESSs), and controllable loads (CLs) in the networked system, with the purpose of minimizing the overall energy cost over a forecasting horizon. The algorithm is tested with practical wind power data from the Woolnorth wind farm in Tasmania, Australia. The load power and electricity price data are obtained from the Australia Energy Market Operator.

\subsection{Problem Statement} \label{c7s2}

\begin{figure}
  \centering
  \includegraphics[width=13 cm,trim=0 0 0 2,clip]{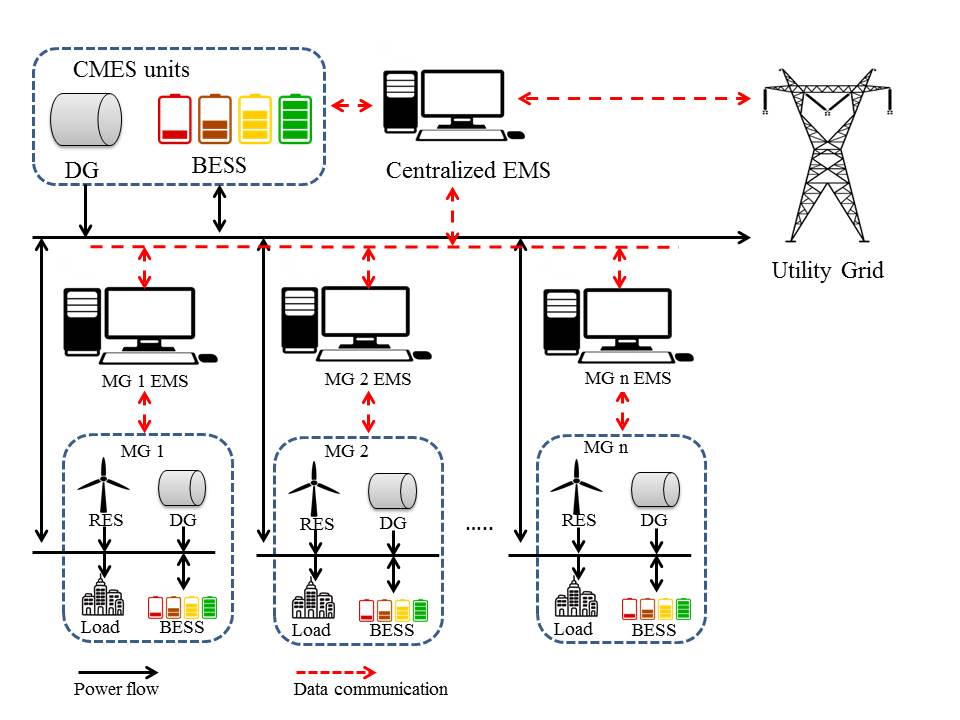}
  \caption{Microgrid network structure.}
  \label{chap7fig1}
\end{figure}

We consider a microgrid network that consists of multiple microgrids supplied by their own renewable energy systems and distributed generators. Battery energy storage systems and controllable loads are used to control the exchange power of all microgrids to the network. In addition, a centralized energy management system is included, which conducts the power trading with the utility grid, and is responsible for the power balancing task on the network. The structure of the networked system can be seen in Figure \ref{chap7fig1}.

We assume a deregulated energy market environment for the microgrid network. Each microgrid can sell the excessive power to the network, or purchase power from the network to meet its power demand. This transaction is conducted between the centralized energy management system and the microgrid, and the electricity price is based on the value provided by the energy market operators. Based on the sum of exchanged power from all microgrids, the centralized energy management system will communicate with the utility grid to purchase or sell electricity, in order to balance the power on the network. As a result, based on predicted renewable power, load power and electricity price over a fixed length of control horizon, the control objective is to minimize the electricity cost and the operational cost from all units in the network over the horizon.

The problem can be considered as controlling a set of resources over multiple time periods under uncertainties \cite{bkpowell1}. The output power of generators, energy stored in the batteries, reduced power consumption from controllable loads, predicted renewable power, load demand and electricity price can be perceived as system states. Changes in these term between each control stage can be considered as the system actions, and the error in the forecasting data will be the exogenous information to the system.

Due to the high dimensional state and action spaces, we use the approximate dynamic programming algorithm to solve the problem. In the ADP algorithm, instead of computing the value function, or the cost-go-function, by enumerating over all state spaces, it is replaced with some forms of statistical approximation. 

\subsection{System Model and Problem Formulation} \label{c7s3}
\subsubsection{Distributed Generator} \label{c7s3.1}
Suppose there are $K$ microgrids connected in the microgrid network, then the total number of distributed generators (DG) in the microgrid grid network is $K+1$. We refer to the generator model introduced in \cite{r13}, and the cost of generation is described by the following equation:

\begin{equation}
  \label{c7eq1} 
 C_{DG,i}(t)=a_GP_{G,i}(t)+b_G
\end{equation}

where $P_{G,i} (t)$ is the power output of $ith$ DG at time $t$; $a_G$ and $b_G$ are generation cost coefficients of DG. The DG indexed as $K+1$ is the generator used in the centralized energy management system.

We consider our system as a discrete-time system with the sampling rate $\delta>0$ so the function $P_{G,i}(t)$ is supposed to be a piecewise function with constant values over periods $[k\delta,(k+1)\delta)$. Hence, the cost of generation in one DG over a control horizon of $N$ stages will be: 

\begin{equation}
  \label{c7eq2} 
 \sum_{t=0}^{N-1} (a_GP_{G,i}(t)+b_G)
\end{equation}

Also, the power output of DG should always satisfy the constraints:

\begin{equation}
  \label{c7eq3} 
 P_{G,i}^{min} \leq P_{G,i}(t) \leq  P_{G,i}^{max}
\end{equation}

with some given constants $0<P_{G,i}^{min}\leq P_{G,i}^{max}$. To avoid the start-up and shut-down costs, the minimum output is non-zero. 

Furthermore, we consider a control $u_{PG,i} (t)$ for the DG output, which is the change in the DG output between $t$ and $t+1$, and the output of DG will satisfy the following dynamics:

\begin{equation}
  \label{c7eq4} 
P_{G,i} (t+1)=P_{G,i} (t)+u_{PG,i} (t)
\end{equation}

And the change in the DG output will be constrained by ramping capabilities of the generator:

\begin{equation}
  \label{c7eq5} 
 u_{PG,i}^{min} \leq u_{PG,i}(t) \leq  u_{PG,i}^{max},
\end{equation}

with some constants $u_{PG,i}^{max}>0$ and $u_{PG,i}^{min}<0$ to represent the maximum change in the DG output at each control step.


\subsubsection{Battery Energy Storage System} \label{c7s3.2}

We refer to the battery model proposed in \cite{r17}, and the dynamics of BESS is described by the following equation: 

\begin{equation}
  \label{c7eq6} 
E_{B,i}(t+1)=E_{B,i}(t)-u_{PB,i}(t)\Delta\delta-d|{u_{PB,i}(t)\Delta\delta}| \\
\end{equation}

where $E_{B,i} (\cdot)$ is the energy state of the ith BESS, $u_{PB,i} (\cdot) $is the charging/discharging power of the $ith$ BESS, $\Delta\delta$is the time period to convert power to energy, $d>0$ is the charging/discharging loss factors of the BESS. Based on the model, $u_{PB,i} (t)>0$ indicates the discharging action and u$_{PB,i} (t)<0$ indicates the charging action of BESS. Furthermore, BESS with the index of $K+1$ is used in the centralized energy management system.

The cost of BESS, $C_{B,i} (t)$, over the time interval $[k_0 \delta,(N-1)\delta]$ is modelled as:

\begin{equation}
  \label{c7eq7} 
\sum_{t=0}^{N-1} C_{B,i}(t)=\sum_{t=0}^{N-1} (\gamma_1 |u_{PB,i}(t)|\Delta\delta+\gamma_2E_{B,i}(t))
\end{equation}

where $\gamma_1>0$ and $\gamma_2>0$ are some given constants. Also, the following constraints should be satisfied: 

\begin{equation}
  \label{c7eq8} 
 u_{PB,i}^{min} \leq u_{PB,i}(t) \leq  u_{PB,i}^{max}
\end{equation}

\begin{equation}
  \label{c7eq9} 
 E_{B,i}^{min} \leq E_{B,i}(t) \leq  E_{B,i}^{max}
\end{equation}

where $u_{PB}^{min}<u_{PB}^{max}$ and $0 \leq E_{B,i}^{min} \leq E_{B,i}^{max}$.


\subsubsection{Controllable Loads} \label{c7s3.3}

Controllable Loads (CLs) used in the microgrid is considered as a part of the demand response program, in which users of the microgrid are provided with some incentives, such as a reduced electricity price, to cut their electricity consumption with CLs during peak hours. There are no controllable loads in the central energy management system.  

We use $P_{CL,i} (t)$ to indicate the power of CLs, namely the reduced power consumption, in the $i$th microgrid, which is also a piecewise function with constant values over intervals $[k/delta,(k+1)/delta)$. We refer to the model introduced in \cite{r13}, and let $C_{CL,i}$ be the cost resulted from the controllable load power $P_{CL,i} (t)$ in the $i$th microgrid, which can be defined as follows: 

let $C_{CL,i}$ be the cost resulted from the controllable load power $P_{CL,i}(t)$ in the $i$th microgrid, which can be defined as follows: 
\begin{equation}
  \label{c7eq10} 
\sum_{t=0}^{N-1} C_{CL,i}(t)=\sum_{t=0}^{N-1} (a_{CL}+b_{CL}P{CL,i}(t))
\end{equation}

where $a_{CL}>0$ and $b_{CL}>0$ are some given constants. The controllable load power $P_{CL,i} (t)$ is subjected to the following constraint.

\begin{equation}
  \label{c7eq11} 
P_{CL}^{min}\leq P_{CL,i}(t)\leq P_{CL}^{max}
\end{equation}

where $0\leq P_{CL}^{min}<P_{CL}^{max}$. 

Change in the controllable load power is considered as the control/decision for $P_{CL,i} (t)$, which is indicated as $u_{CL,i} (t)$. The dynamics of the controllable load is described by the following equation:

\begin{equation}
  \label{c7eq12} 
P_{CL,i}(t+1)=P_{CL,i}(t)+u_{CL,i}(t)
\end{equation}

And $u_{CL,i} (t)$ should always satisfy the following constraint:

\begin{equation}
  \label{c7eq13} 
u_{CL,i}^(min)\leq u_{CL,i}(t)+\leq u_{CL,i}^(max)
\end{equation}

with some given constants $u_{CL,i}^{max}>0$ and $u_{CL,i}^{min}<0$.


\subsubsection{Forecasting Data} \label{c7s3.4}

Predicted data used in the microgrid network include renewable power, load demand and electricity price over the control horizon. As the rolling horizon approach is utilized, they will be updated at every control step. Let $P_{RES,i} (t)$ and $P_{L,i} (t)$ be the renewable power generated and the load power demand in the $i$th microgrid between $t$ and $t+1$. Suppose that both $P_{RES,i} (0)$ and $P_{L,i} (0)$ are known at the beginning of each control horizon, and we can forecast the changes in them over the control horizon. We use $\hat{W}_{RES,i}(t+1)$ and $\hat{W}_{L,i}(t+1)$) to represent those predicated changes, and they can be described with the following dynamics:  

\begin{equation}
  \label{c7eq14} 
P_{RES,i}(t+1)=P_{RES,i}(t)+\hat{W}_{RES,i}(t+1)
\end{equation}

\begin{equation}
  \label{c7eq15} 
P_{L,i}(t+1)=P_{L,i}(t)+\hat{W}_{L,i}(t+1)
\end{equation}

$\hat{W}_{RES,i}(t+1)$ and $\hat{W}_{L,i}(t+1)$) are stochastic variables that can be considered as the exogenous information to the system, and we could have knowledge on the distribution patterns of these forecasting errors. $P_{RES,i} (t)$ and $P_{L,i} (t)$ will then be the system states/parameters affected by the exogenous information.

Regarding the electricity price, we use $EP(t)$ to represent the price for both selling and purchase electricity between $t$ and $t+1$ in all sectors, which include transactions between the CEMS and the unity grid, as well as those made between microgrids in the network. In practice, those electricity prices could be different in selling and purchasing, and they can be deterministic among the network based on bilateral contracts between the CEMS and the microgrids.

The dynamics of the electricity price is described as:

\begin{equation}
  \label{c7eq16} 
EP(t+1)=EP(t)+\hat{W}_{EP,i}(t+1)
\end{equation}

where $\hat{W}_{EP,i}(t+1)$ is the predicted change in the electricity price.


\subsubsection{Individual Microgrid} \label{c7s3.5}

Each microgrid in the network is comprised of a battery energy storage system, a distributed generator, a renewable energy system, controllable loads and local loads. A significant element of the whole system is the power exchanged from the individual microgrid to the network, which determines the energy cost in each microgrid and affects the power balancing on the network. 

As the objective of our problem is to minimize the total energy cost in the microgrid network, we use $P_{exc,i} (t)$ to indicate the difference between the power consumed and the power generated in the $i$th microgrid between $t$ and $t+1$. In other words, it will the amount of power imported from the network if $P_{exc,i} (t)>0$, and $P_{exc,i} (t)<0$ indicates the $i$th microgrid is transmitting power to the network.

Based on previous sections, $P_{exc,i} (t)$ will the linear combination of the BESS action, DG output, controllable load power, generated renewable power and power demand, which is described in the following equation:

\begin{equation}
  \label{c7eq17} 
P_{exc,i}(t)=P_{L,i}(t)-P_{CL,i}(t)-P_{RES,i}(t)-P_{G,i}(t)-u_{PB,i}(t)
\end{equation}

Also, we will limit the amount of exchanged power at each control stage to meet the ramp rates limit on the network, with the following constraint:

\begin{equation}
  \label{c7eq18} 
P_{exc,i}^{min}\leq P_{exc,i}(t)+\leq P_{exc,i}^{max}
\end{equation}

with some given constants $P_{exc,i}^{max}>0$ and $P_{exc,i}^{min}<0$.

As a result, the energy cost to meet the power demand in the $i$th microgrid between $t$ and $t+1$ will be the product of $EP(t)$ and $P_{exc,i} (t)$.  And a positive value of $P_{exc,i} (t)$ indicates the microgrid is transferring power to the network while a negative value indicates a reversed transmission. Also, $\sum_{i=1}^{K} P_{exc,i}(t)$ will be the overall exchanged power from all microgrids in the network. 

\subsubsection{Centralized Energy Management System} \label{c7s3.6}

The centralized energy management system (CEMS), that consists of a BESS and a DG, is responsible for the power balance task in the microgrid network. It also communicates with the utility grid to conduct the power transaction.

The sum of $P_{exc,i}$ from all microgrids in the network, as introduced in the previous section, should be balanced with the combination of DG output, BESS action, and the power exchanged with the utility grid from the CEMS. Suppose there are $K$ microgrids connected in the network, and the index of the CEMS is $K+1$, and let $P_{ug,i}$ indicate the power imported from the utility grid when $P_{ug,i}>0$, while $P_{ug,i}<0$ indicates the CEMS is selling power to the grid. As a result, the power balancing constraint in the network can be described with the following equation:

\begin{equation}
  \label{c7eq19} 
(\sum_{i=1}^{K} P_{exc,i}(t))-P_{G,K+1}(t)-u_{PB,K+1}(t)-P_{ug}(t)=0
\end{equation}

It is assumed that the DG and the BESS in the CEMS have the same dynamics and the constraints introduced in Section \ref{c7s3.1} and \ref{c7s3.2}. Therefore, the cost resulted from the CEMS include the operational costs from its components, and the electricity cost from $P_{ug} (t)$.

\subsubsection{Control System Formulation} \label{c7s3.7}

Based on these models, we can define a set of states for the network system, which includes DG power output, controllable load power, energy stored in BESS, forecasted renewable power, load demand and electricity price. Let $S_t$ be the set of microgrid network system states at $t$, it can be described as follows:

\begin{equation}
  \label{c7eq20} 
  \begin{split}
S_t= [EP(t),&E_{B,1}(t),P_{G,1}(t),P_{CL,1}(t),P_{RES,1}(t),P_{L,1}(t),\\ &E_{B,2}(t),P_{G,2}(t),P_{CL,2}(t),P_{RES,2}(t),P_{L,2}(t),\cdots \cdots,\\
& E_{B,K}(t),P_{G,K}(t),P_{CL,K}(t),P_{RES,K}(t),P_{L,L}(t),E_{B,K+1}(t),P_{G,K+1}(t)]
\end{split}
\end{equation}

The number of elements in this set is $5K+3$. Note that $P_{exc,i}$ and $P_{ug}$ are not included in this set as they can be expressed as functions of states. Also, the corresponding control set, indicated as $u_t$, will be:

\begin{equation}
  \label{c7eq21} 
  \begin{split}
u_t= [&u_{PG,1}(t),u_{PB,1}(t),u_{CL,1}(t),u_{PG,2}(t),u_{PB,2}(t),u_{CL,2}(t),\cdots \cdots\,\\
&u_{PG,K}(t),u_{PB,K}(t),u_{CL,K}(t),u_{PG,K+1}(t),u_{PB,K+1}(t)]
\end{split}
\end{equation}

where the number of controls is $3K+2$. 

Furthermore, the transition of system states is affected by a set of stochastic variables, which includes forecasting errors in renewable power, load demand and electricity price. Let $W_(t+1)$ be the corresponding set, it will be:

\begin{equation}
  \label{c7eq22} 
  \begin{split}
W_{t+1}= [&W_{EP}(t+1),W_{RES,1}(t+1),W_{L,1}(t+1),W_{RES,2}(t+1),W_{L,2}(t+1)\cdots \cdots,\\ 
&W_{RES,K}(t+1),W_{L,K}(t+1)]
\end{split}
\end{equation}

where the number of stochastic variables is $2K+1$.

We use $S^M(\cdot)$to represent the transition from state $S_t$ to $S_{t+1}$, and it can be written as:

\begin{equation}
  \label{c7eq23} 
S_{t+1}=S^M(S_t,u_t,W_{t+1})
\end{equation}


\subsubsection{Objective Function} \label{c7s3.8}

Two major costs are required to be minimized in our problem, which are the operational cost from all components in the network and the electricity cost from all power transactions.

Let $C_{opr} (t)$ be the overall operational costs of all components in the microgrid network over one control stage, it can be described with the following equation:

\begin{equation}
  \label{c7eq24} 
C_{opr}(t)=\sum_{i=1}^{K+1}C_{B,i}(t)+\sum_{i=1}^{K+1}C_{DG,i}(t)+\sum_{i=1}^{K}C_{CL,i}(t)
\end{equation}

Regarding the electricity cost, we use $C_{ele} (t)$ to indicate the total electricity cost from all microgrids and the CEMS. Based on equation (\ref{c7eq19}), it can be defined as follows:

\begin{equation}
  \label{c7eq25} 
  \begin{split}
C_{elec}(t) &= EP(t)((\sum_{i=1}^{K}P_{exc,i}(t))+P_{ug}(t))\\ 
&=EP(t)(2(\sum_{i=1}^{K}P_{exc,i}(t))-P_{G,K+1}(t)-u_{PB,K+1}(t))
\end{split}
\end{equation}

As a result, the objective function is to minimize the additive cost of them over the control horizon of $N$ stages, which is:

\begin{equation}
  \label{c7eq26} 
\min_{u_t}\epsilon[\sum_{t=0}^{N-1}C_{opr}(t)+C_{ele}(t)],
\end{equation}

where$\epsilon[\cdot]$ is the expectation function.

Since $C_{opr} (t)$ and $C_{ele} (t)$ can be written as functions of  $S_t$ and $u_t$, we can write the objective function as:

\begin{equation}
  \label{c7eq27} 
\min_{u_t}\epsilon[\sum_{t=0}^{N-1}C_{\tau}(S_t,u_t)],
\end{equation}

and the problem is subjected to constraints (\ref{c7eq3}) (\ref{c7eq5}) (\ref{c7eq8}) (\ref{c7eq9}) (\ref{c7eq11}) (\ref{c7eq13}) (\ref{c7eq18}) (\ref{c7eq19}).

\subsection{Solution Technique} \label{c7s4}

The problem can be perceived as controlling a set of resources over multiple time periods under uncertainty, which can be considered as a Markov decision process (MDP) and solved with dynamic programming (DP) algorithms \cite{bkpowell1}. However, the high dimensional state and control spaces greatly increase the computation time and the memory space of the algorithm; the existence of random variable further increases the computational complexity. This is known as the ‘curse of dimensionality’ in dynamic programming. To overcome this issue, we utilize the approximate dynamic programming (ADP) algorithm that will be explained in detail in this section.
\subsubsection{Bellman's Equation} \label{c7s4.1}

We first formulate the minimization problem (\ref{c7eq27}) with Bellman’s equation \cite{bk1}, which could be used to solve MDP with DP and ADP algorithms. Let $V_t(S_t)$ be the minimum cost starting from $t$ onwards to the last stage of the control horizon, the Bellman equation will be:

\begin{equation}
  \label{c7eq28} 
V_t(S_t)=\min_{u_t}(C_{\tau}(S_t,u_t)+\epsilon[V_{t+1}(V_{t+1}(S^M(S_t,u_t,W_{t+1}))]),
\end{equation}

It can be perceived that $V_0(S_0)$ and the corresponding $u_0$ are the solutions of our problem.

The Bellman equation, or the ‘cost-to-go’ function that used in the standard DP algorithm, will work in a recursive order and enumerate over all states to determine the ‘cost-to-go’ values for every possible state.


\subsubsection{Post-Decision State} \label{c7s4.2}

Equation \ref{c7eq28} includes solving the minimum of an expected value, which can be difficult to compute even if we replace $V_t(S_t)$ with an estimated value function $\bar{V}_t(S_t)$. To address this, the concept called ‘post-decision state’ is utilized. According to \cite{bkpowell1}, it is the state immediately after the decision is made before any exogenous information arrives, and $S_t$ can be considered as the ‘pre-decision state’. Let $S_t^u$ indicate the post-decision state, the transition between the pre-decision state and the post-decision state can be described as:

\begin{equation}
  \label{c7eq29} 
S_t^u=S^{M,u}(S_t,u_t),
\end{equation}

\begin{equation}
  \label{c7eq30} 
S_{t+1}=S^{M,W}(S_t^u,W_{t+1})
\end{equation}

Based on equation (\ref{c7eq28}), and let $V_t^u (S_t^u)$ be the value of being in the post-decision state $S_t^u$, it will become:

\begin{equation}
  \label{c7eq31} 
V_t(S_t)=\min_{u_t}(C_{\tau}(S_t,u_t)+V_t^u(S_t^u)),
\end{equation}

\begin{equation}
  \label{c7eq32} 
V_t(S_t^u)=\epsilon[V_{t+1}(S_{t+1})|S_t],
\end{equation}

It can be perceived that by substituting (\ref{c7eq32}) into (\ref{c7eq31}), we obtain equation (\ref{c7eq28}), and equation (\ref{c7eq31}) is a deterministic optimization problem. According to \cite{c7adpr1}, with suitable functions to approximate $V_t^u(S_t^u)$, problems with large state and control spaces can be solved with some nonlinear optimization algorithms that are readily available in many software. 


\subsubsection{Approximate Dynamic Programming Algorithm} \label{c7s4.3}

The ADP algorithm introduced in \cite{c7adpr1} is utilized to solve our problem, which is described with the following pseudocode:

\begin{itemize}
  \item[] \textbf{Step $0$}. Initialization:
  \begin{itemize}
  \item[] Step $0a$. Initialize $\bar{V}_t^0,\forall t$.
  \item[] Step $0b$. Choose an initial state $S_0^1$.
  \item[] Step $0c$. Set $n=1$.
  \end{itemize}
  \item[] \textbf{Step $1$}. Choose a sample path $\omega^n$.
  \item[] \textbf{Step $2$}. For $t=0,1,2,\cdots,N-1$ do:
  \begin{itemize}
  \item[] Step $2a$. Solve $\hat{v}_t^n=\min_{u_t}(C_{\tau}(S_t^n,u_t)+\bar{V}_t^{n-1}(S^{M,u}(S_t^n,u_t)))$, and let $u_t^n$ be the value of $u_t$ that solves the minimization problem.
  \end{itemize}
  \begin{itemize}
  \item[] Step $2b$. Update the value function using:\\
$\bar{V}_{t-1}^n (S_{t-1}^{u,n})=(1-a_{n-1})\bar{V}_{t-1}^{n-1}(S_{t-1}^{u,n})+a_{n-1}\hat{v}_t^n$
\end{itemize}
  \begin{itemize}
   \item[] Step $2C$. Update the state:\\
   $S_t^{u,n}=S^{M,u}(S_t^n,u_t^n)$\\
   $S_t^n=S^{M,W}(S_{t-1}^{u,n},W_t(\omega ^n))$
     \end{itemize}
  \item[] \textbf{Step $3$}. Let $n=n+1$. If $n<M$, go to \textbf{step $1$}.
    \item[] \textbf{Step $4$}. Return the value function approximations $\bar{V}_t^M,\forall t$
\end{itemize}

In this algorithm, $n$ indicates the number of iterations and M is the maximum number of iterations. \textbf{Step $2a$}, we solve the Bellman equation (\ref{c7eq30}) with an approximated $\bar{V}_t^u(S_t^u)$ to obtain a decision $u_t^n$, and $\hat{v}_t^n$, which is an observation of the value function. 

In the next step, $u_t^n$ and $\hat{v}_t^n$ are used to update the previous post-decision state $\bar{V}_{t-1}^n(S_{t-1}^{u,n})$. And $a_{n-1}$ is the step size, or the learning rate schedules \cite{bkpowell1} used to smooth between old and new estimates. In this paper, the Harmonic step size sequence introduce in \cite{c4powel1} is used, which is:

\begin{equation}
  \label{c7eq33} 
a_{n-1}=\dfrac{\epsilon}{\epsilon+n^{\beta}-1}
\end{equation}

Values of $\epsilon$ and $\beta$ are chosen to adjust the rate at which the step size drops to zero.

The sample path $\omega^n$ is the random observation of exogenous information. At each iteration, a sample realization of the random variable $W_{t+1}$, indicated as $W_t(\omega^n)$, is chosen to update the system transition between the post-decision state and the pre-decision state. 


\subsubsection{Value Function Approximation} \label{c7s4.4}

We use the basis functions approach to approximate our value functions in this paper. According to \cite{c7adpr1}, basis function are features drawn from the post-decision state variables, which could have a measurable impact on the value function. In our case, it could be the state variables, the exchanged power in each microgrid, and some nonlinear functions of the state variables.  Let $\mathcal{F}$ be a set of features drawn from the post-decision state $S_{t-1}^{u,n}$, and each feature $f\in \mathcal{F}$. Suppose the approximated value function can be expressed as a weighted linear combination of features with a vector of parameter $\theta$, it can be written as:

\begin{equation}
  \label{c7eq34} 
\bar{V}_t^{u,n}(S_t^{u,n})=\sum_{f\in \mathcal{F}} \theta_f^n\phi_f(S_t^{u,n}),
\end{equation}

where $\theta_f^n\phi_f(S_t^{u,n})$ is the basis function or the features.

To estimate the value of the parameter vector $\theta$, the following standard linear regression problem will be solved:

\begin{equation}
  \label{c7eq35}
  \min_{\theta}\min_{n=1}^{n_{max}}(\bar{V}_t^{u,n}-(\sum_{f=1}^{f_{max}}\theta_f^n\phi_f(S_t^{u,n})))^2,
\end{equation}

where $n_{max}$is the maximum number of observations and $f_{max}$ is the maximum number of features.

To solve this problem, we use the recursive least squares estimation method. A detailed explanation of the method can be viewed in \cite{bkpowell1} with a case study on freight consolidation in \cite{c7adpr2}. The algorithm is described as follows:

Let $\theta^n$ be the parameter vector, and $\phi^n$ be the vector created by computing $\phi_f(S_t^{u,n})$ for each feature $f\in \mathcal{F}$. The parameters vector $\theta^n$ can be updated with:

\begin{equation}
  \label{c7eq36}
\theta^n=\theta^{n-1}-H^n\phi^n\hat{\epsilon}^n,
\end{equation}

where 

\begin{equation}
  \label{c7eq37}
  \hat{\epsilon}^n=\bar{V}_{t-1}^{n-1}(S_{t-1}^{u,n}|\theta_f^{n-1})-\hat{v}_t^n,
\end{equation}

is the error in the estimation of the value function. The matrix $H^n$ is derived using:

\begin{equation}
  \label{c7eq38}
H^n=\dfrac{1}{\gamma^n}B^{n-1},
\end{equation}

where $B^{n-1}$ is an $\mathcal{F}$ by $\mathcal{F}$ matrix that is updated recursively with: 

\begin{equation}
  \label{c7eq39}
B^n=\dfrac{1}{\gamma^n}(B^{n-1}-\dfrac{1}{\gamma^n}(B^{n-1}\phi^n(\phi^n)^TB^{n-1})),
\end{equation}

and $\gamma^n$ is a scalar given by: 

\begin{equation}
  \label{c7eq40}
\gamma^n=\lambda^n+(\phi^n)^TB^{n-1}\phi^n.
\end{equation}

$\lambda^n$ can be considered as the parameter used to determine the weight between observation. According to \cite{bkpowell1}, setting $\lambda^n=1$ indicates equal weight on all observations, and lower values of $\lambda^n$ means we put a higher weight on more recent observations. 

Furthermore, features used in our basis function will include the following types: (1) All state variables. (2) All post-decision state variable. (3) Product of electricity price and all post-decision state variables. (4) Product of electricity price and all renewable power and load power demand. (5) Constants. Therefore, in a microgrid network with $K$ microgrids, the number of parameters in each type of features will be $5K+3$, $3K+2$, $3K+2$, $2K$, and $1$ respectively. 

\subsection{Simulation} \label{c7s5}
\subsubsection{System Parameters} \label{c7s5.1}

In our simulation, we consider a similar microgrid network structure as the one introduced in \cite{r13}, which consists of three microgrids with different BESS and DG ratings. The actual time of each control stage is $5$ minutes, and we use a one-hour predicting horizon that is equivalent to $12$ steps.

Parameters of DGs in each microgrid and the CEMS is summarized in Table \ref{c7t1}. 

\begin{table}
\caption{Parameters of DGs in the network.}\label{c7t1}
\centering

\begin{tabular}{ccccc}
\hline
{}	& MG1	& MG2 & MG3	& CEMS \\
\hline
$a_G,b_G(\$/kWh)$		& 0.3,0.05 & 0.22,0.03  & 0.43,0.04	& 0.31,0.06\\
\hline
$P_{G,i}^{min},P_{G,i}^{max}(kW)$	& 50,20	& 40,180   & 30,160	& 100,500  \\
\hline
$u_{pg,i}^{min},u_{pg,i}^{max}(kW)$		& -20,20 & -20,20 & -20,20	& -50,50\\
\hline
\end{tabular}
\end{table}

The controllable loads used in the microgrids are generally thermostatically controlled loads such as air conditioners, heaters and ventilation systems. According to \cite{r13}, the maximum CL power is set to 20\% of the maximum loads in the microgrid, and the cost parameters $a_{CL}$ and $b_{CL}$ are $0.33\$/kW$ and $0.05\$/kW$ respectively. The maximum rate of change between each control stage is 5\% of the maximum loads in the microgrid.

Regarding the BESS, we refer to the models introduced in \cite{r13} and \cite{c5fengji1}, and the parameters of BESSs used in the network in summarized in Table \ref{c7t2}.

\begin{table}[H]
\caption{Parameters of BESSs in the network.}\label{c7t2}
\centering

\begin{tabular}{ccccc}
\hline
{}	& MG1	& MG2 & MG3	& CEMS \\
\hline
$\gamma_1,\gamma_2(\$/kWh)$	& 0.08,0.08 & 0.08,0.08  & 0.08,0.08	& 0.08,0.08\\
\hline
$E_{B,i}^{min},E_{B,i}^{max}(kWh)$	& 40,160	& 30,160   & 50,180	& 80,360  \\
\hline
$u_{B,i}^{min},u_{B,i}^{max}(kW)$ & -150,150 & -125,125 & -160,160	& -300,300\\
\hline
$d$ & 0.95 & 0.98 & 0.95	& 0.98\\
\hline
\end{tabular}
\end{table}

The initial BESS energy state used in our simulation are $100kWh$, $120kWh$,$140kWh$, and $240kWh$ respectively. Furthermore, the limits of exchanged power $P_{exc,i} (t)$ over each control stage is $\pm500kW$.

\subsubsection{Database} \label{c7s5.2}

The electricity price and the load demand data are obtained from the Australia energy market operator (AEMO), and the wind power data is retrieved from the Woolnorth wind farm in Tasmania Australia. As the data used are actual observations, we include some forecasting errors based on the forecasting technique introduced in \cite{c4windprediction} in our simulation. We also use it to generate the sample paths in the ADP algorithm. The actual and predicted data used in our simulation are summarized in the following figures:

\begin{figure}
  \centering
  \includegraphics[width=13 cm,trim=0 0 0 2,clip]{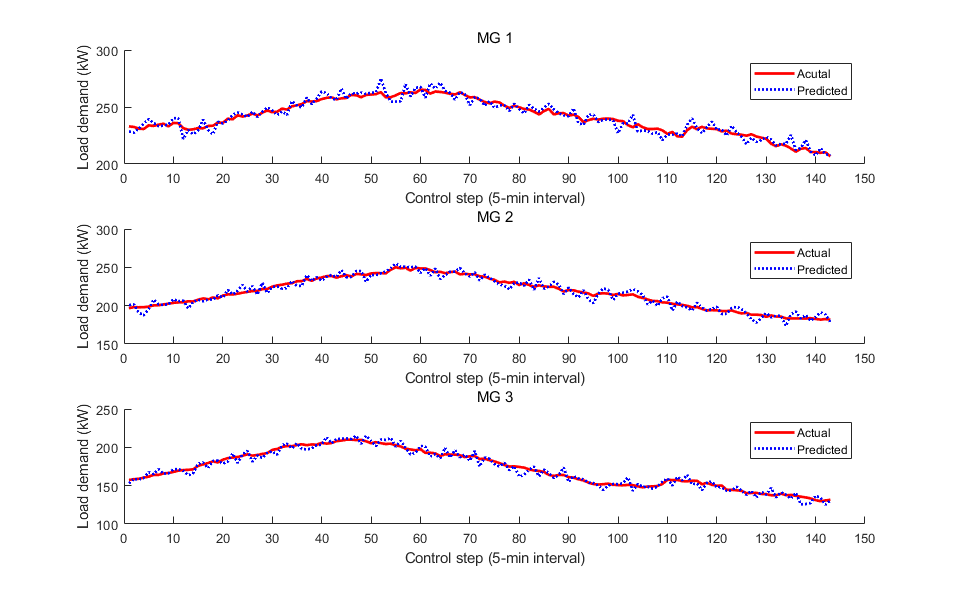}
  \caption{Actual and predicted load demand power in the three MGs.}
  \label{chap7fig2}
\end{figure}

\begin{figure}
  \centering
  \includegraphics[width=13 cm,trim=0 0 0 2,clip]{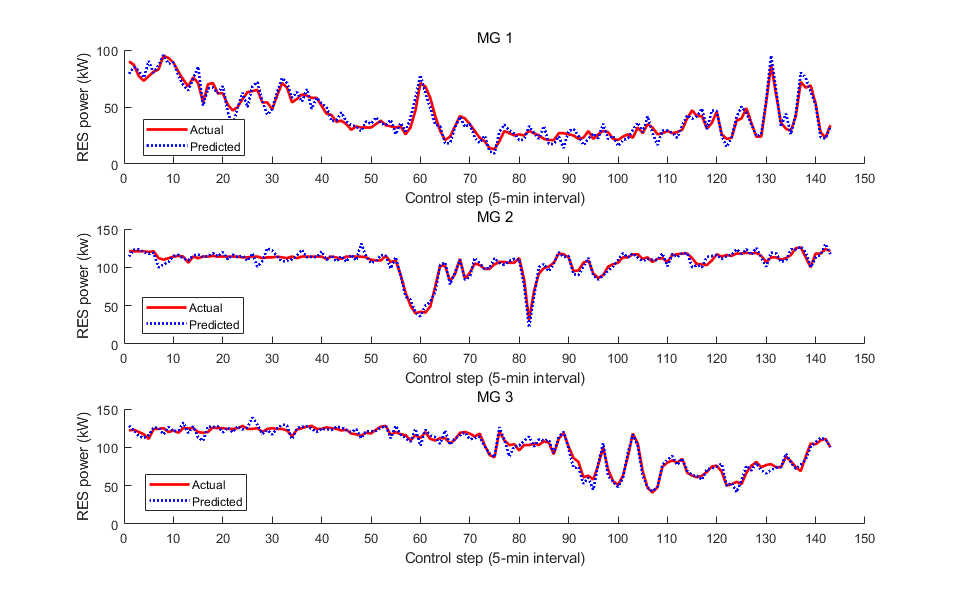}
  \caption{Actual and predicted RES power in the three MGs.}
  \label{chap7fig3}
\end{figure}

\begin{figure}
  \centering
  \includegraphics[width=13 cm,trim=0 0 0 2,clip]{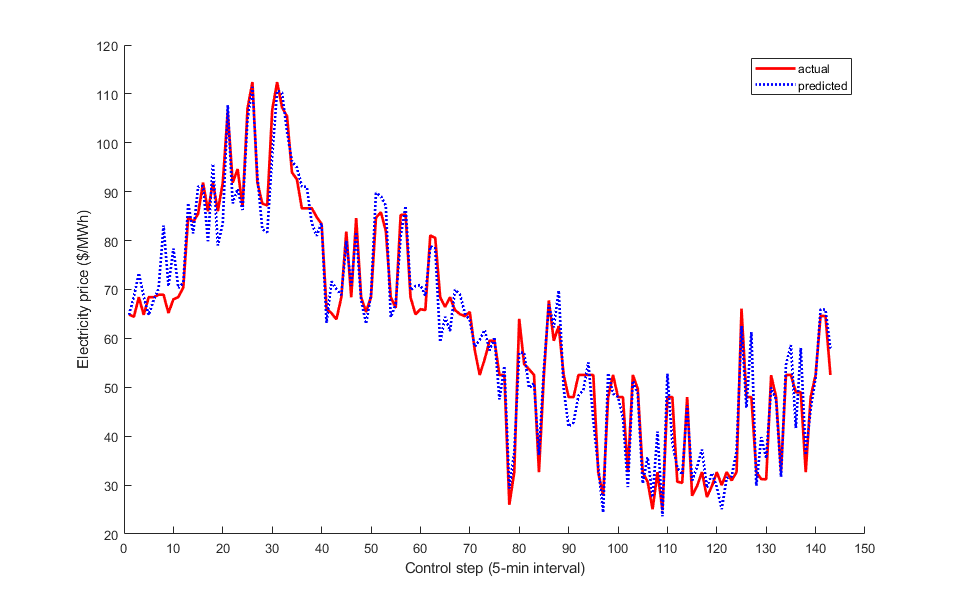}
  \caption{Actual and predicted electricity price.}
  \label{chap7fig4}
\end{figure}


\subsubsection{Simulation Results} \label{c7s5.3}

The amount of exchanged power in the three MGs are summarized in the following figure, which are calculated using results solved with the ADP algorithm and the equation (\ref{c7eq17}).

\begin{figure}
  \centering
  \includegraphics[width=13 cm,trim=0 0 0 2,clip]{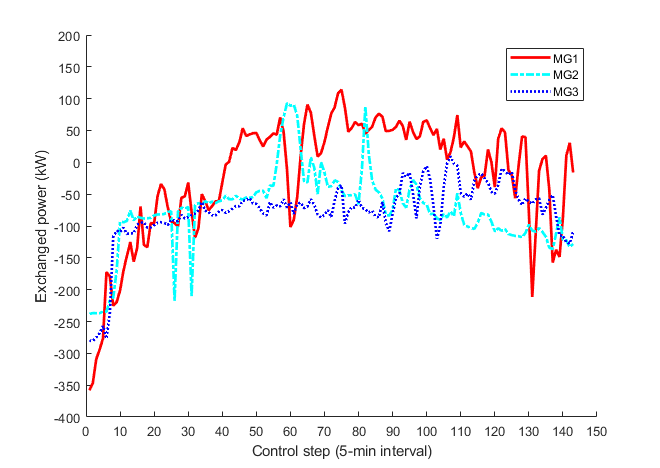}
  \caption{Exchanged power in the three MGs.}
  \label{chap7fig5}
\end{figure}

Similar patterns in the exchanged power can be observed in all three MGs, as they use the same electricity price to determine the actions for components. 

To evaluate the effectiveness of our algorithm, we compare it with a ‘short-sighted’ strategy: the optimization problem (\ref{c7eq27}) is solved in one-step instead of minimizing over multiple time periods and utilizing the rolling horizon approach. The cost computed at each step under the two schemes are presented in the following figure:

\begin{figure}
  \centering
  \includegraphics[width=13 cm,trim=0 0 0 2,clip]{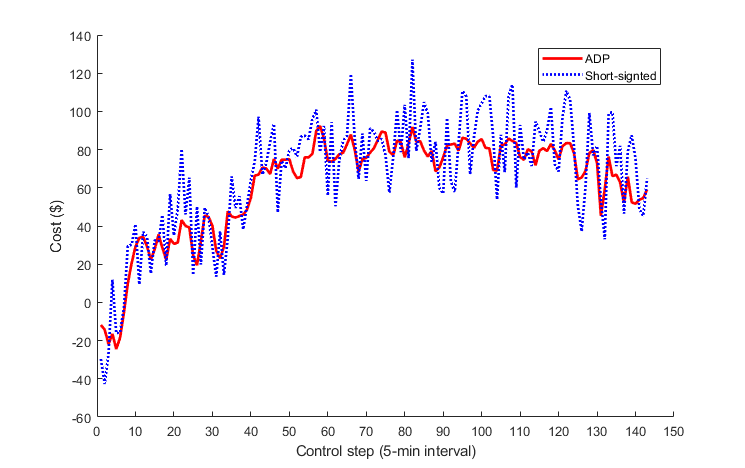}
  \caption{Simulation Results.}
  \label{chap7fig6}
\end{figure}

The negative values represent the profit gained by selling the excessive power to the grid, and the overall cost solved with the ADP algorithm and the one-step algorithm are $8821.27$ and $9555.8$ respectively, which is around $8\%$ improvement in the 12-hour simulation.

\subsection{Summary} \label{c7s6}

In this chapter, an approximate dynamic programming algorithm is produced to solve the energy management problem in a interconnected microgrid network. The use of an approximate technique is to increase the computation speed and meet the operating constraints in real-time. A networked microgrid system Incorporated with many renewable power generations can be considered a system with high dimensional states and stochastic variables, which is difficult to optimize with commercial optimization solvers. Our proposed algorithm can simplify the problem and solve it efficiently. 

\section{Conclusion and Future Work\label{cha:conclusion}}

This report has introduced multiple optimal control strategies that are suitable for modern electrical grids. The storage units, renewable energy, and controllable loads will play a key role in achieving a sustained and reliable grid. The objective of our work is to use advanced control techniques to address some of the issues that existed in previous studies.  

In Section 3, a novel cost model of batteries based on battery lifetime degradation is introduced. The model is recursive and additive that can be used with dynamic programming. It is also applied in a microgrid environment to verify its profitable on a deregulated energy market.  The control algorithm introduced can guarantee the global optimum in the planning horizon, which could be more reliable and consistent in real-time operating. Also, our proposed cost model of battery eliminates the need to execute the cycle counting algorithm at the cost of extra storage space, which reduces the computation complexity and can be used as the benchmark to assess some learning-based algorithms.

In Section 4, an approximate dynamic programming algorithm that can be used for energy trading in wind farms or microgrids is presented. An approximate technique is motivated by the fact that the receding horizon approach is utilized in many studies to alleviate the effects of forecasting errors, which will require a strict time constraint in real-time operation. The proposed algorithm can maximize the income and minimize the consumption of battery lifetime based on predicted data of wind power and electricity price in both short and long period, and the well-known 'curse of dimensionality' in large-scale BESS systems can be avoided.

In Section 5, a microgrid power scheduling strategy to minimize the cost required to maintain the desired indoor temperature is proposed. The control scheme we developed is fully decentralized. The~controller for each TCL measures only indoor temperature in the corresponding area, while the~control signals for the generator and the BESS do not need to have any measurements from TCL units, which eliminates any significant data communication subsystem. The~conducted computer simulations showed
that the proposed control scheme significantly outperforms other control algorithms.

In Section 6, a control strategy to smooth the wind power fluctuation is introduced. The thermostatically controlled loads and batteries are used as the controllers to regulate the actual wind power delivered to the main electrical grid. The objective is to minimize the operating cost of the storage unit while meeting the grid ramp rates requirements and keep the room temperature under the desired level. The simulation results show that the proposed control scheme significantly outperforms other algorithms.

In Section 7, an approximate dynamic programming algorithm is developed to solve the energy control problem in a microgrid network. With the aim to minimize the overall cost to meet the energy demand on the network, the problem is formulated as a high dimensional MDP. The basis function approach is used to approximate the values of Bellman function, or the 'cost-to-go', and the results obtained are compared with a 'short-sighted' strategy and significant improvements can be observed.

It should be pointed out that most of the control strategies introduced are high-level control schemes, and~the basic idea is to determine the optimal actions of the components such as batteries, generators, and TCLs. In future studies, the importance of power electronics interconnect those units and the power system should be investigated. For instance, in~the microgrid environment, a~large amount of distributed generation units (e.g., solar panels, microturbines), storage units, and~non-linear loads will be integrated, and~a network of power inverters connected in parallel will be necessary in order to obtain good power sharing  and stabilize system frequency. In~our case, the~distributed BESS is used, and~multiple batteries connected in parallel should be coordinated and synchronized, which raises concerns on power sharing and frequency. In other words, a low-level control on the power inverters would be necessary for the future, some of the research on this topic can be viewed in \mbox{\cite{fut1,fut2,fut3,fut4}}.

In addition, the optimal sizing of BESS can be a challenging direction for future studies. In order to achieve that, we believe that an economic cost model of the BESS considering the degradation of batteries from actions under different conditions, such as temperature, depth of discharge, and rates of charge/discharge should be constructed. In addition, the~long-term historical data on temperature and solar power should be investigated to determine the charge/discharge patterns over different seasons, and a possible objective function to determine the optimal capacity can be the minimization of power curtailments, battery lifetime consumption, and the unused BESS capacity over a period. Some of the research on this field can be viewed in \mbox{\cite{size1,size2,size3,size4}}.

It can be expected that more electric vehicles(EVs) will be integrated into the existing power grid, which can affect the security and reliability of the grid due to the uncontrolled charging/discharging procedure of EVs. Microgrids supported with appropriate control strategies have the potential to incorporate the intermittent renewable energy sources and EVs through the concept of Vehicle-to-Grid, namely the bilateral power flow between the power banks in EVs and the microgrids \cite{conclusion1}. In this way, EVs can be considered as distributed storage units that can participate in energy management problems. For instance, depending on the electricity price and the demand on microgrids, EV owners can make a profit while parking, while the intermittent renewable energy can be smoothed at a lower cost with sufficient amount of EVs integrated. This can be a topic for further research in the field of microgrids. This can be a challenging topic for future research in microgrids as more uncertainties are involved with these uncontrolled actions of EV battery banks. There are some previous works conducted on this area, see, \mbox{\cite{ev1,ev2,ev3,ev4}}.

Since most of the control problems in this report are formulated as MDP, the dynamic programming based-algorithms are used as the main optimization solvers to optimize the costs related to microgrids.  Therefore, one direction for future research is to apply some other advanced methods for control and state estimation in microgrid systems with renewable generation and BESS, which include H-infinity based robust control \mbox{\cite{hinf1,hinf2,hinf3,hinf4}}, communication constrained control \mbox{\cite{com1,com2,com3,com4,com5}}, and robust Kalman state estimation \mbox{\cite{kalman1,kalman2,kalman3,kalman4,kalman5,kalman6}}.

\bibliographystyle{unsrt}  
\bibliography{references}  



\end{document}